\documentclass[a4paper,fleqn]{cas-sc}

\usepackage[numbers]{natbib}
\usepackage{subcaption}
\usepackage{rotating}
\usepackage{multirow}
\usepackage[ruled,vlined]{algorithm2e} 
\SetArgSty{textnormal}
\usepackage{amsmath}
\DeclareMathOperator{\sign}{sign}
\DeclareMathOperator{\tr}{tr}

    \def\tsc#1{\csdef{#1}{\textsc{\lowercase{#1}}\xspace}}
\tsc{WGM}
\tsc{QE}

\begin{document}
\let\WriteBookmarks\relax
\def\floatpagepagefraction{1}
\def\textpagefraction{.001}

\shorttitle{Two-stage 2D-to-3D reconstruction of realistic microstructures}

\shortauthors{Seibert~et~al.}  

\title [mode = title]{Two-stage 2D-to-3D reconstruction of realistic microstructures: Implementation and numerical validation by effective properties}  

\author[1]{Paul Seibert}[orcid=0000-0002-8774-8462] 
\credit{Conceptualization, Formal analysis, Investigation, Methodology, Software, Validation, Visualization, Writing - original draft, Writing - review and editing} 

\author[1]{Alexander Raßloff}[orcid=0000-0002-9134-4874]
\credit{Conceptualization, Investigation, Methodology, Writing - review and editing} 

\author[1]{Karl A. Kalina}[orcid=0000-0001-6170-4069]
\credit{Conceptualization, Methodology, Visualization, Writing - review and editing} 

\author[2]{Joachim Gussone}[orcid=0000-0002-3036-5188] 
\credit{Resources, Writing - original draft, Writing - review and editing} 

\author[2]{Katrin Bugelnig}[orcid=0000-0003-0932-1751]
\credit{Resources, Writing - review and editing} 

\author[3,4]{Martin Diehl}[orcid=0000-0002-3738-7363]
\credit{Software, Writing - review and editing} 

\author[1,5]{Markus Kästner}[orcid=0000-0003-3358-1545]
\credit{Funding acquisition, Resources, Writing - review and editing} 
\cormark[1]

\affiliation[1]{organization={Institute of Solid Mechanics},
    addressline={TU Dresden}, 
    city={Dresden},
    postcode={01069}, 
    country={Germany}}
\affiliation[2]{organization={Institute of Materials Research, German Aerospace Center (DLR)},
    addressline={Lindner Hoehe}, 
    city={Cologne},
    postcode={51147}, 
    country={Germany}}
\affiliation[3]{organization={Department of Computer Science, KU Leuven},
    addressline={Celestijnenlaan 200A}, 
    city={Leuven},
    postcode={3001}, 
    country={Belgium}}
\affiliation[4]{organization={Department of Materials Engineering, KU Leuven},
    addressline={Kasteelpark Arenberg 44}, 
    city={Leuven},
    postcode={3001}, 
    country={Belgium}}
\affiliation[5]{organization={Dresden Center for Computational Materials Science},
    addressline={TU Dresden}, 
    city={Dresden},
    postcode={01069}, 
    country={Germany}}

\cortext[cor1]{Corresponding author}


\begin{abstract}
  Realistic microscale domains are an essential step towards making modern multiscale simulations more applicable to computational materials engineering.
  For this purpose, 3D computed tomography scans can be very expensive or technically impossible for certain materials, whereas 2D information can be easier obtained.
  Based on a single or three orthogonal 2D slices, the recently proposed differentiable microstructure characterization and reconstruction (DMCR) algorithm is able to reconstruct multiple plausible 3D realizations of the microstructure based on statistical descriptors, i.e., without the need for a training data set.
  Building upon DMCR, this work introduces a highly accurate two-stage reconstruction algorithm that refines the DMCR results under consideration of microstructure descriptors. 
  Furthermore, the 2D-to-3D reconstruction is validated using a real {computed tomography (CT)} scan of a recently developed $\beta$-Ti/TiFe alloy as well as anisotropic "bone-like" spinodoid structures.
  After a detailed discussion of systematic errors in the descriptor space, the reconstructed microstructures are compared to the reference in terms of the numerically obtained effective elastic and plastic properties.
  Together with the free accessibility of the presented algorithms in \emph{MCRpy}, the excellent results in this study motivate interdisciplinary cooperation in applying numerical multiscale simulations for computational materials engineering.\\ 
\end{abstract}

\begin{graphicalabstract}
\includegraphics{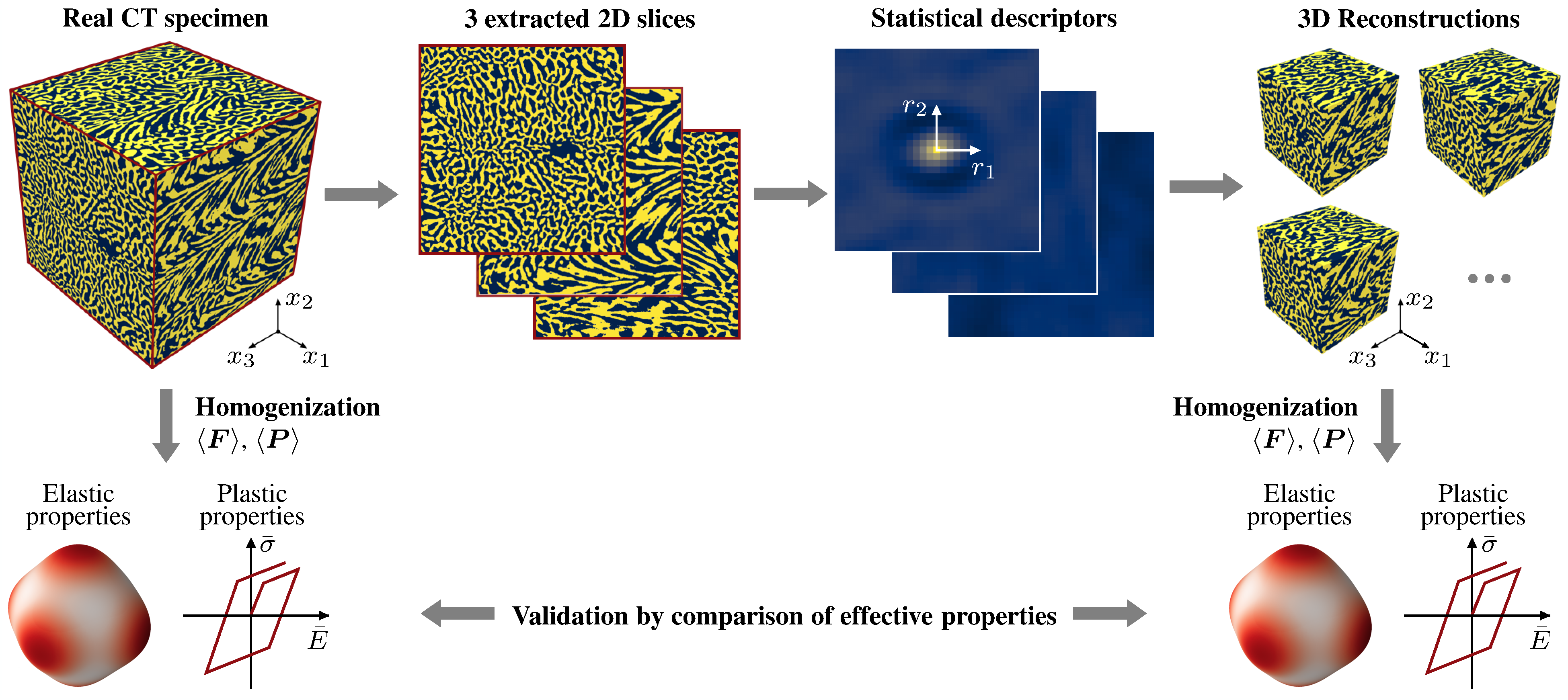}
\end{graphicalabstract} 

\begin{highlights}
\item Two-stage gradient-based algorithm for reconstructing realistic 3D microstructures from 2D slices
\item Validation based on real computed tomography scan of recent Ti-Fe system and synthetic spinodoid structures
\item Error analysis in terms of statistical descriptors as well as simulated anisotropic elastic and plastic effective behavior
\end{highlights}


\maketitle

\section{Introduction}
\label{sec:introduction}
Computational micromechanics has experienced significant progress in the last decade. 
Fast multiscale schemes~\cite{le_computational_2015,li_multiscale_2020,fuhg_model-data-driven_2021-1,masi_multiscale_2022-1,kalina_feann_2022} alleviate the computational burden of {FE${}^2$ (finite element square)}.
Based on the {fast Fourier transform~(FFT)}, efficient approaches have been presented for the homogenization of {the effective behavior of heterogeneous} media~\cite{moulinec_numerical_1998} or for simulation using composite voxels~\cite{kabel_use_2015,schneider_convergence_2015,schneider_review_2021,keshav_fft-based_2022}.
{Furthermore, fast surrogates for the computationally intensive homogenization are given by convolutional neural networks~\cite{wessels_computational_2022,eidel_deep_2022,henkes_deep_2021-1} or deep material networks~\cite{liu_exploring_2019,liu_deep_2019,gajek_fe-dmn_2022} trained on the basis of suitable data sets.}
In summary, extensive research is conducted on the constitutive and multiscale aspects of computational micromechanics, with a focus on high accuracy or performance.
In contrast, a very relevant field that attracts relatively little attention is the generation of realistic 3D volume elements, especially if these are solely based on 2D image data~\cite{bargmann_generation_2018}. 
While special cases like fiber composites have relatively well-defined microscale geometries, many engineering materials are random heterogeneous media.
For these complex materials, the  elastic~\cite{zeghadi_ensemble_2007-1} and plastic~\cite{zeghadi_ensemble_2007} behavior can be very sensitive\footnote{For instance, determining yield surfaces by virtual testing is a relevant and open problem~\cite{geers_multi-scale_2010-1,habraken_analysis_2022} with ongoing developments~\cite{kraska_virtual_2009,zhang_virtual_2016-1,do_nascimento_machine_2022}.} to the exact 3D micro-geometry~\cite{diehl_neighborhood_2016-1,diehl_coupled_2017-1}, which is, however, stochastic and unknown.
Computed tomography (CT) scans are a common strategy for directly obtaining this 3D information~\cite{basista_micro-ct_2017}.
However, compared to microscopy images, CT scans require intricate computer post-processing, where the chosen parameters of filters and thresholds strongly influence the microstructure and the predicted material behavior~\cite{gebhardt_influence_2022}.
Furthermore, for some materials and length scales, CT scans are either very expensive~\cite{gussone_ultrafine_2020} or technically impossible due to a lack of phase contrast.
The alternative, serial sectioning, becomes extremely expensive and time-consuming.
In contrast, obtaining microscopy images of one or few cross-sections can be significantly more feasible and requires fewer impactful parameters to be chosen manually. {These images can serve as a 2D basis for the reconstruction of 3D microstructures.}

{A brief introduction on microstructure reconstruction techniques with a focus on recent techniques for random heterogeneous media is given in the following. For this purpose, a distinction is made between \emph{descriptor-based} and \emph{data-based} methods. The reader is referred to the reviews~\cite{bargmann_generation_2018,bostanabad_computational_2018,sahimi_reconstruction_2021} for a more detailed overview.}

Inspired from the advent of machine learning, {data-based methods} have received much attention lately.
The central idea is to train a generative machine learning model on a data set of microstructures\footnote{Sometimes, a single, very large microstructure is used and split into smaller sections or cut into slices. However, it could be argued that this is more similar to starting with a set of training data than to really requiring a single example only.} and to then use the trained model to sample new realizations of the same structure.
Noticeable examples are generative adversarial networks for 2D-to-3D~\cite{kench_generating_2021} and 3D-to-3D~\cite{henkes_three-dimensional_2022} reconstruction as well as hybrid methods with autoencoders~\cite{zhang_slice--voxel_2021} or transformers~\cite{phan_size-invariant_2022}.
While these methods achieve excellent results throughout a variety of material classes, the disadvantage of data-based methods is the general necessity of a data set.

Descriptor-based methods, in contrast, do not require a training data set.
Instead of training a generative machine learning model by optimizing its weights, the optimization problem is directly carried out in the space of possible microstructures.
For this purpose, the utility of a microstructure is computed by means of a descriptor that quantifies the morphology of the structure.
These descriptors can range from simple volume fractions to high-dimensional $n$-point statistics and are discussed in more detail in Section~\ref{sec:characterization}.
The desired value of the descriptor can be prescribed directly or computed from a reference microstructure.
Here, a \emph{single} microstructure example is sufficient for descriptor-based reconstruction, as opposed to machine learning methods, which usually require a data \emph{set} of structures.
While the latter need data to learn how to constrain their own flexibility, the former harness expert knowledge about microstructure descriptors to reduce the required amount of data to a minimum.
One of the most well-known descriptor-based reconstruction methods is the Yeong-Torquato algorithm~\cite{yeong_reconstructing_1998}, which directly solves the optimization problem using a specially adapted version of a common stochastic optimizer.
While this is very elegant, it becomes computationally challenging at high resolutions and in 3D, where billions of iterations are required for convergence~\cite{adam_efficient_2022}.
A common method of simplifying the optimization problem is to approximate the structure by ellipsoidal~\cite{xu_descriptor-based_2014-1,scheunemann_design_2015} inclusions or by directly incorporating the grain structure of metallic materials~\cite{groeber_dream3d_2014}.
Alternatively, without any approximations on the microstructure morphology, \emph{differentiable} descriptors can be used in order to directly solve the underlying optimization problem using a gradient based optimizer.
This is known as differentiable microstructure characterization and reconstruction (DMCR)~\cite{seibert_reconstructing_2021-1,seibert_descriptor-based_2022} and has been made publicly available by the authors in the \emph{MCRpy} package~\cite{seibert_microstructure_2022}.

Based on this work, the present contribution proposes a two-stage reconstruction algorithm that improves upon DMCR by (i) guaranteeing the correct volume fractions and (ii) producing smoother results without introducing descriptor discrepancies.
With DMCR constituting the first stage, the need for a descriptor-based smoothing procedure is stressed and the implementation of this post-processing step is discussed in detail.
{Furthermore, the aim of this work is to rigorously validate the reconstruction procedure, i.e., to systematically check for the physical plausibility of 3D microstructures reconstructed from one or few 2D slices. Thereby, a special feature of this contribution is to use real 3D CT data of a novel titanium alloy that was developed for additive manufacturing~\cite{gussone_ultrafine_2020} as a reference, which shows a much higher morphological complexity than common synthetic structures.}
For this, a single slice is extracted from the 3D data to mimic the case that only a 2D microscopy image is available.
The reconstruction from this slice is compared to the original 3D data regarding
(i) the microstructure morphology, quantified by statistical descriptors, and
(ii) the simulated and homogenized elastic and plastic properties.
Using the homogenized properties, the same procedure is repeated on a higher length scale for synthetic spinodoid "bone-like" metamaterials~\cite{kumar_inverse-designed_2020} to apply the algorithm in the highly anisotropic regime.
While some previous works on microstructure reconstruction list errors for the Young's modulus~\cite{chandrasekhar_integrating_2021,henkes_three-dimensional_2022}, to the authors' best knowledge, the present work is the first to (i) consider the full anisotropic stiffness tensor, (ii) also determine the plastic response and (iii) validate a reconstruction procedure based on DMCR.
Overall, the present work not only proposes a new microstructure reconstruction workflow, but also aims at demonstrating its range of applicability, allowing the {multiscale modeling and simulation} community to assess its potential.
This creates a basis for a plethora of possible future works that harness both, microstructure reconstruction and numerical homogenization, to advance scale-bridging simulations of complex materials.

{The organization of the paper is as follows:} After an introduction to the underlying methods in Section~\ref{sec:methods}, the experimental and synthetic 3D data used for validation are presented in Section~\ref{sec:experiment}.
The results are presented and discussed in Section~\ref{sec:results} and a conclusion is drawn in Section~\ref{sec:summary}.

The notation {within this work} is as follows: Tensors of first and second order are denoted as bold and italic letters in lower and uppercase, i.e.,~$\boldsymbol{a}$ and $\boldsymbol{A}$, respectively. Furthermore, fourth-order tensors are given by~$\mathbb{A}$. 
{Single and double contractions of  tensors are denoted by $\boldsymbol A \cdot \boldsymbol B$ and $\boldsymbol A : \boldsymbol B$, respectively.}
Arrays, in contrast, are bold but not italic, i.e.,~$\textbf{A}$, and the {dimension} is generally not specified.
The coordinates of a tensor are given by~$[A_{kl}]$, whereas the entries of an array are given by~$A_{i,j,k}$.
This allows to simply express the $i$-th slice in the first dimension of an array as~$\textbf{A}_{i,:,:}$.
More detail on this notation is given in Section~\ref{sec:reconstrucion}. 
The Einstein summation convention is not used.

\section{Methods}
\label{sec:methods}
\begin{figure}[t]
    \centering
    \includegraphics[width=\textwidth]{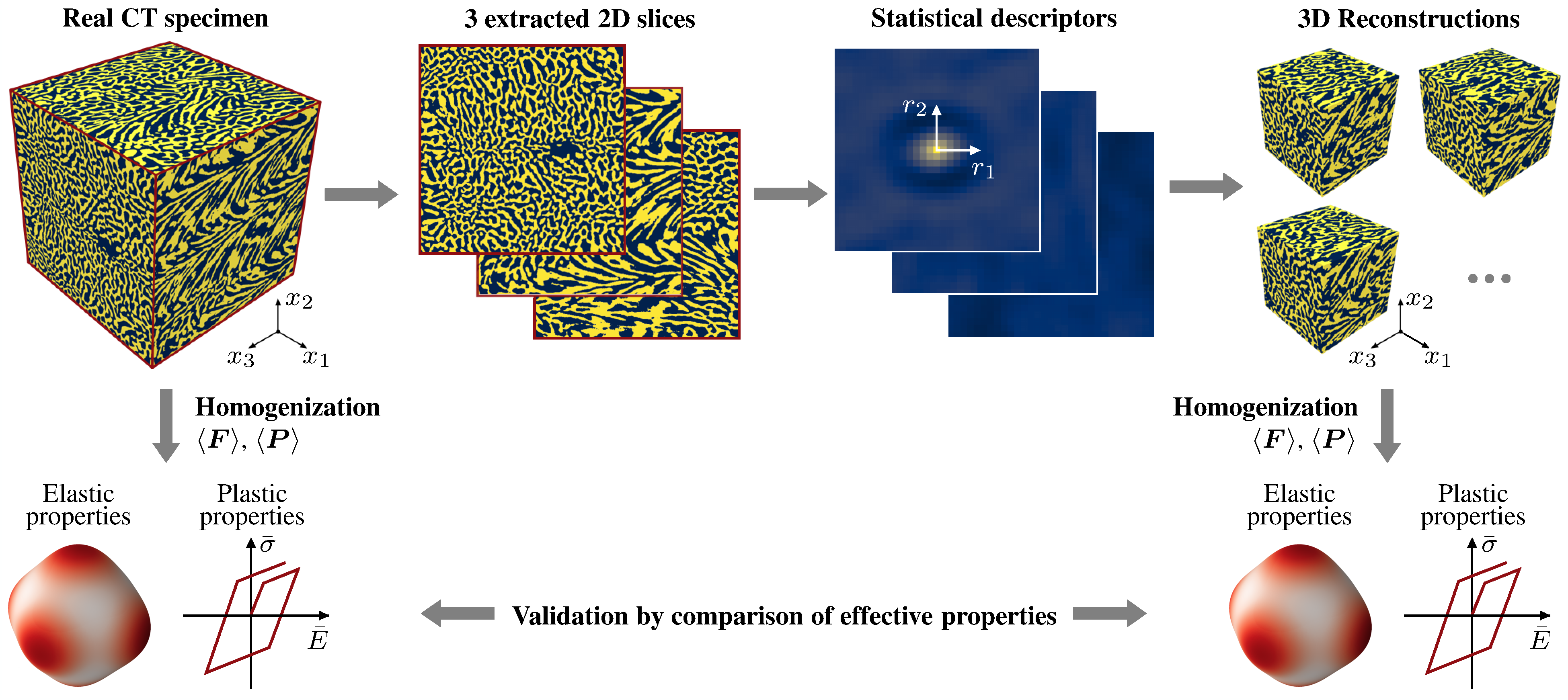}
    \caption{Schematic overview {of the validation procedure}: Three orthogonal slices are extracted from a real 3D specimen to mimic the situation that only 2D data is available. From these slices, multiple synthetic structures are reconstructed based on statistical descriptors and compared to the reference in terms of morphology as well as the effective elastic and plastic response. Herein, $r_1$ and $r_2$ denote the coordinates of the vector $\boldsymbol{r}$ for the computation of the spatial two-point correlation. \label{fig:schema}}
\end{figure}

In this work, 2D-to-3D microstructure reconstruction is numerically validated as shown in Figure~\ref{fig:schema}:
From a 3D reference microstructure, three orthogonal slices are extracted.
These slices are characterized by statistical descriptors as described in Section~\ref{sec:characterization}.
Based on these descriptors, multiple 3D realizations of the structure are reconstructed using a new, two-stage procedure.
With DMCR as a first step being introduced in Section~\ref{sec:reconstrucion}, a descriptor-based post-processing and smoothing is presented in Section~\ref{sec:postprocessing} as a second step.
In Section~\ref{sec:simulation}, the reconstructed structures serve as volume elements for {computational} homogenization in order to obtain the effective elasto-plastic properties.
These values are compared to the material response of the original 3D data in order to quantify the reconstruction error.

\subsection{Microstructure Characterization}
\label{sec:characterization}
The characterization function
\begin{equation}
    f_C : \textbf{M} \mapsto \{ \textbf{D}_i \}_{i=1}^{n_\text{d}}
\end{equation}
maps the pixel- or voxel-based representation of a microstructure~$\textbf{M}$ to a set of~{$n_\text{d}$} different descriptors~$\textbf{D}_i$. 
These descriptors quantify the microstructure morphology in a stationary and translation-invariant manner.
This allows for quantitative comparisons of microstructures, for example the computation of differences.
Volume fractions and pore size distributions are simple examples of such descriptors, however, in this work, more abstract and high-dimensional descriptors are employed. 
Specifically, we use spatial three-point correlations, Gram matrices of the VGG-19 network~\cite{simonyan_very_2015} and the variation descriptor as laid out in the following.

\subsubsection{Spatial Correlations}
Spatial $n$-point correlations are one of the most commonly used descriptors for random heterogeneous media~\cite{yeong_reconstructing_1998,bostanabad_computational_2018}.
An excellent introduction is given in~\cite{jiao_modeling_2007} and briefly summarized in the following. 
Consider an indicator function
\begin{equation}
I^{p}(\boldsymbol{x}) = 
\begin{cases}
1 \qquad & ,\text{ if} \; \boldsymbol{x} \in V_p  \\
0 \qquad & ,\text{ else}
\end{cases}
\label{eqn:indicator_function}
\end{equation} 
of the spatial coordinates~$\boldsymbol{x}$, where $V_p$ is the region occupied by phase $p$. 
Spatial $n$-point correlations stochastically quantify the outcome of simultaneously probing this indicator function at different spatial locations $(\boldsymbol{x}^1, \boldsymbol{x}^2, ..., \boldsymbol{x}^n)$. 
Considering only statistically homogeneous media, the $n$-point correlation function does not depend on the absolute positions, but only their relative displacements $\boldsymbol{r}^{ij} = \boldsymbol{x}^j - \boldsymbol{x}^i$.
Thus, the $n$-point correlation function of phases $p_1, p_2, ..., p_m$ can then be written as\footnote{While these equation serve as a concise and comprehensive definition, FFT-based algorithms are often used in practice to efficiently compute spatial correlations~\cite{brough_materials_2017}.}
\begin{equation}
S_n^{p_1, p_2, ..., p_m}(\boldsymbol{r}^{12}, \boldsymbol{r}^{13}, ..., \boldsymbol{r}^{1n}) = \lim_{n_\text{r} \to \infty} \langle I^{p_1}(\boldsymbol{x}^1)\,I^{p_2}(\boldsymbol{x}^2)\, ... \, I^{p_n}(\boldsymbol{x}^n) \rangle^{n_\text{r}},
\label{eqn:definition}
\end{equation}
where the $\langle ... \rangle^{n_\text{r}}$ denotes the average over an ensemble of $n_\text{r}$ realizations of placing~$\boldsymbol{x}^1$ randomly in the microstructure such that~$\boldsymbol{x}^2\dots \boldsymbol{x}^n$ follow from~$\boldsymbol{x}^1$ via~$(\boldsymbol{r}^{12} \dots \boldsymbol{r}^{1n})$. 
The special case of all phases being equal ($p_1 = p_2 = .. = p_m = p$) is called auto-correlations and is given in a simplified notation as
\begin{equation}
	S_n^{p \to p}(\boldsymbol{r}^{12}, \boldsymbol{r}^{13}, ..., \boldsymbol{r}^{1n}) = S_n^{p, p, ..., p}(\boldsymbol{r}^{12}, \boldsymbol{r}^{13}, ..., \boldsymbol{r}^{1n})\; .
\end{equation}
For example, the spatial two-point autocorrelation function  $S_2^{1 \to 1}(\boldsymbol{r}^{12})$ can be interpreted as the probability of both ends of the vector $\boldsymbol{r}^{12}$ being in phase 1 when randomly placed into the microstructure.
While most applications focus on~$S_2$ only and ignore higher-order correlations for computational efficiency~\cite{brough_materials_2017}, a differentiable formulation of spatial two- and three-point correlations are given in~\cite{seibert_reconstructing_2021-1} and implemented in \emph{MCRpy}~\cite{seibert_microstructure_2022}.
The higher computational effort is compensated by the higher information content and differentiability, which allow for an efficient reconstruction.
In the following, the notation~$\textbf{S}(\textbf{M})$ refers to an array comprising the values of~$S_3^{1 \to 1}(\boldsymbol{r}^{12}, \boldsymbol{r}^{13})$ for the microstructure~$\textbf{M}$, where each array entry corresponds to distinct values of~$\boldsymbol{r}^{12}$ and~$\boldsymbol{r}^{13}$ as described in~\cite{seibert_reconstructing_2021-1}.

\subsubsection{Gram Matrices}
Gram matrices are a recent alternative approach to characterizing microstructures using the internal activations of a pre-trained convolutional neural network (CNN), which are often called {feature maps}~\cite{lubbers_inferring_2017}. 
The activation of the~${p}$-th channel in layer~${n}$ at spatial position~${s}$ is denoted as~{$F_{s,p}^n$}. 
In each layer~${n}$, the activations of all channels at all positions must contain relevant information about the image, because otherwise the classification head of the network would have no basis for its predictions. 
These feature maps provide a richer representation than the image because each layer combines low-level features to obtain condensed, higher-order information. 
To harness this representation, it is rendered approximately translation-invariant by computing the Gram matrix
\begin{equation}
G_{p,r}^n = \sum_{{s}} F_{s,p}^n F_{s,r}^n \; .
\end{equation}
It can be seen that the summation eliminates the spatial index~${s}$, and an analogy between the Gram matrices and spatial three-point correlations of feature maps is drawn in~\cite{seibert_reconstructing_2021-1}.
Gram matrices are used for microstructure classification in~\cite{lubbers_inferring_2017} and for reconstruction in~\cite{li_deep_2018,bostanabad_reconstruction_2020,seibert_descriptor-based_2022,seibert_microstructure_2022}.

\subsubsection{Variation}
The variation was originally introduced to microstructure reconstruction as a denoising parameter~\cite{bostanabad_reconstruction_2020}.
The total variation accumulates the absolute deviations between neighboring pixels over the entire microstructure.
Later, it was shown that the normalized total variation, referred to as variation~$\mathcal{V}$ in the following, can be interpreted as a conventional microstructure descriptor quantifying the amount of phase boundary per unit volume~\cite{seibert_descriptor-based_2022}. 
Incorporating the variation to the loss function drastically reduces noise, especially in 3D reconstruction~\cite{bostanabad_reconstruction_2020,seibert_descriptor-based_2022}.

With the correlations~$\textbf{S}$, the Gram matrices~$\textbf{G}$ and the variation~$\mathcal{V}$ computed on three orthogonal 2D slices, a 3D microstructure is reconstructed as described in the following section.

\subsection{Gradient-based Microstructure Reconstruction}
\label{sec:reconstrucion}
Differentiable microstructure characterization and reconstruction (DMCR) constitutes the first step in generating microstructures from the descriptors in Section~\ref{sec:characterization}.
In contrast to other reconstruction frameworks like the Yeong-Torquato Algorithm~\cite{yeong_reconstructing_1998}, DMCR formulates microstructure reconstruction as a \emph{differentiable} optimization problem, facilitating significant speedups by means of highly efficient, gradient-based optimizers.
DMCR is implemented in the open-source software package \emph{MCRpy}~\cite{seibert_microstructure_2022}, which is used in this work.
While the reader is referred to~\cite{seibert_reconstructing_2021-1,seibert_descriptor-based_2022} for a detailed description to DMCR, a brief summary is given in the following.
For this purpose, the procedure is first introduced in 2D in order to define the 3D process as a composition of parallel and coupled 2D reconstructions.

\subsubsection{DMCR - 2D Reconstruction}
In 2D, microstructure reconstruction is approached by solving an optimization problem
\begin{align}
    \textbf{M}^\text{rec} =&  \underset{\mathbf M \, \in \, \mathcal M^{2\text{D}}}{\arg\min} \; \mathcal{L}\left( \left\{ (\textbf{D}_i(\textbf{M}), \; \textbf{D}_i^\text{des}) \right\}_{i=1}^{n_\text{d}} \right) \; \text{with}  \label{eqn:optgeneral}\\
    \mathcal M^{2\text{D}} :=& \left\{ \textbf m \in \mathbb R^{N_1 \times N_2} \,|\, \forall i \in \{1, 2, \dots, N_1\} , j \in \{1, 2, \dots, N_2\} \,  0 \leq m_{i,j} \leq 1 \right\} \;,
     \label{eqn:spaceM2D}
\end{align}
where the loss~$\mathcal{L}$ quantifies the difference between the desired descriptor~$\textbf{D}^\text{des}$ and its current value~$\textbf{D}(\textbf{M})$.
For example, the present results are achieved with the loss function
\begin{equation}
    \mathcal{L}(\textbf{M}) = \lambda_S || \textbf{S}(\textbf{M}) - \textbf{S}^\text{des} ||_\text{MSE} + \lambda_{G} || \textbf{G}(\textbf{M}) - \textbf{G}^\text{des} ||_\text{MSE} + \lambda_\mathcal{V} || \mathcal{V}(\textbf{M}) - \mathcal{V}^\text{des} ||_\text{MSE}
\label{eqn:loss}
\end{equation}
where~$\lambda_S$, $\lambda_G$ and $\lambda_\mathcal{V}$ are scalar weights for the spatial correlations~$\textbf{S}$, the Gram matrices~$\textbf{G}$ and the variation~$\mathcal{V}$, respectively and~{$||\textbf A||_\text{MSE}$} is the mean squared error norm of {an array~$\textbf A$}, i.e. the average of the squared values of all entries.
In this context, it is worth noting that the search space~$\mathcal{M}^{\text{2D}}$ of DMCR defined in Eq.~\ref{eqn:spaceM2D} not only contains integer-valued microstructures as in the Yeong-Torquato algorithm, but also real-valued "intermediate states".
The reason for this choice is rooted in the gradient-based optimization of Eq.~\ref{eqn:optgeneral}.
In other words, while the Yeong-Torquato algorithm updates the microstructure in every iteration by applying a random mutation, DMCR utilizes the gradient $\partial_{\textbf M} \mathcal L$ of the loss function~$\mathcal{L}$ with respect to the microstructure $\textbf M$ for a more informed modification of the microstructure.
The existence of this gradient is ensured by using only differentiable descriptors to compose the loss function.
In practical application, this requires 
\begin{enumerate}
    \item[\textit{(i)}] the descriptors to be defined not only for integer-valued indicator functions, but also for real-valued "intermediate states",
    \item[\textit{(ii)}] the value of the descriptor not to jump discontinuously as the value of individual pixels in the microstructure gradually change, and
    \item[\textit{(iii)}] in order to avoid plateaus during optimization, the gradient of the descriptor is required to be non-zero for the widest possible range of arguments.
\end{enumerate}
More details on the theoretical foundations of defining differentiable descriptors as well as a practical example for spatial correlations are given in~\cite{seibert_reconstructing_2021-1}.
Despite the constraint of differentiability, the range of possible descriptors to choose from is still very large and currently far from fully exploited~\cite{seibert_microstructure_2022}.
Most importantly, the gradient-based optimizers that can be accessed by DMCR significantly outperform the previously used stochastic methods, reducing the number of required iterations by several orders of magnitude.

\subsubsection{DMCR - 3D Reconstruction}
In 3D, Eq.~\eqref{eqn:loss} is called on all possible 2D slices in all directions. 
To express this, we introduce the notation \mbox{$\textbf{M}_{i,:,:} \in \mathcal M^\text{2D}$} to refer to the $i$-th slice in the $x_1$-direction of the 3D microstructure array~$\textbf M \in \mathcal M^\text{3D}$, where $\mathcal M^\text{3D} \subsetneq \mathbb R^{N_1\times N_2 \times N_3}$ is the space of two-phase 3D microstructures in analogy to Eq.~\eqref{eqn:spaceM2D}.
Similarly,~\mbox{$\textbf{M}_{:,j,:} \in \mathcal M^\text{2D}$} and~$\textbf{M}_{:,:,k} \in \mathcal M^\text{2D}$ denote the~$j$-th and~$k$-th slice in~$x_2$ and~$x_3$, respectively.
Furthermore, for anisotropic structures, the desired descriptor values differ between orthogonal slices, for example~$\textbf{S}^{\text{des,}x_1}\neq\textbf{S}^{\text{des,}x_2}\neq\textbf{S}^{\text{des,}x_3}$.
This is accounted for by introducing different loss functions~$\mathcal{L}^{x_1}$, $\mathcal{L}^{x_2}$ and~$\mathcal{L}^{x_3}$ for different dimensions.
With this notation, 3D reconstruction is expressed as
\begin{equation}
    \textbf{M}^{\text{rec}} = \underset{\textbf{M} \, \in \, \mathcal M^\text{3D}}{\arg\min} \; \sum_i \mathcal{L}^{x_1}(\textbf{M}_{i,:,:}) + \sum_j \mathcal{L}^{x_2}(\textbf{M}_{:,j,:}) + \sum_k \mathcal{L}^{x_3}(\textbf{M}_{:,:,k}) \; .
    \label{eq:slicing}
\end{equation}

\subsubsection{MCRpy Software Tool}
\emph{MCRpy} allows to compose loss functions like Eq.~\eqref{eqn:loss} by choosing the type of loss function and providing a list of descriptors and corresponding weights.
If all descriptors are differentiable, automatic differentiation allows to use gradient-based optimizers such as L-BFGS-B~\cite{byrd_stochastic_2015} for the reconstruction.
Furthermore, a simple multi-grid procedure is implemented that can be applied to any descriptor, where a low-resolution approximation to~$\textbf{M}$ is first computed on a coarse grid and then iteratively refined in a multi-level hierarchical pyramid scheme similar to~\cite{pant_multigrid_2015,karsanina_hierarchical_2018}.
The present work uses the current version of \emph{MCRpy} with the settings given in Table~\ref{tab:reconstructionsettings}.
\begin{table*}[h]
	\centering
	\caption{Settings for the microstructure reconstruction using \emph{MCRpy}~\cite{seibert_microstructure_2022}.}
	\label{tab:reconstructionsettings}
	\begin{tabular}{l | l  }
\toprule
Parameter & Value \\ 
\midrule
Resolution & $128^3$ \\ 
Optimizer & L-BFGS-B \\ 
Multiphase descriptors & False \\ 
Multigrid reconstruction & True \\ 
Iterations per multigrid level & $800$ \\ 
Descriptors & $\textbf{S}, \textbf{G}, \mathcal{V}$ \\ 
Descriptor weights & $0.1, 0.1, 10$ \\ 
Correlation limit & $16$ \\ 
\bottomrule
	\end{tabular}
\end{table*}

\subsection{Microstructure Post-Processing}
\label{sec:postprocessing}
\subsubsection{Motivation for Specialized Algorithm}
Despite the effort of reducing noise in reconstructed structures by means of the total variation~\cite{bostanabad_reconstruction_2020,seibert_descriptor-based_2022}, the reconstructed structures are still noticeably less smooth than the original one.
This motivates the second step of the two-stage procedure, namely a post-processing algorithm that can reduce spurious noise while at the same time keep the descriptors associated with the structure at the desired values.
This is especially relevant when investigating the effective plastic behavior of the reconstructed structures: If a post-processing procedure eliminates sharp corners that are actually observed in the real structure, this reduces the intensity of stress concentrations and consequently increases the effective yield strength.
Because simple, conventional smoothing procedures such as Gaussian filters fail at distinguishing between real and spurious corners and edges, this issue requires further attention.
Although a variety of smoothing algorithms have been developed~\cite{fan_brief_2019} and their applicability to reconstructed microstructures has not been studies thoroughly yet, descriptor-based microstructure reconstruction constitutes a special case as morphological information for distinguishing between real and spurious corners is available in form of the descriptors.

\subsubsection{Proposed Post-Processing}
In order to make use of this information, we propose Algorithm~\ref{alg:post}: 
First, the reconstructed microstructure~$\textbf{M}^\text{rec} \in \mathcal{M}^\text{3D}$ is projected to~$\hat{\textbf{M}} \in \mathcal{\hat{M}}^\text{3D}$ by element-wise rounding, where~$\mathcal{\hat{M}}^\text{3D} \subsetneq \mathcal{{M}}^{3\text{D}}$ is the set of integer-valued 3D microstructures
\begin{equation}
    \mathcal{\hat{M}}^{3\text{D}} := \left\{ \textbf m \in \mathbb Z^{N_1 \times N_2 \times N_3} \,|\, \forall i \in \{1, 2, \dots, N_1\} , j \in \{1, 2, \dots, N_2\} , k \in \{1, 2, \dots, N_3\} \, m_{i,j,k} \in \{0, 1\} \right\} \; .
\end{equation}
Secondly, the structure is adjusted to the correct volume fraction by selecting random pixels from an over-represented phase and swapping them to under-represented phases.
Thirdly, randomly selected pixels from different phases are swapped if, and only if, this reduces the error in terms of the desired descriptor.
Hereby, the error is identical to the loss function~\eqref{eqn:loss} used during gradient-based reconstruction in Section~\ref{sec:reconstrucion}.
Most importantly, in the second and third stage of the algorithm, the probability of selecting a pixels depends on the number of neighboring pixels are of a different phase.
For this purpose, a sparse operator~$P_{i,j,k,l}$ is introduced that maps microstructure entries from a triple index~$(i, j, k)$ to a single index~$l$ as
\begin{equation}
    \hat{{m}}_l' = \sum_{i,j,k} P_{i,j,k,l} \hat{{M}}_{i,j,k} \; .
\end{equation}
With this, the probability of selecting index~$l$ is defined as
\begin{equation}
    p(l) \propto \max \left( 0, \, \sum_{\Tilde{l} \in \mathcal{N}(l)} \left[ 1 - |\hat{{m}}_l' - \hat{{m}}_{\Tilde{l}}'| \right] - c \right)^2 \; ,
    \label{eqn:prob}
\end{equation}
where $\mathcal{N}(l)$ yields the six indices of the non-diagonal neighbors of~$l$ and~$c$ is a hyperparameter which is chosen as~$c=2.9$.
This means that isolated pixels such as spurious noise or valid sharp corners are significantly more likely to be swapped than pixels within a phase cluster.
At the same time, the zero-tolerance in the acceptance criterion ensures that valid corners are not smoothed.
\begin{algorithm}[htb]
\DontPrintSemicolon
\SetAlgoLined
\KwIn{Reconstructed microstructure ${\textbf{M}^{\text{rec}}}$; desired volume fractions $v_\mathrm{f}^\text{des}$; number of iterations $n_\text{iter}^\text{max}$}
 $\hat{\textbf{M}}^{\text{prev}}, \, v_\mathrm{f}^\text{prev} , \, n_\text{iter} \gets \hat{\textbf{M}}^{\text{rec}} , \, -1 , \, 0$  \tcp*{initialize values}
 $\hat{\textbf{M}}^{\text{rec}} \gets $ projection of ${\textbf{M}^{\text{rec}}}$ to $\hat{\textbf{M}}^{\text{rec}} \in \hat{\textbf{M}}^\text{3D}$ \tcp*{first step: projection} 
 \While(\tcp*[f]{second step: fix volume fractions}){$| v_\mathrm{f}(\hat{\textbf{M}}^{\mathrm{rec}}) - v_\mathrm{f}^\mathrm{des} | < | v_\mathrm{f}^\mathrm{prev} - v_\mathrm{f}^\mathrm{des} |$}{
  $l \gets$ sample index of over-represented phase with different-phase neighbors  \tcp*{use Eq.~(\ref{eqn:prob})}
  $v_\mathrm{f}^\text{prev} , \, \hat{\textbf{M}}^{\text{prev}}  \gets v_\mathrm{f}(\hat{\textbf{M}}^{\mathrm{rec}}) , \, \hat{\textbf{M}}^{\text{rec}}$ \tcp*{store variables}
  $\hat{\textbf{M}}^{\text{rec}} \gets$ mutation of $\hat{\textbf{M}}^{\text{rec}}$ with value at location $l$ flipped \;
 }
 $\hat{\textbf{M}}^{\text{rec}} \gets \hat{\textbf{M}}^{\text{prev}}$  \tcp*{use penultimate state due to loop break criterion}
 \While(\tcp*[f]{third step: smooth microstructure}){$n_\mathrm{iter} \leq n_\mathrm{iter}^\mathrm{max}$}{
  $l_0, l_1 \gets$ sample indices of phases 0 and 1 with different-phase neighbors  \tcp*{use Eq.~(\ref{eqn:prob})}
  $\hat{\textbf{M}}^{\text{prev}} \gets \hat{\textbf{M}}^{\text{rec}}$  \tcp*{store variables}
  $\hat{\textbf{M}}^{\text{rec}} \gets$ mutation of $\hat{\textbf{M}}^{\text{rec}}$ with values at locations $l_0$ and $l_1$ swapped \;
  \If(\tcp*[f]{check if solution worsened, use Eq.~(\ref{eqn:loss}) as loss}){$\mathcal{L}(\hat{\textbf{M}}^{\text{prev}}) < \mathcal{L}(\hat{\textbf{M}}^{\text{rec}})$}{
   $\hat{\textbf{M}}^{\text{rec}} \gets \hat{\textbf{M}}^{\text{prev}}$  \tcp*{undo mutation}
   }
  $n_\mathrm{iter} \gets n_\mathrm{iter} + 1 $ \tcp*{increment counter}
 }
\KwOut{Post-processed microstructure $\hat{\textbf{M}}^{\text{rec}}$}
 \caption{Descriptor-based post-processing and smoothing procedure.}
 \label{alg:post}
\end{algorithm}

The procedure is effectively a version of the Yeong-Torquato algorithm~\cite{yeong_reconstructing_1998} with an acceptance criterion as in the Great Deluge~\cite{dueck_new_1993} and a type of different-phase-neighbor (DPN) sampling rule~\cite{zhao_new_2007,pant_stochastic_2014}.
Indeed, the procedure is practically implemented as an optimizer plugin that is added to~\emph{MCRpy} upon acceptance of this work.
The difference to a full-fledged reconstruction algorithm is that we can expect to be close to the solution and only want to minimize noise.
This motivates the zero-tolerance pixel swap criterion and also allows for the number of iterations as low as $\mathcal{O}(10^4)$, as opposed to up to $\mathcal{O}(10^9)$ as in the Yeong-Torquato algorithm~\cite{adam_efficient_2022}.

The settings used in this work are identical to the reconstruction settings in Table~\ref{tab:reconstructionsettings} except that multigrid reconstruction is not applicable and 20{,}000 iterations are used. 
In this context, it is worth noting that in each iteration, the descriptors only need to be computed on slices that are affected by pixel swaps.
Therefore, although more iterations are used for post-processing than for the actual reconstruction, the computational cost per iteration is much lower.

\subsection{Numerical Simulation and Homogenization}
\label{sec:simulation}
The numerical simulations are conducted in \emph{DAMASK}~\cite{roters_damask_2019}, which comprises efficient FFT-based solvers for regular grids. 
This enables the simulations directly on voxel data, which is readily available as the reconstruction output, without introducing discretization errors that would arise from approximating the reconstruction output by, e.g. a tetrahedral mesh.

\subsubsection{Constitutive Model}
\label{subsec:constitutive}
The assumed elasto-plastic constitutive behavior of the individual phases within the simulations is described in the following. 
A more detailed description of this model is given in~\cite{roters_damask_2019}. 
Therein, DAMASK, a toolbox mainly designed for crystal plasticity simulations, is presented. 
Despite the advanced capabilities of DAMASK, the present work uses it merely in a reduced form with a fully phenomenological plasticity model without considering individual grains.

As usual in finite strain inelasticity, the multiplicative decomposition\footnote{Within the original model \cite{roters_damask_2019}, the split $\boldsymbol F = \boldsymbol F^\text{e} \cdot \boldsymbol F^\text{i} \cdot \boldsymbol F^\text{p}$ into elastic, lattice-distorting inelastic and lattice-preserving inelastic parts is applied. Here, $\boldsymbol F^\text{i}$ is set to $\boldsymbol I$. Due to this, Mandel stress in the plastic configuration $\boldsymbol M^\text{p} = (\boldsymbol F^\text{i})^T \cdot \boldsymbol F^\text{i} \cdot \boldsymbol S$  and second Piola-Kirchhoff stress coincide.} $\boldsymbol F = \boldsymbol F^\text{e} \cdot \boldsymbol F^\text{p}$ of the deformation gradient $\boldsymbol F$ into elastic $\boldsymbol F^\text{e}$ and plastic $\boldsymbol F^\text{p}$ parts is assumed, where $\boldsymbol F^\text{p}$ is purely isochoric. Thus, the elastic Green-Lagrange strain tensor $\boldsymbol E^\text{e}$ is given by $\boldsymbol E^\text{e} := \frac{1}{2}\left[ (\boldsymbol F^\text{e})^T \cdot \boldsymbol F^\text{e} - \boldsymbol I \right]$, where $\boldsymbol{I}$ denotes the unit tensor. Assuming a Saint Venant-Kirchhoff model for the stress-strain relation, the second Piola-Kirchhoff stress tensor $\boldsymbol{S}$ is linked to $\boldsymbol E^\text{e}$ by
\begin{equation}
    \boldsymbol{S} = \boldsymbol{\mathbb{C}} : \boldsymbol{E}^\text{e} \; ,
    \label{eq:elasticconst}
\end{equation}
where  $\boldsymbol{\mathbb{C}}$ denotes the fourth-order stiffness tensor. 

Plasticity is approached in a regularized manner by a rate-dependent model.
The evolution of the plastic deformation gradient is given in terms of the plastic velocity gradient 
\begin{equation}
    \boldsymbol L^\text{p} = \frac{\dot \gamma_\text{p}}{3} \frac{\boldsymbol S^\text{dev}}{\|\boldsymbol S^\text{dev}\|_\text{F}} \; .
\end{equation}
by the flow rule $\dot{\boldsymbol F}^p = \boldsymbol L^\text{p} \cdot \boldsymbol F^\text{p}$. For the case of isotropic plasticity and purely isochoric plastic deformation, the plastic strain rate $\dot \gamma_\text{p}$ is modeled by the phenomenological power law
\begin{equation}
    \dot{\gamma}_\text{p} = \dot{\gamma}_0 \left( \sqrt{\frac{3}{2}} \frac{\|\boldsymbol S^\text{dev}\|_\text{F}}{M \xi} \right)^n \; , 
    \label{eq:gamma}
\end{equation}
where~$M=3$ and~$\dot{\gamma}_\text{p}$ depends on the initial strain rate $\dot{\gamma}_0$, the stress exponent $n$, the Frobenius norm $\|(\bullet)\|_\text{F}$ of the deviatoric part $\boldsymbol S^\text{dev} := \boldsymbol S - \frac{1}{3}\boldsymbol I \tr(\boldsymbol S)$ of the second Piola-Kirchhoff stress and the material resistance~$\xi$. Herein,the evolution of~$\xi$ from its initial value~$\xi_0$ towards its final value $\xi_\infty$ is given by
\begin{equation}
    \dot{\xi} = \dot{\gamma}h_0 \left| 1 - \frac{\xi}{\xi_\infty} \right|^a \; \sign \left( 1 - \frac{\xi}{\xi_\infty} \right) \; ,
    \label{eq:xi}
\end{equation}
where $h_0$ denotes the initial hardening and $a$ is a fitting parameter. 
The numerical values of the material properties of each phase are summarized in Table~\ref{tab:materialparameters} in Section~\ref{sec:experiment}.

\subsubsection{Numerical Homogenization}
To determine the effective elastic and plastic properties of the considered heterogeneous materials, a \emph{computational homogenization} framework is applied. Following common practice, the effective deformation gradient~$\bar{\boldsymbol{F}}$ and first Piola-Kirchhoff stress tensor~$\bar{\boldsymbol{P}}$ are defined by
\begin{equation}
    \bar{\boldsymbol{F}} := \langle \boldsymbol{F} \rangle \quad \text{and} \quad \bar{\boldsymbol{P}} := \langle \boldsymbol{P} \rangle \; ,
\end{equation}
where $\langle (\bullet) \rangle$ is the volume averaging operator and $\bar{(\bullet)}$ labels effective quantities in the following. Other effective quantities such as the Green-Lagrange strain $\bar{\boldsymbol{E}} := \frac{1}{2}\left[ \bar{\boldsymbol F}^T \cdot \bar{\boldsymbol F} - \boldsymbol I \right]$ and~$\bar{\boldsymbol{S}}$ cannot be obtained directly by averaging the local fields, but need to be computed from~$\bar{\boldsymbol{F}}$ and~$\bar{\boldsymbol{P}}$. To fulfill the Hill-Mandel condition, periodic boundary conditions are applied.

The effective elastic properties of the material, i.e., the case that $\boldsymbol F^\text{p} = \boldsymbol I$, are given by means of the full stiffness tensor~$\bar{\mathbb{C}}$.
As a simple way to determine~$\bar{\mathbb{C}}$, six load cases are chosen such that only one or two entries of~$\bar{\boldsymbol{E}}$ are non-zero.
This is achieved by prescribing
\begin{equation}
    [\bar{{E}}_{kl}] = \dfrac{1}{2}\begin{bmatrix}
    \bar \lambda^2 - 1 & 0 & 0  \\
    0 & 0 & 0 \\
    0 & 0 & 0 \\
    \end{bmatrix} , \quad \text{i.e.,} \quad 
    [\bar{{F}}_{kl}] = \begin{bmatrix}
    \bar \lambda & 0 & 0 \\
    0 & 1 & 0 \\
    0 & 0 & 1 \\
    \end{bmatrix} \; ,
\end{equation}
for tension and 
\begin{equation}
    [\bar{{E}}_{kl}] = \dfrac{1}{2}\begin{bmatrix}
    0 & \bar \gamma / 2 & 0  \\
    \bar \gamma / 2 & 0 & 0 \\
    0 & 0 & 0 \\
    \end{bmatrix} , \quad \text{i.e.,} \quad 
    [\bar{{F}}_{kl}] = \begin{bmatrix}
    \bar \gamma / (2\delta) & \delta & 0 \\
    \delta & \bar \gamma / (2\delta) & 0 \\
    0 & 0 & 1 \\
    \end{bmatrix} \; ,
\end{equation} 
for shear, where
\begin{equation}
    \delta = \sqrt{\dfrac{1}{2}\left( 1 - \sqrt{1 - \bar \gamma^2} \right)} \, .
\end{equation}
Herein, $\bar \gamma$ denotes the shear and $\bar \lambda$ is the stretch. 
Analogous load cases are chosen for tension and shear in the other two directions.
It follows that
\begin{equation}
    \bar{{C}}_{ij11} = \dfrac{\bar{T}_{ij}^{11}}{\bar{E}_{11}^{11}} \qquad
    \bar{{C}}_{ij22} = \dfrac{\bar{T}_{ij}^{22}}{\bar{E}_{22}^{22}} \qquad
    \bar{{C}}_{ij33} = \dfrac{\bar{T}_{ij}^{33}}{\bar{E}_{33}^{33}}  
\end{equation}
and
\begin{equation}
    \bar{{C}}_{ij23} = \dfrac{\bar{T}_{ij}^{23}}{2\bar{E}_{23}^{23}} \qquad
    \bar{{C}}_{ij13} = \dfrac{\bar{T}_{ij}^{13}}{2\bar{E}_{13}^{13}} \qquad
    \bar{{C}}_{ij12} = \dfrac{\bar{T}_{ij}^{12}}{2\bar{E}_{12}^{12}} \; , 
\end{equation}
where the subscript denotes the tensor component and the superscript denotes the load case.

Furthermore, in order to characterize the effective plastic properties of the considered materials, stress-deformation states, e.g., 
\begin{equation}
    [\bar F_{kl}] = \begin{bmatrix}
    \bar F_{11} & 0 & 0 \\
    0 & - & 0 \\
    0 & 0 & - \\
\end{bmatrix}
\label{eq:uniax}
\end{equation}
are prescribed within the computational homogenization. Analogous states are applied into the $x_2$- and $x_3$-directions.

\subsubsection{Visualization of Effective Stiffness}
\begin{figure}[t]
	\centering
	\subfloat[Isotropic]{\includegraphics[height=0.25\textwidth]{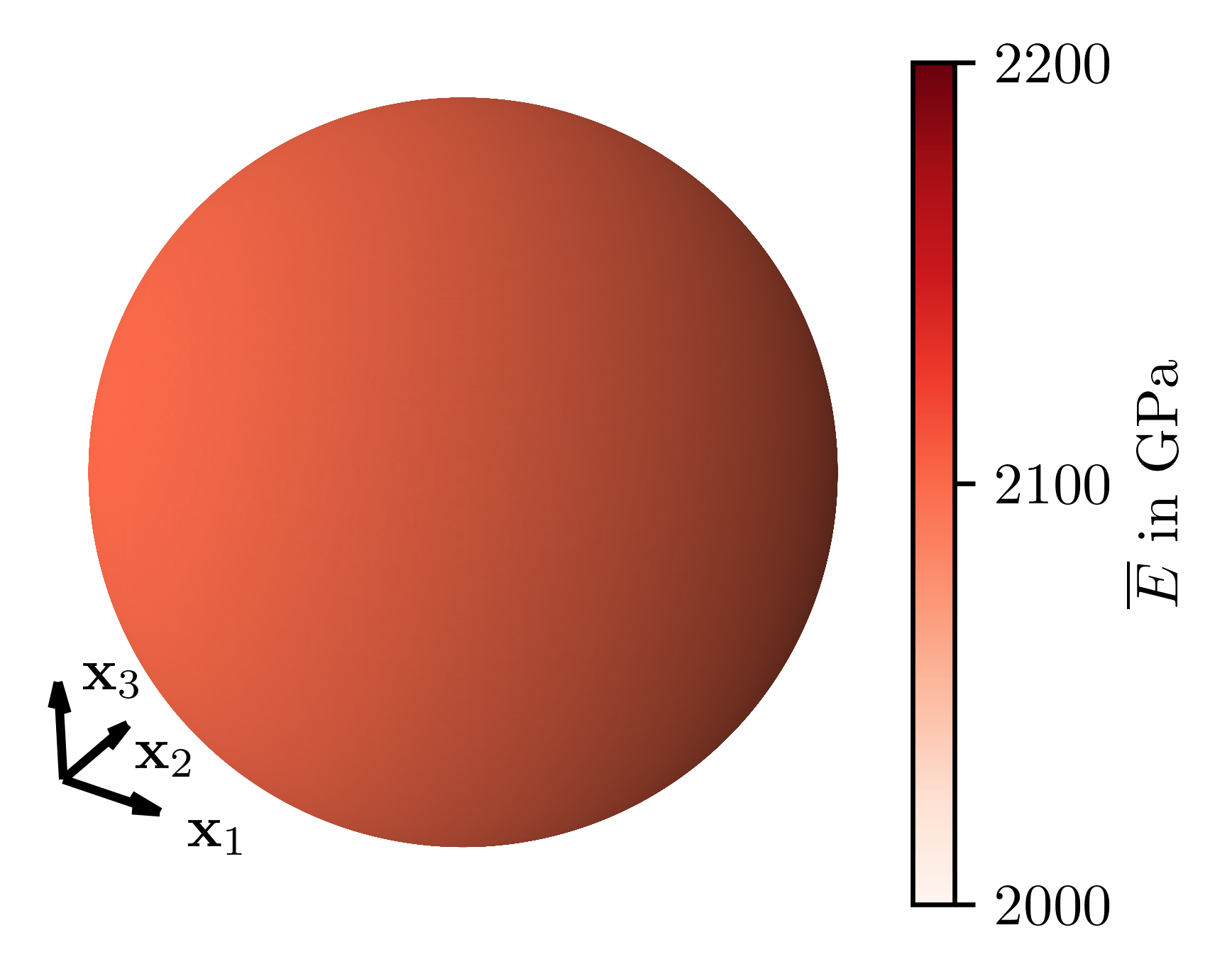}}
	\hfill
	\subfloat[Transversally isotropic]{\includegraphics[height=0.25\textwidth]{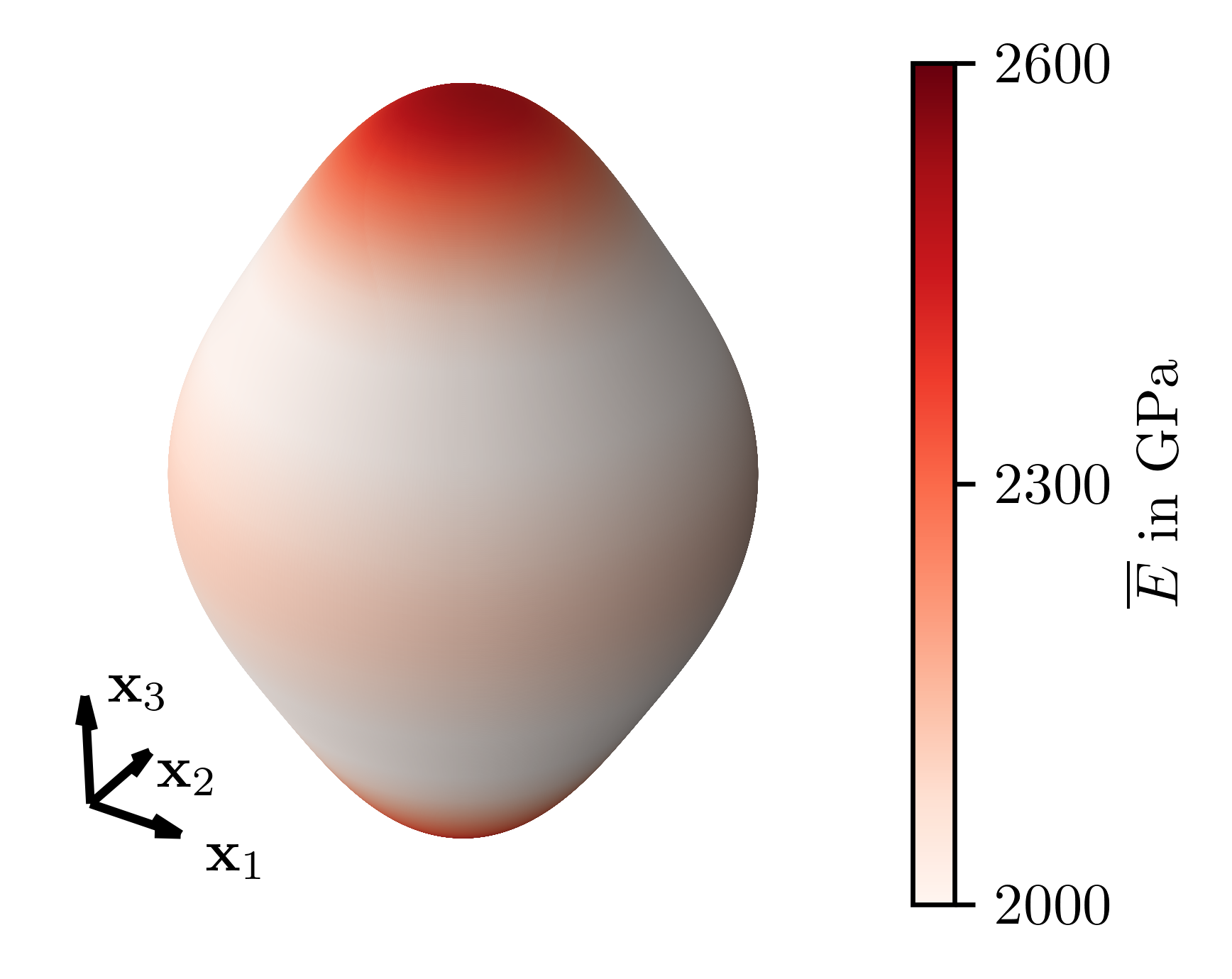}}
	\hfill
	\subfloat[Cubic]{\includegraphics[height=0.25\textwidth]{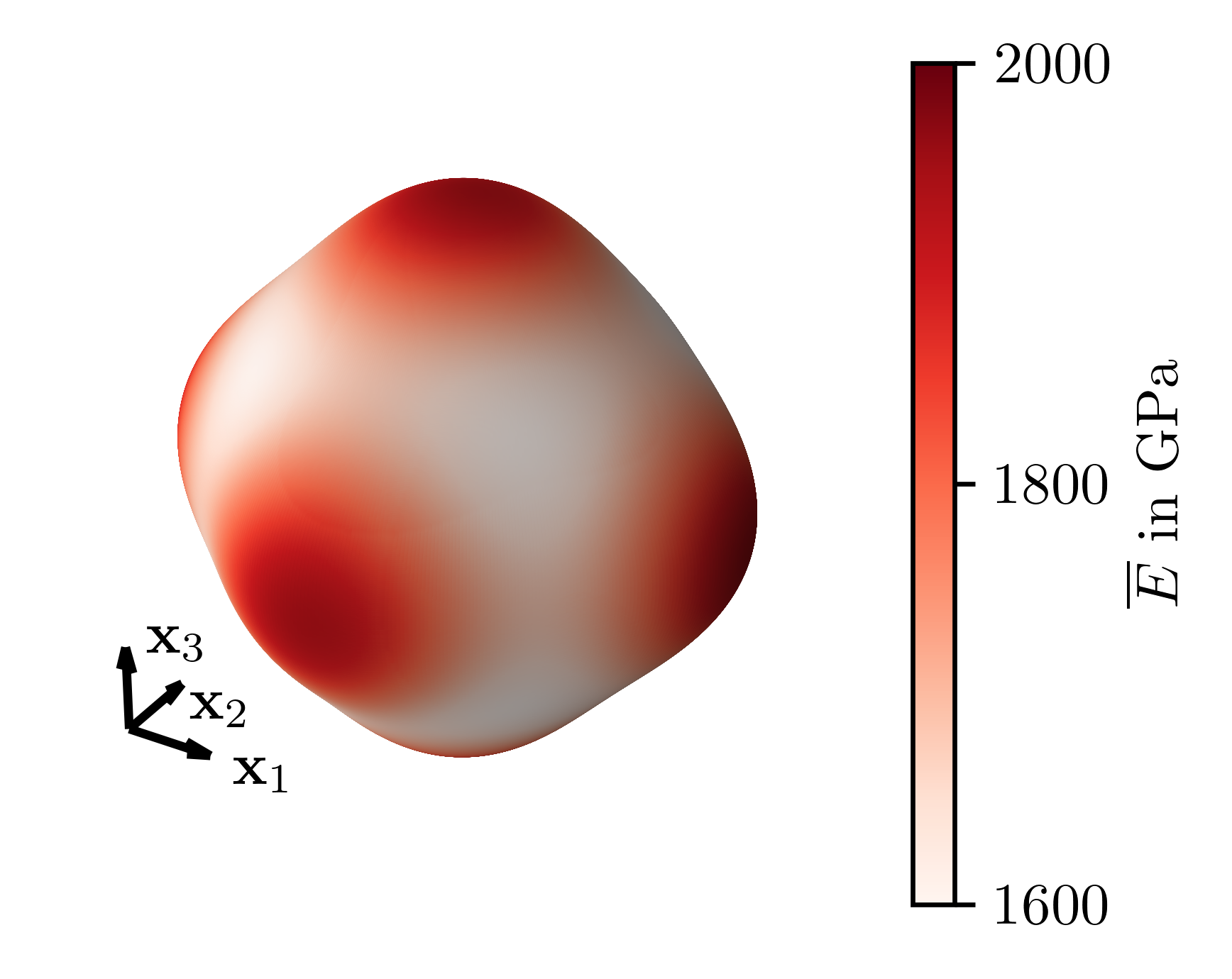}}
	\caption{Exemplary elastic surface plots for an isotropic (a), a transversally isotropic (b) and a cubic (c) material model. \label{fig:elasticsurfaceexamples}}
\end{figure}

Instead of numerically listing all components of~$\bar{\mathbb{C}}$, elastic surface plots are used for visualization~\cite{bohlke_graphical_2001,nordmann_visualising_2018}.
As shown in Figure~\ref{fig:elasticsurfaceexamples} for isotropic, transversally isotropic and cubic material behavior, for each orientation $(\varphi, \theta)$ in a spherical coordinate system\footnote{Herein, $\varphi \; \in \; [0, \; 2 \pi]$ quantifies the azimuthal angle, i.e., the rotation around the $\textbf{z}$-axis, whereas $\theta \; \in \; [0, \; \pi]$ is the polar angle.}, the radius of the elastic surface is given by the directional Young's modulus $\bar{E}(\varphi, \theta)$ that would be measured under tension in this orientation.
Using the effective stiffness tensor in Voigt notation~$\bar{\textbf{C}}^\text{V}$ , it can be extracted from the transformed compliance matrix
\begin{equation}
\bar{\textbf{N}}' = \left( 
\textbf{Q} \cdot \bar{\textbf{C}}^\text{V} \cdot \textbf{Q}^\text{T}
\right)^{-1}
\end{equation}
as
\begin{equation}
\bar{E}(\varphi, \theta) = \dfrac{1}{\bar{N}_{11}'} \quad ,
\end{equation}
where
\begin{equation}
    [\textbf{Q}]_{kl} = 
    \begin{bmatrix}
    b_{11} b_{11} & b_{12} b_{12} & b_{13} b_{13} & 2 b_{12} b_{13} & 2 b_{11} b_{13} & 2 b_{11} b_{12} \\
    b_{21} b_{21} & b_{22} b_{22} & b_{23} b_{23} & 2 b_{22} b_{23} & 2 b_{21} b_{23} & 2 b_{21} b_{22} \\
    b_{31} b_{31} & b_{32} b_{32} & b_{33} b_{33} & 2 b_{32} b_{33} & 2 b_{31} b_{33} & 2 b_{31} b_{32} \\
    b_{21} b_{31} & b_{22} b_{32} & b_{23} b_{33} & b_{22} b_{33} + b_{32} b_{23} & b_{21} b_{33} + b_{31} b_{23} & b_{21} b_{32} + b_{31} b_{22} \\
    b_{11} b_{31} & b_{12} b_{32} & b_{13} b_{33} & b_{12} b_{33} + b_{32} b_{13} & b_{11} b_{33} + b_{31} b_{13} & b_{11} b_{32} + b_{31} b_{12} \\
    b_{11} b_{21} & b_{12} b_{22} & b_{13} b_{23} & b_{12} b_{23} + b_{22} b_{13} & b_{11} b_{23} + b_{21} b_{13} & b_{11} b_{22} + b_{21} b_{12} \\
    \end{bmatrix}
\end{equation}
with
\begin{equation}
    [\textbf{b}]_{kl} = \begin{bmatrix}
    \sin{\theta} \cos{\varphi} & \sin{\theta} \sin{\varphi} & \cos{\theta} \\
    \cos{\theta} \cos{\varphi} & \cos{\theta} \sin{\varphi} & - \sin{\theta} \\
    -\sin{\varphi} & \cos{\varphi} & 0 \\
    \end{bmatrix}
    \quad.
\end{equation}

\subsubsection{Plastic Properties}
The effective plastic properties investigated in this work are limited to the computation of the effective yield strength in three directions, since the determination of the full yield surface would exceed the scope of this work. 
For this purpose, uniaxial tension according to Eq.~\eqref{eq:uniax} is applied and the yield strength~$\bar{\sigma}_\text{y}$ is defined by a plastic strain of $\bar{E}^\text{p} = 0.2 \%$, where~$\bar{E}^\text{p}$ is obtained from~$\bar{E}$ by subtracting the elastic part. 

\section{Data for Validation}
\label{sec:experiment}
Motivated by the advent of additive manufacturing and metamaterials and the complex resulting structure-property linkages, a novel titanium alloy and a "bone-like" spinodoid structure are considered in Sections~\ref{sec:experimenttitanium} and~\ref{sec:experimentspinodal}, respectively.

\subsection{Titanium Alloy on Microscale}
\label{sec:experimenttitanium}
\begin{figure}[t]
	\centering
	\subfloat[Real Ti-Fe scan~\cite{gussone_ultrafine_2020}~(10 {\textmu}m)]{\includegraphics[width=0.3\textwidth]{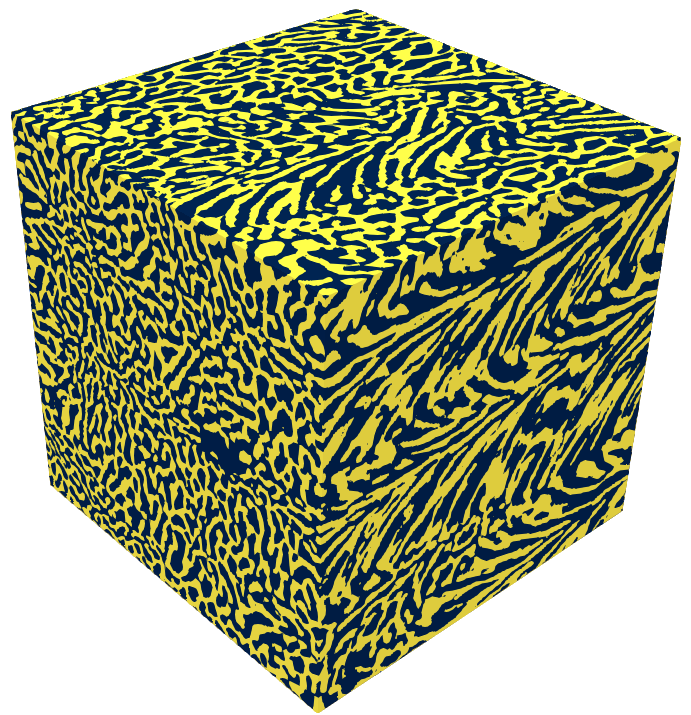}}
	\hfill
	\subfloat[Synthetic columnar structure~(10 cm)]{\includegraphics[width=0.3\textwidth]{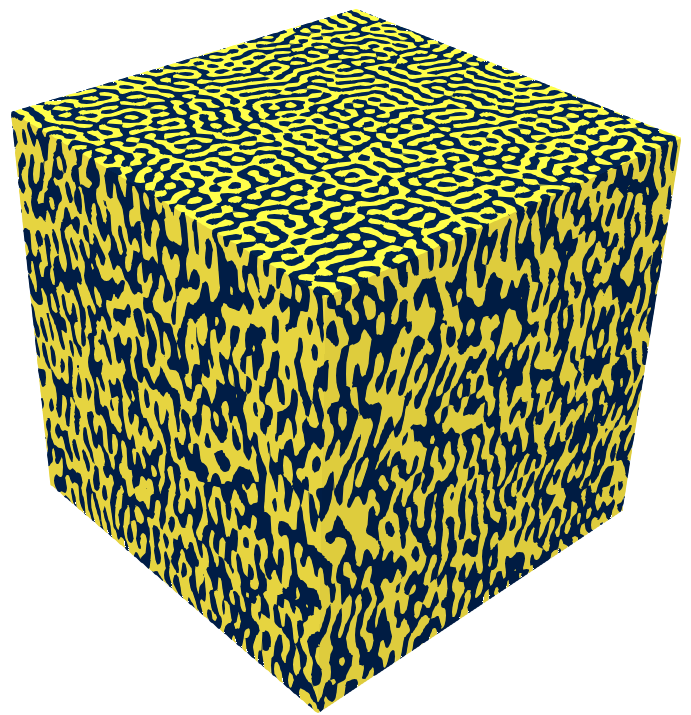}}
	\hfill
	\subfloat[Synthetic lamellar structure~(10 cm)]{\includegraphics[width=0.3\textwidth]{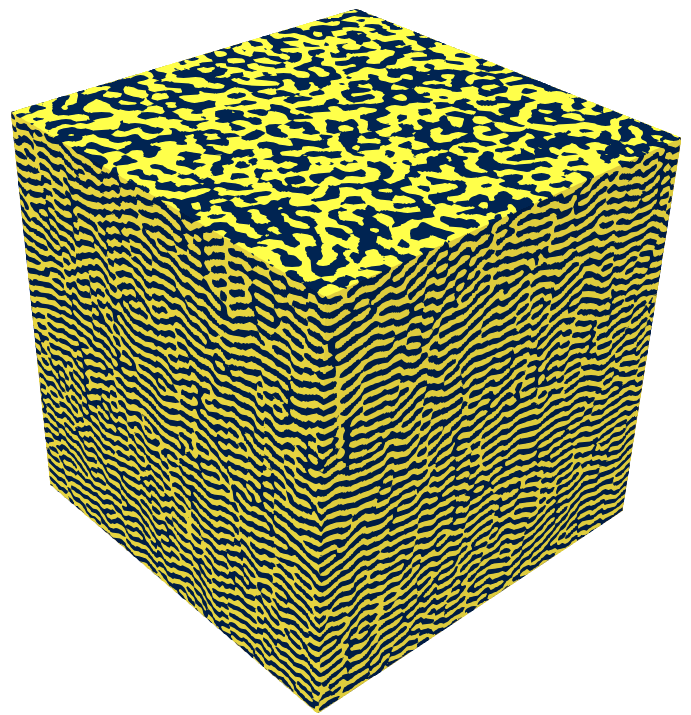}}
	\caption{Original 3D structures used for the validation with the length of the samples given in parentheses. The Ti-Fe alloy (a) is a real CT scan in the microscale, whereas (b) and~(c) are synthetically generated mesoscale structures.\label{fig:ct}}
\end{figure}

On the microscale, a new alloy developed for laser powder bed fusion is used as a representative microstructure. 
The alloy and experimental methods are described in detail in~\cite{gussone_ultrafine_2020} and a brief summary is given in the following.
The titanium-rich binary eutectic material (Ti-32.5Fe, wt.$\%$) was produced by in-situ alloying using elemental powder blends during high temperature laser powder bed fusion at a temperature of $\theta \approx 600$ °C in an SLM Solutions 280 HL machine. 
The near-field ptychographic X-ray tomography experiments were performed with micrometer-sized cylinders (diameter $d = 18$ {\textmu}m and height $h = 40$ {\textmu}m, extracted from the material by focused ion beam milling) at the ID16A nano-imaging beamline of the European Synchrotron Radiation Facility (ESRF)~\cite{noauthor_id16a_nodate}.
The processing of the tomographic volume of each sample was carried out in three steps: 
\emph{(i)} phase retrieval of the near-field ptychographic imaging scan at each tomographic angle, 
\emph{(ii)} pre-processing of the retrieved tomographic projections and 
\emph{(iii)} tomographic reconstruction. 
The segmentation of the two major microstructural constituents (hereafter called $\beta$-Ti and TiFe) was carried out after pre-processing the reconstructed volumes using bandpass filters available in Fiji~\cite{schneider_nih_2012} and Avizo Fire $9.5$ to enhance their quality.
As a reference structure for the present work, a statistically homogeneous cuboid of length~$l = 10$ {\textmu}m is extracted from the CT scan and shown in Figure~\ref{fig:ct}.

The material parameters of the individual phases for the numerical simulations are taken from the literature~\cite{schlieter_anisotropic_2011,zhu_first-principles_2012,zhu_ab_2014} and are summarized in Table~\ref{tab:materialparameters}.
\begin{table*}[h]
	\centering
	\caption{Material parameters for the micro- and mesoscale simulations.}
	\label{tab:materialparameters}
	\begin{tabular}{c | c  c | c  c }
\toprule
\multirow{2}{*}{Parameter} & \multicolumn{2}{c}{Microscale} & \multicolumn{2}{c}{Mesoscale} \\ 
 & $\beta$-Ti & TiFe & Ti alloy & Void\\ 
\midrule
$\mathbb{C}_{1111}$ in GPa & 174.6 & 298.0 & 218.3 & 0.2 \\ 
$\mathbb{C}_{1122}$ in GPa & 82.2 & 135.8 & 97.9 & 0.0 \\  
$\mathbb{C}_{1123}$ in GPa & 46.2 & 81.1 & 60.2 & 0.1 \\ 
$\xi_0$ in MPa & 535.1 & 572.0 & 530 & -\\  
$\xi_\infty$ in MPa & 1426.0 & 1525.6 & 1400.0 & -\\  
$h_0$ in MPa & 713.3 & 762 & 700 & -\\ 
$\dot{\gamma}_0$ & 0.001 & 0.001 & 0.001 & -\\
n & 20 & 20 & 20 & -\\
a & 2 & 2 & 2 & -\\ 
\bottomrule
	\end{tabular}
\end{table*}

\subsection{Spinodoid Structure on Mesoscale}
\label{sec:experimentspinodal}
On the mesoscale, "bone-like" spinodoid structures are reconstructed and homogenized using the effective material parameters from the microscale to demonstrate the ability to accurately capture highly anisotropic stiffness tensors.
Spinodoid structures are very resilient due to the absence of notches~\cite{hsieh_mechanical_2019} and have received much attention lately in inverse design of anisotropic stiffness~\cite{kumar_inverse-designed_2020} in combination with topology optimization~\cite{zheng_data-driven_2021}.
Potential applications are the design of synthetic bones with tunable stiffness or the general design of resilient metamaterials.

In the absence of data, the reference structures do not stem from a CT scan, but are generated from \emph{GIBBON}~\cite{moerman_gibbon_2018}.
Based on the work of Kumar~et~al.~\cite{kumar_inverse-designed_2020}, \emph{GIBBON} allows to generate spinodoid structures from a low-dimensional parametrization using Gaussian random fields.
For the validation of the reconstruction, we use this parametrization to generate two reference structures and pretend they might stem, e.g., from a bone~\cite{kumar_inverse-designed_2020}.
As shown in Figure~\ref{fig:ct}, a columnar and a lamellar structure are considered in this work, both with a volume fraction of $50\,\%$.

The properties of the bulk material are taken from the homogenized response of the microscale domain, which is presented in Section~\ref{sec:results}.
While the elastic properties can be obtained directly, the plastic properties~$\xi_0$, $\xi_\infty$ and~$h_0$ are estimated from the effective yield strength by a rule of proportion.
Due to the inability of the chosen numerical solver to represent voids, a purely elastic material model is chosen with a phase contrast of approximately $1000$ in the stiffness.
The parameters of both material models are summarized in Table~\ref{tab:materialparameters}.

\section{Results and Discussion}
\label{sec:results}
Section~\ref{sec:reconstructionresults} presents the result of the microstructure reconstruction.
After a qualitative and quantitative analysis of the statistical descriptors of the reconstructed structure in Section~\ref{sec:resultsdescriptors}, a validation of the effective properties is carried out for the microscale alloy and the mesoscale metamaterial in Sections~\ref{sec:resultsalloy} and~\ref{sec:resultsspinodal}, respectively.

\subsection{Reconstruction results}
\label{sec:reconstructionresults}
\begin{figure}[t]
    \centering
    \includegraphics[width=\textwidth]{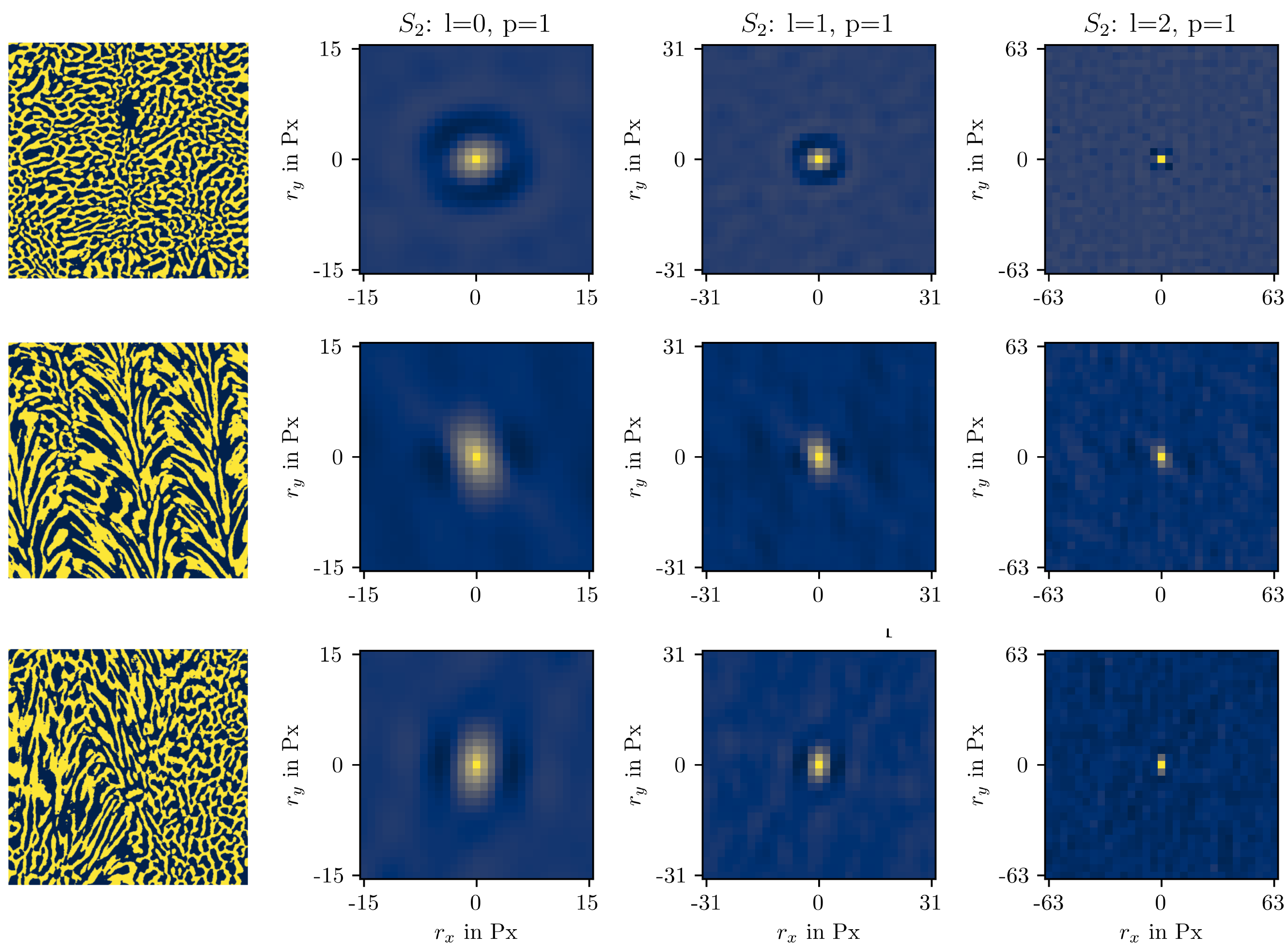}
    \caption{Schematic visualization of the computation of statistical descriptors from the three orthogonal sections extracted from the 3D reference in Figure~\ref{fig:ct}. In this example, the spatial two-point autocorrelation of phase~$p=1$ is shown. These descriptors are used in this work and complemented by higher-order correlations, Gram matrices and the variation. From left to right, it can be seen that short-distance correlations are computed with the highest accuracy~(multigrid level $l=0$), whereas longer-range correlations are computed on down-sampled versions~($l=1,2$) of the structure as described in~\cite{seibert_reconstructing_2021-1}.\label{fig:characterization}}
\end{figure}

For each of the original structures in Figure~\ref{fig:ct}, three orthogonal slices are extracted, down-sampled\footnote{This is done to allow for the reconstructed microstructures to "only" have a resolution of $128^3$ voxels and still be large enough to adequately represent the morphology. Although resolutions of up to $512^3$ voxels have already been reconstructed with \emph{MCRpy}~\cite{seibert_descriptor-based_2022}, this currently requires amounts of computational resources that we find impractical for the large number of structures considered in this work.} to $256 \times 256$ pixels and used for the computation of microstructure descriptors as shown schematically for the Ti-Fe structure in Figure~\ref{fig:characterization}.
As described in Section~\ref{sec:characterization}, the spatial three-point correlations, the Gram matrices and the variation are used as microstructure descriptors.

\begin{figure}[t]
	\centering
	\subfloat[Unsmoothed Ti-Fe]{\includegraphics[width=0.32\textwidth]{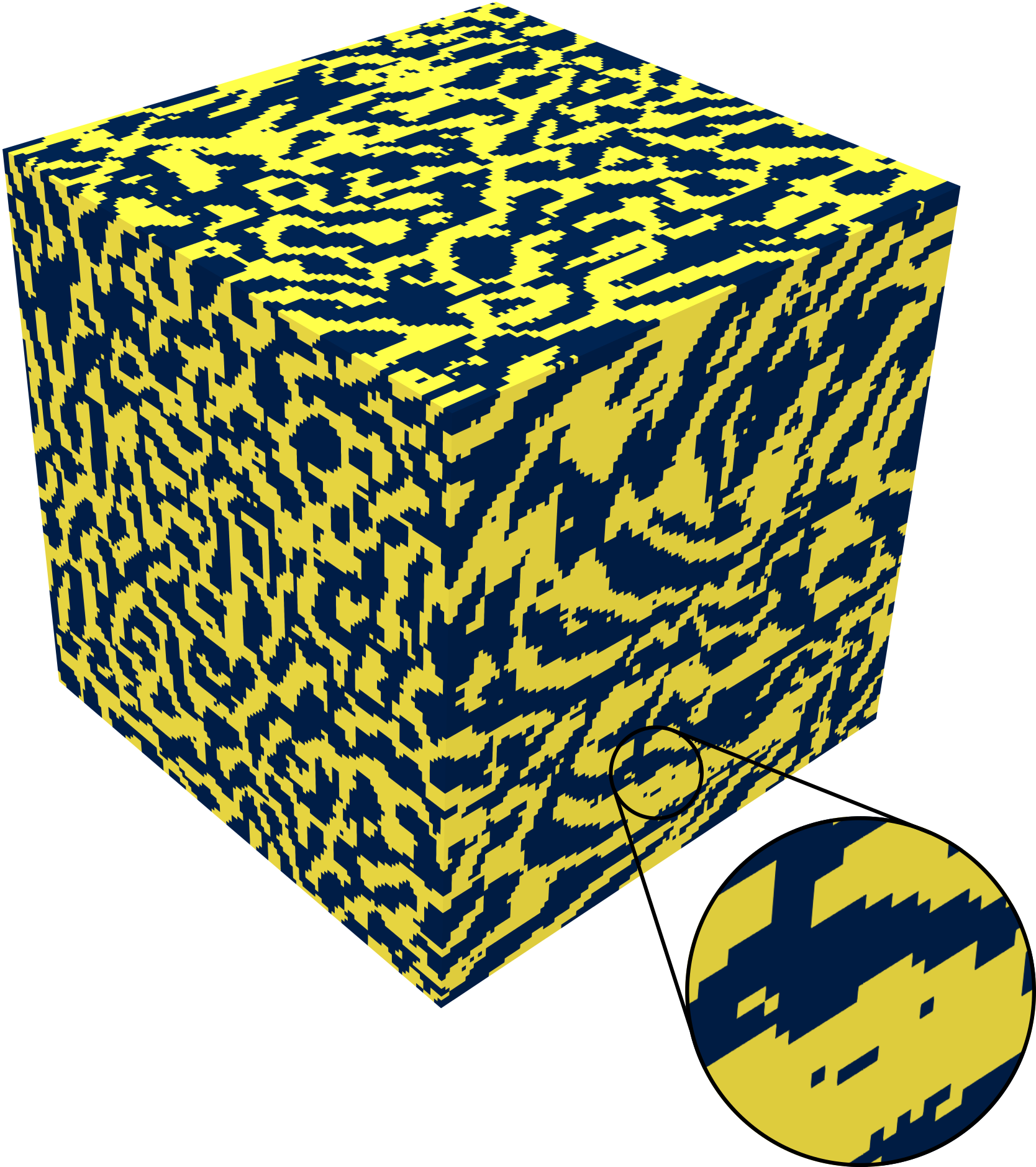}}
	\hfill
	\subfloat[Unsmoothed columnar]{\includegraphics[width=0.32\textwidth]{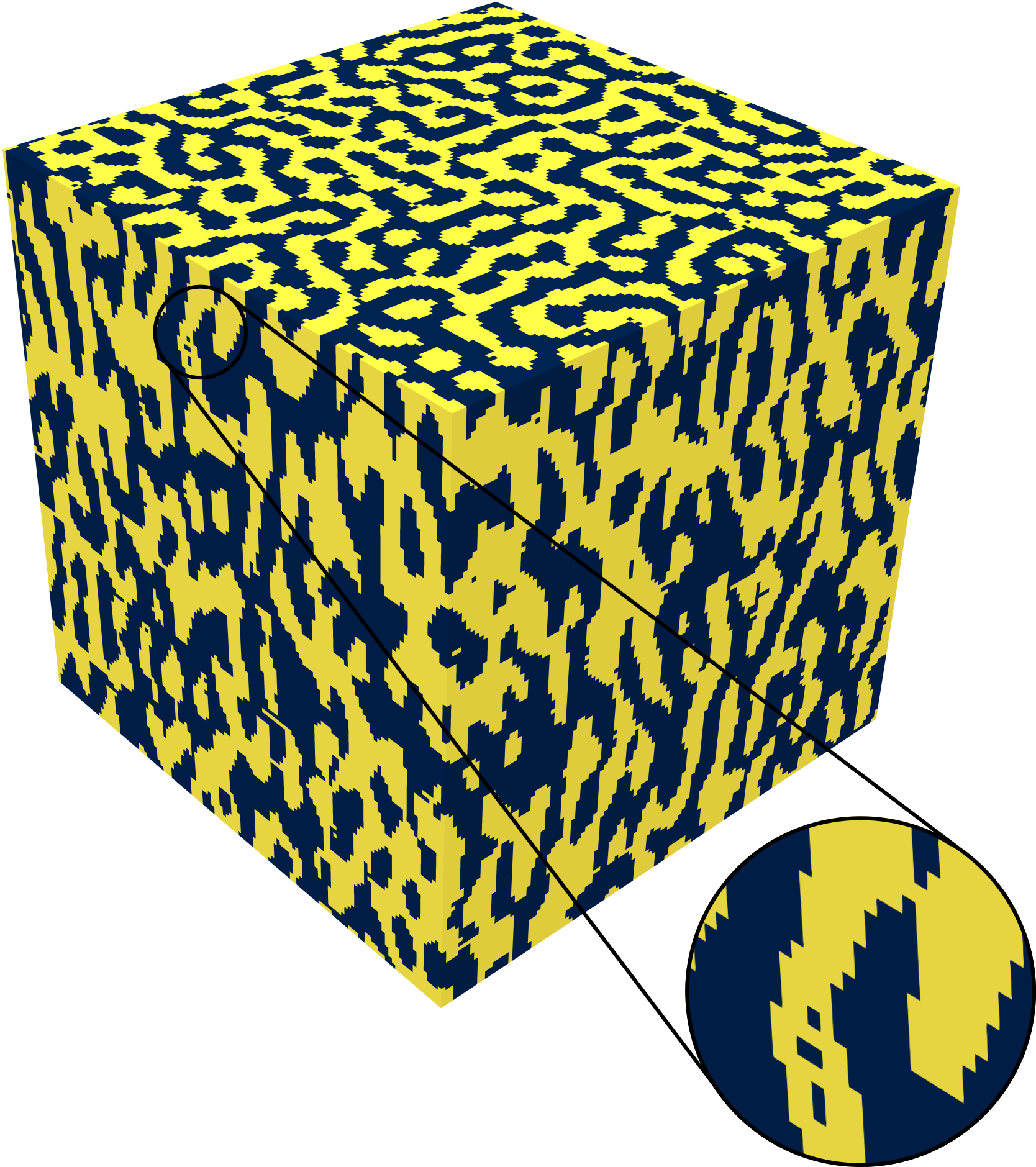}} 
	\hfill
	\subfloat[Unsmoothed lamellar]{\includegraphics[width=0.32\textwidth]{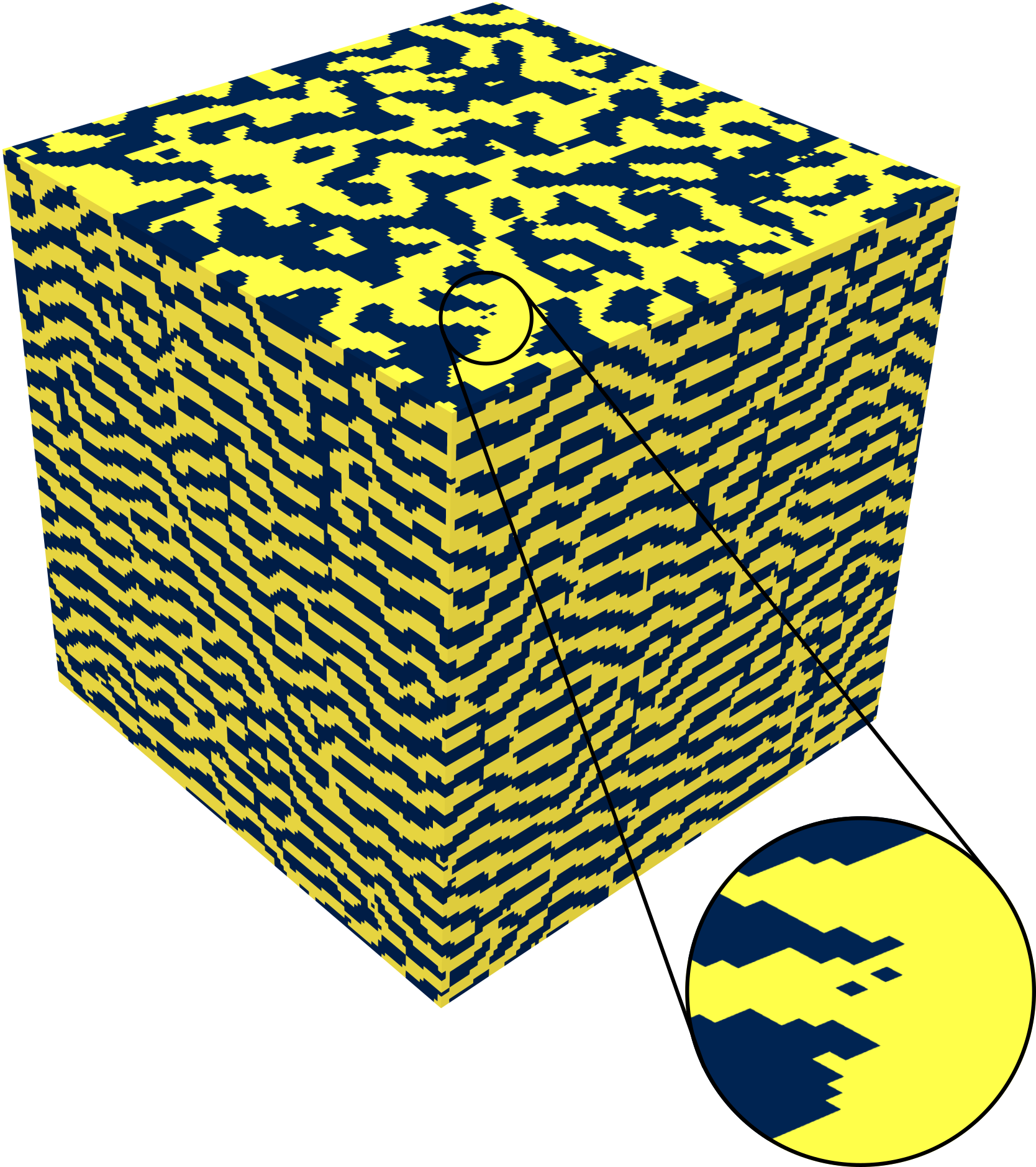}} 
	\vfill
	\subfloat[Smoothed Ti-Fe]{\includegraphics[width=0.32\textwidth]{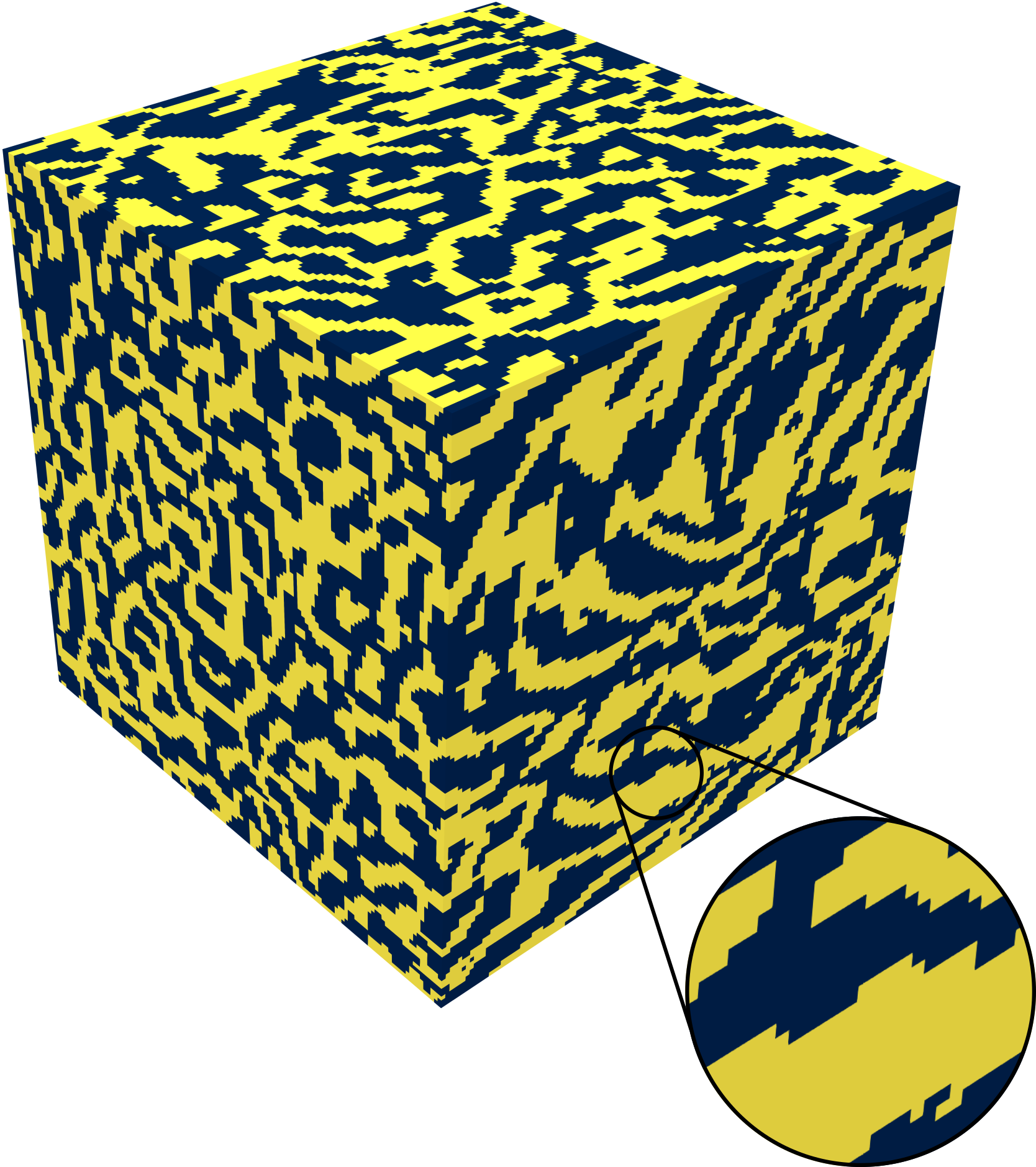}}
	\hfill
	\subfloat[Smoothed columnar]{\includegraphics[width=0.32\textwidth]{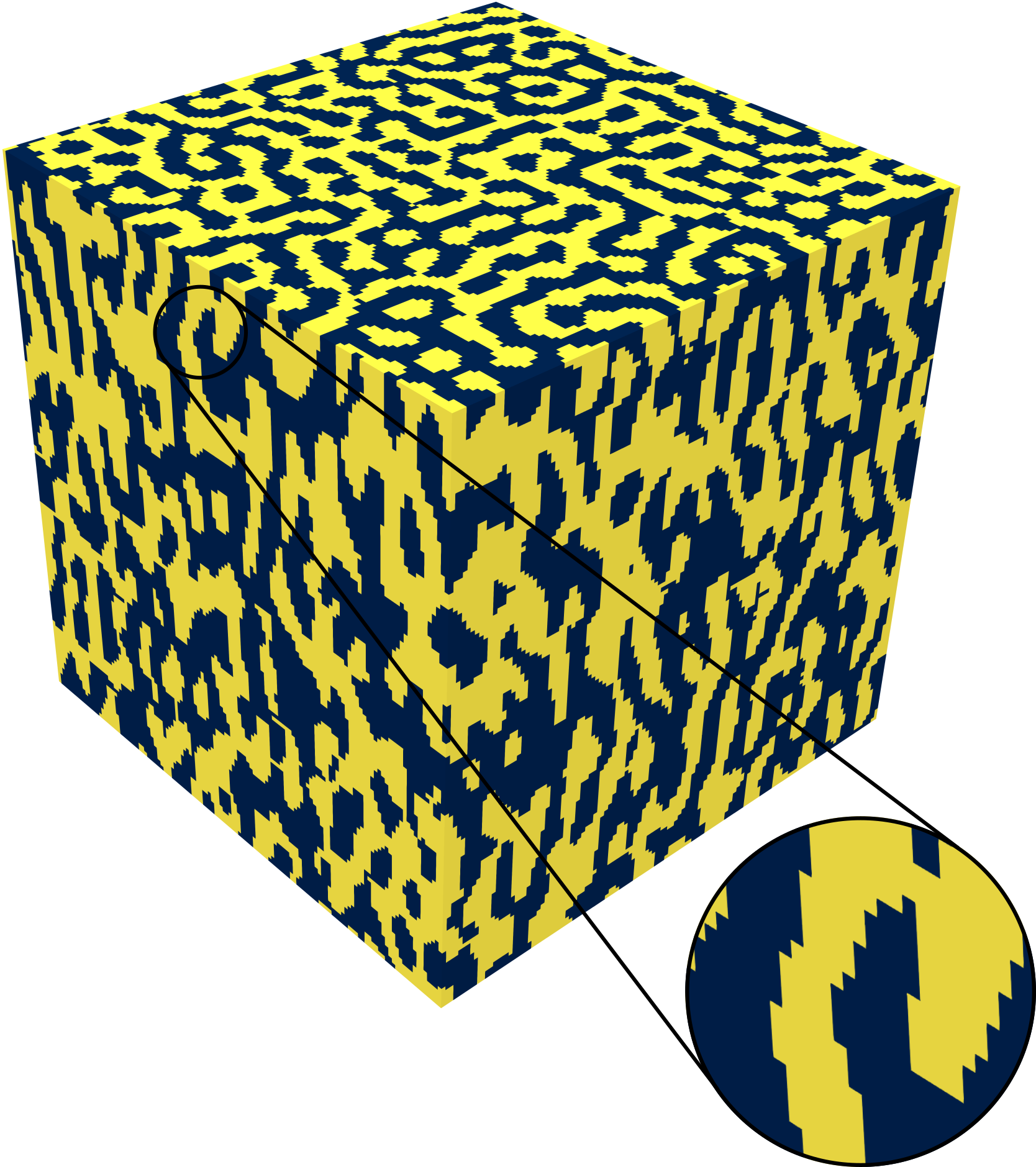}} 
	\hfill
	\subfloat[Smoothed lamellar]{\includegraphics[width=0.32\textwidth]{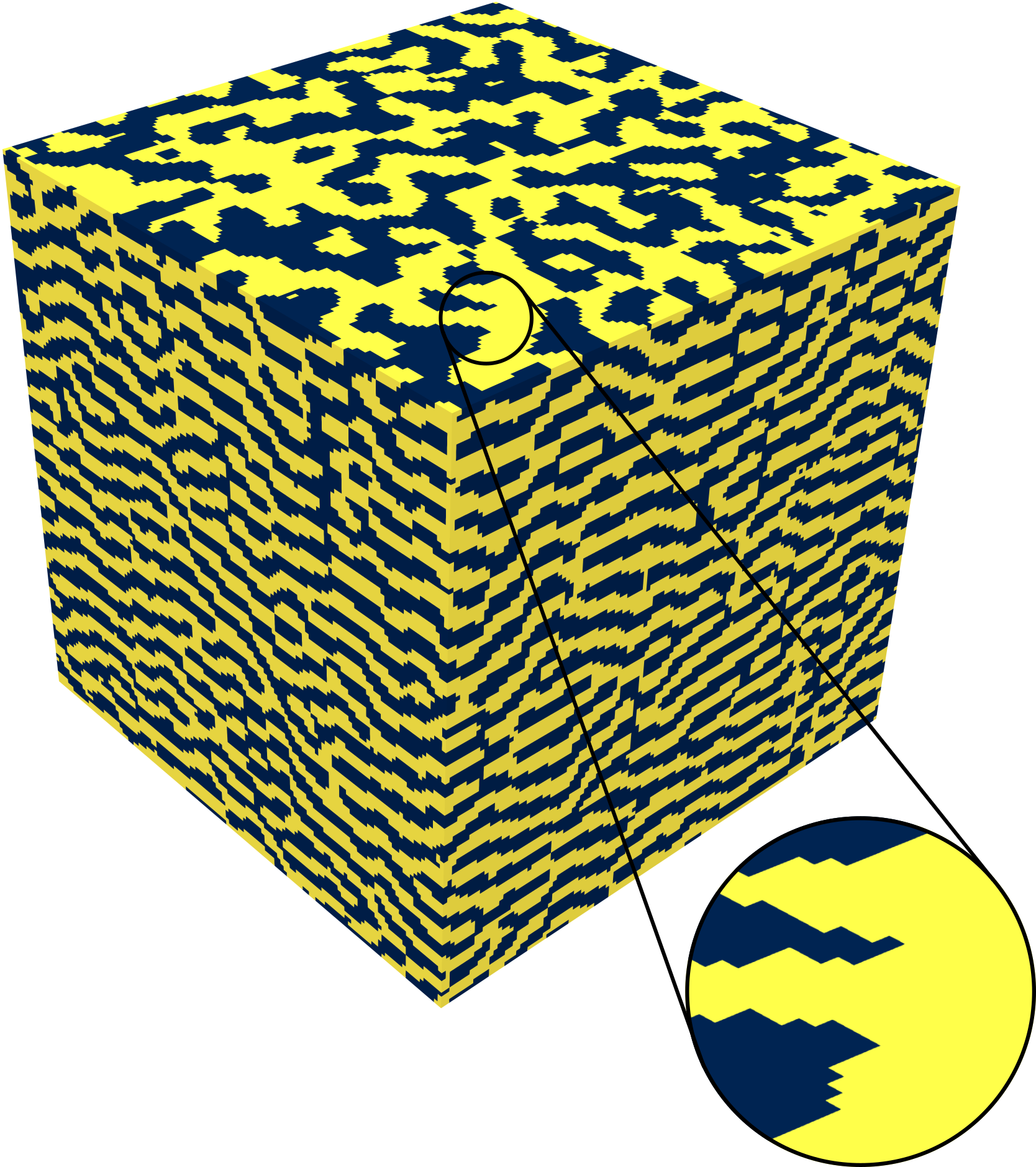}}
	\caption{The effect of descriptor-based smoothing on the reconstructed structures: It can be seen that spurious noise is eliminated while sharp microstructural features are preserved. \label{fig:smoothingcomparison}}
\end{figure}

\emph{MCRpy} is used for the reconstruction as described in Section~\ref{sec:reconstrucion} as well as for the smoothing procedure outlined in Section~\ref{sec:postprocessing}.
With a resolution of $128^3$ voxels, the former took $7$ hours for $800$ iterations\footnote{More precisely, $800$ iterations are performed per multigrid level. The three multigrid levels took $20$ minutes, $1.5$ hours and $5$ hours respectively. The last multigrid level on the highest resolution could have been aborted after $400$ iterations with a very similar outcome, leading to a total wall-clock time of $4.5$ hours, but this was not done.} on a single \emph{Nvidia A100} GPU, whereas the latter required $3$ hours for $20,000$ iterations on the same hardware.
The effect of the smoothing is illustrated in Figure~\ref{fig:smoothingcomparison}.
Although for the present microstructures the authors found that simpler and more efficient smoothing procedures based on Gaussian filtering also work (not shown here), this is not always the case as discussed in Appendix~\ref{sec:descriptorbasedsmoothing}.
Per structure, 20 instances are created to compare the variation between random realizations\footnote{As mentioned in~\cite{seibert_descriptor-based_2022}, the solution found by the optimizer in 3D reconstruction depends on the initial value. Hence, a different random seed can be used to generate different, but statistically equivalent structures, herein called realizations.} to the deviation from the reference structure.

\begin{figure}[t]
    \centering
	\subfloat{\includegraphics[width=0.18\textwidth]{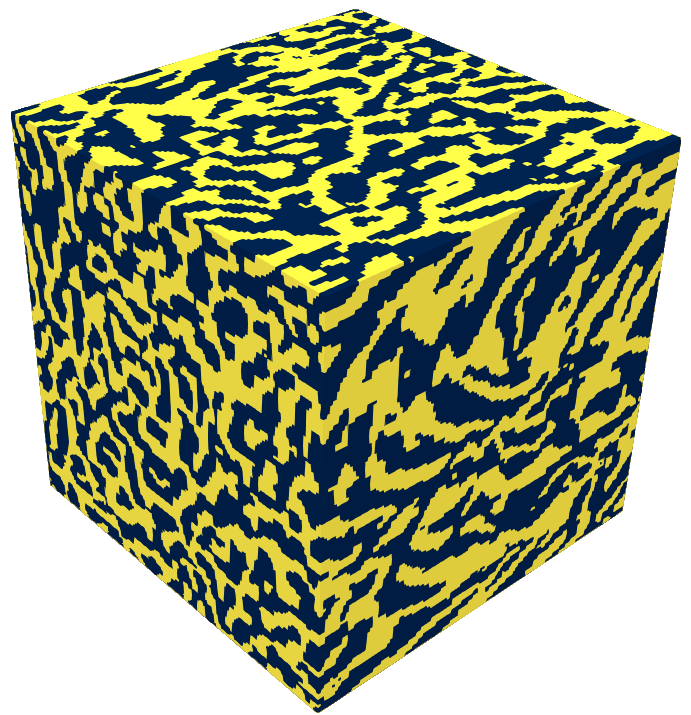}}
	\hfill
	\subfloat{\includegraphics[width=0.18\textwidth]{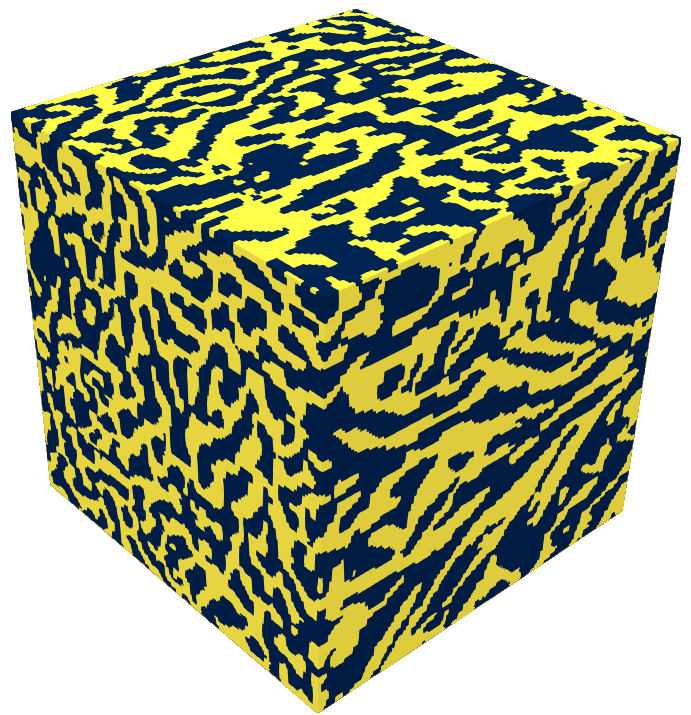}}
	\hfill
	\subfloat{\includegraphics[width=0.18\textwidth]{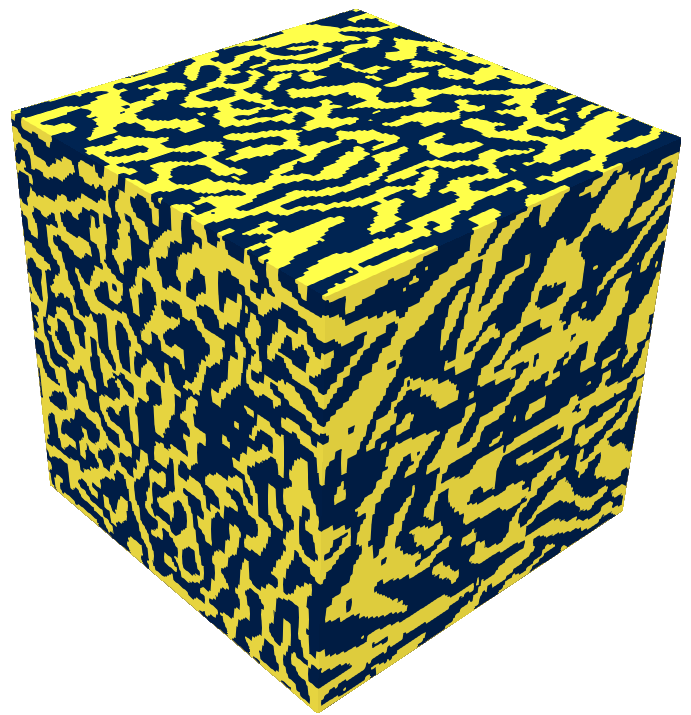}}
	\hfill
	\subfloat{\includegraphics[width=0.18\textwidth]{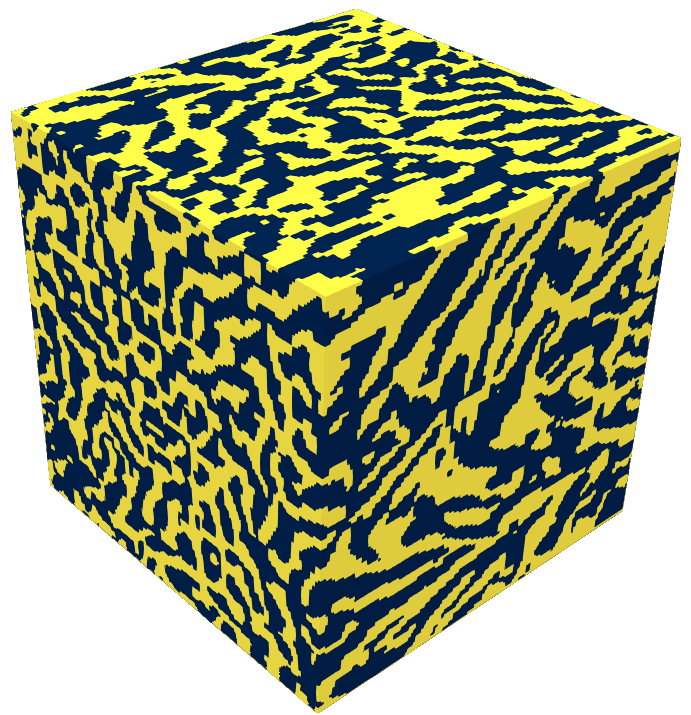}}
	\hfill
	\subfloat{\includegraphics[width=0.18\textwidth]{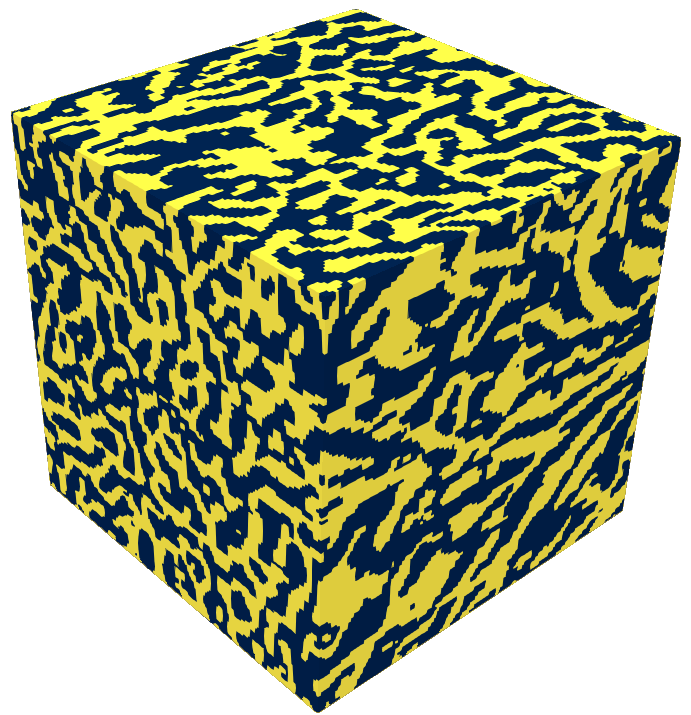}}
	\vfill
	\subfloat{\includegraphics[width=0.18\textwidth]{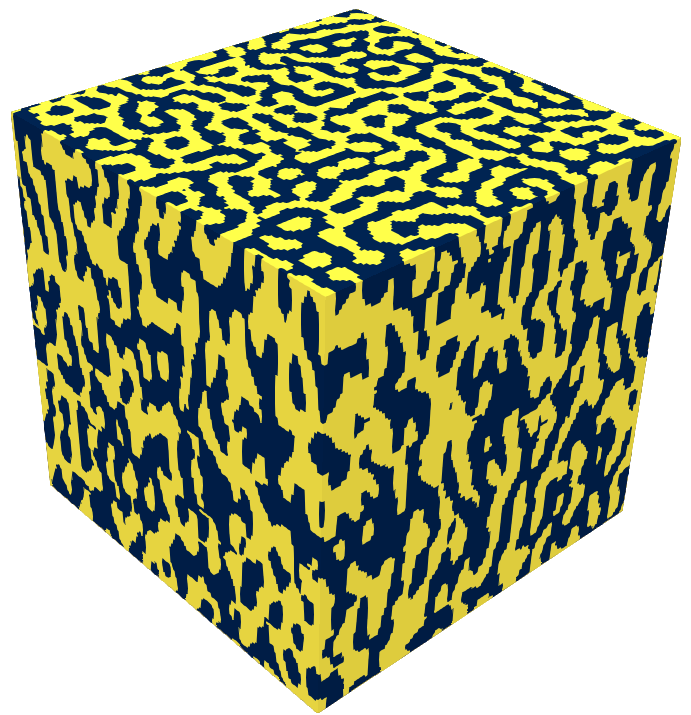}}
	\hfill
	\subfloat{\includegraphics[width=0.18\textwidth]{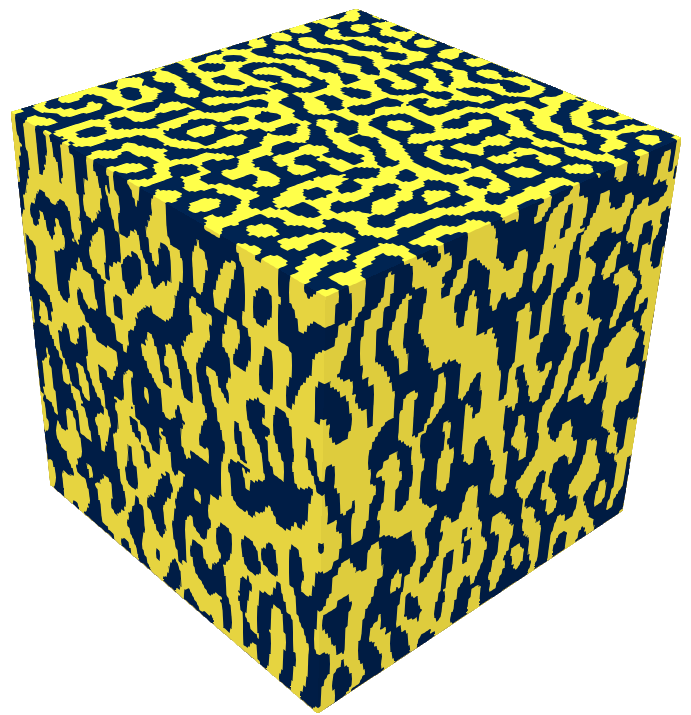}}
	\hfill
	\subfloat{\includegraphics[width=0.18\textwidth]{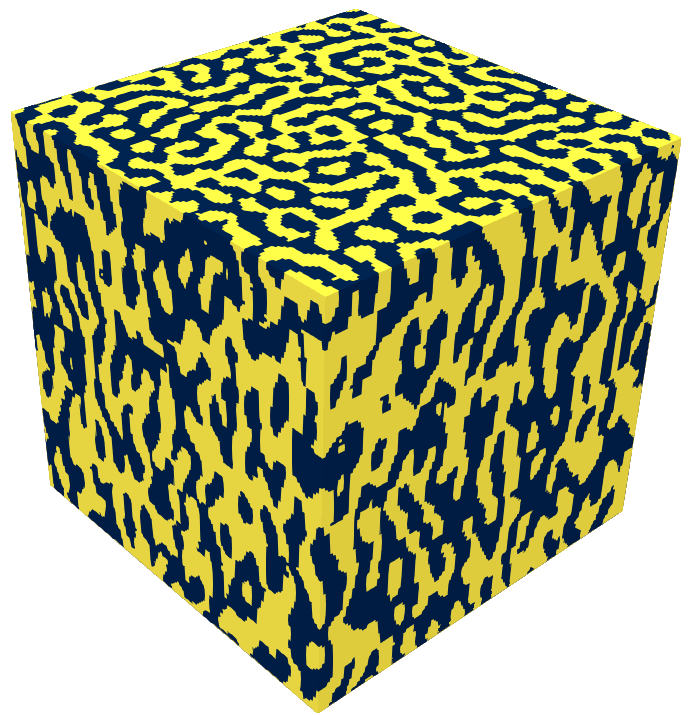}}
	\hfill
	\subfloat{\includegraphics[width=0.18\textwidth]{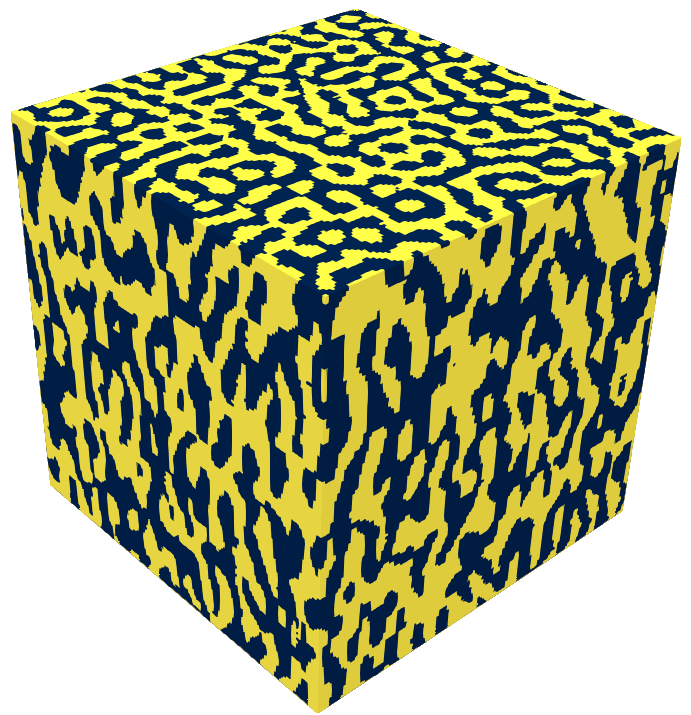}}
	\hfill
	\subfloat{\includegraphics[width=0.18\textwidth]{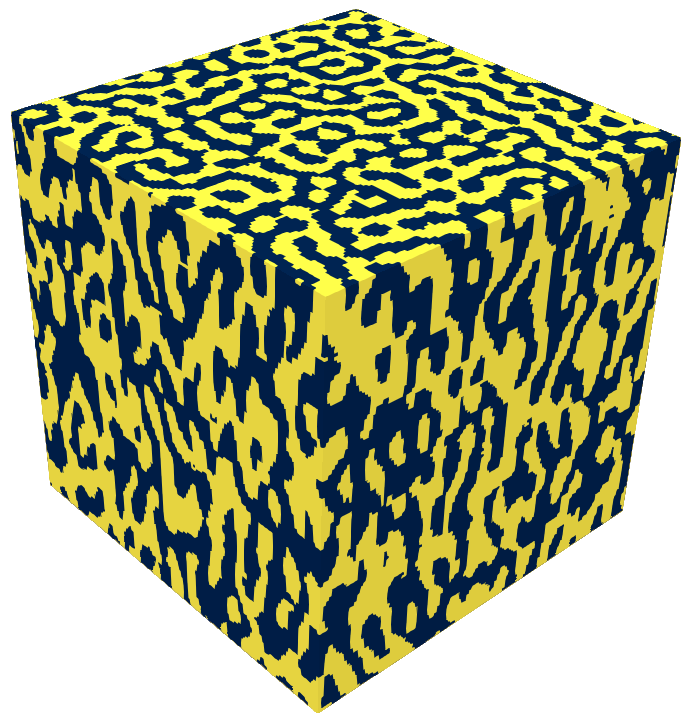}}
	\vfill
	\subfloat{\includegraphics[width=0.18\textwidth]{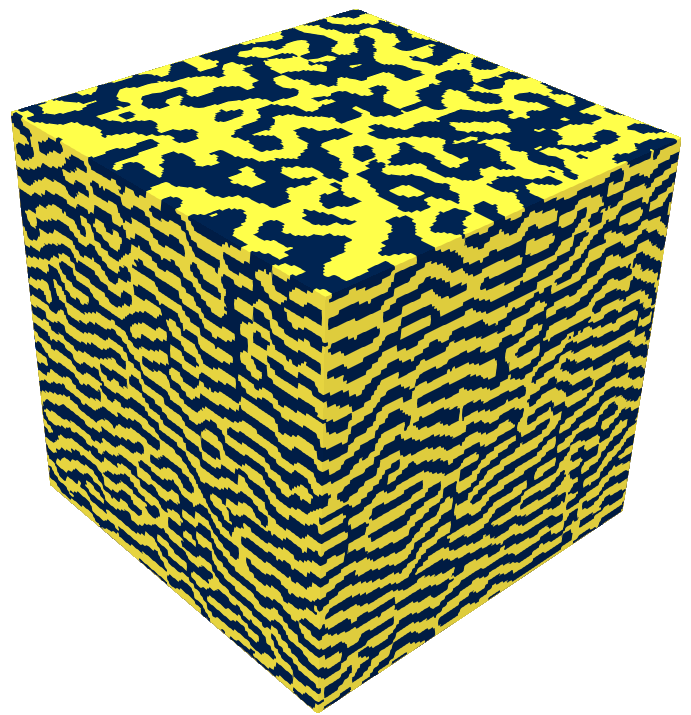}}
	\hfill
	\subfloat{\includegraphics[width=0.18\textwidth]{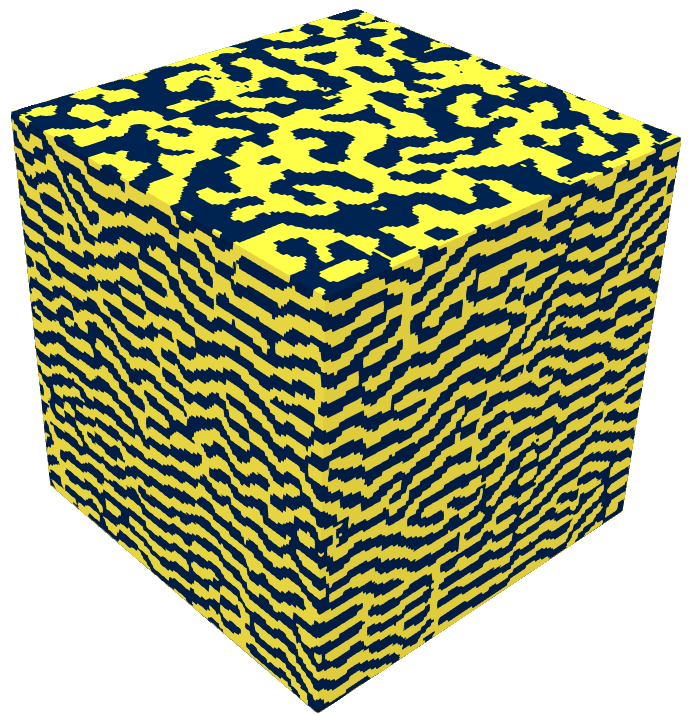}}
	\hfill
	\subfloat{\includegraphics[width=0.18\textwidth]{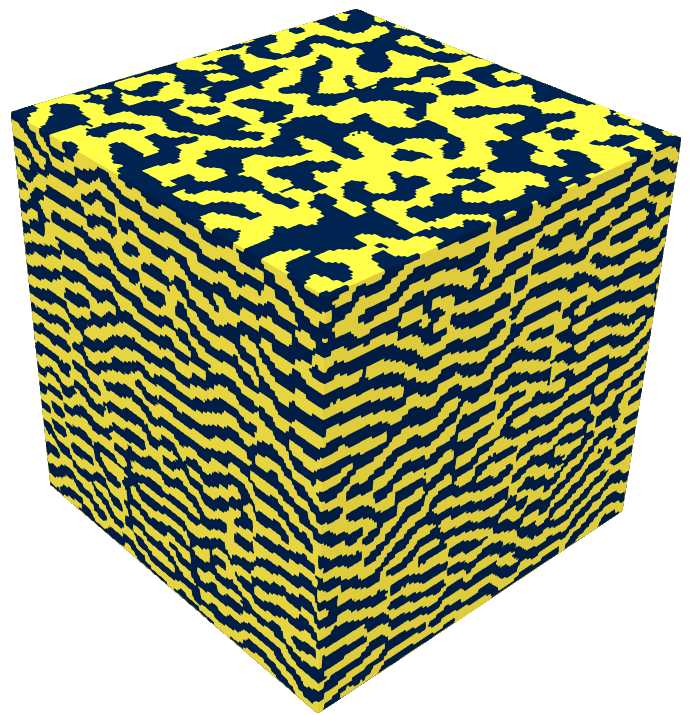}}
	\hfill
	\subfloat{\includegraphics[width=0.18\textwidth]{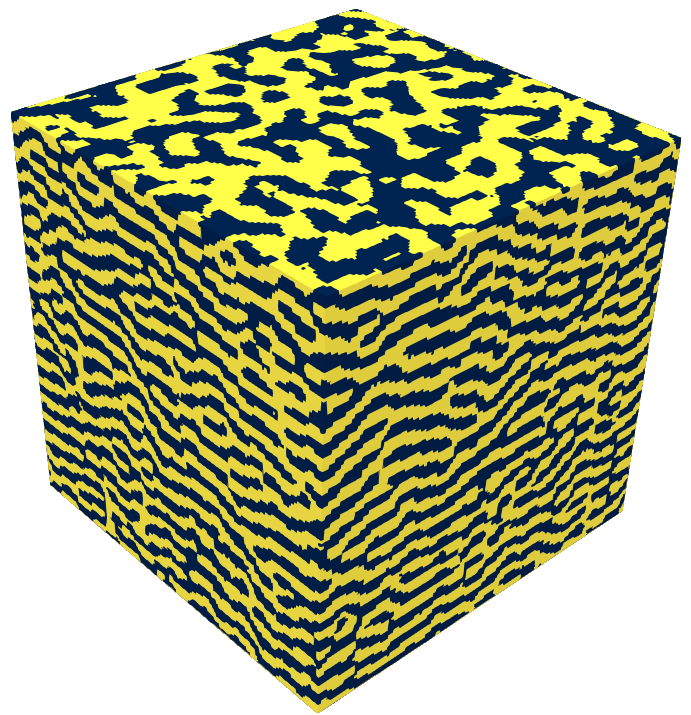}}
	\hfill
	\subfloat{\includegraphics[width=0.18\textwidth]{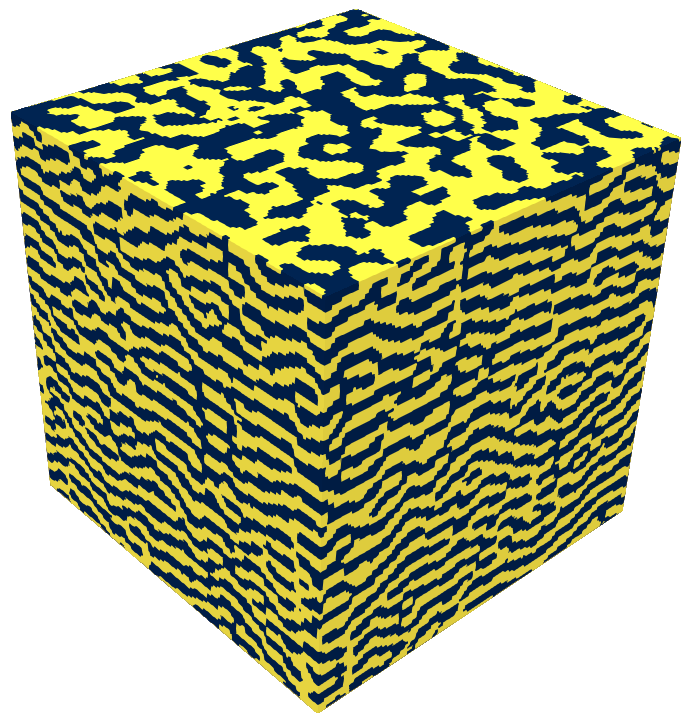}}
    \caption{Five reconstruction examples per structure for the real Ti-Fe material (top) and the columnar (center) and lamellar (bottom) synthetic structure.\label{fig:recexamples}}
\end{figure}

After the reconstruction and subsequent smoothing, the generated structures are visually very similar to the references as shown in Figure~\ref{fig:recexamples}.
To support this qualitative statement, a quantitative morphology analysis is carried out based on spatial correlations.
First, the statistical distribution of 2D correlations over all slices is analyzed for the original and reconstructed structure.
Then, the fully 3D two-point correlations are directly compared.

\subsection{Descriptor errors}
\label{sec:resultsdescriptors}
\begin{figure}[t]
	\centering
	\subfloat[Slicing in $x_1$-direction]{\includegraphics[height=0.3\textwidth]{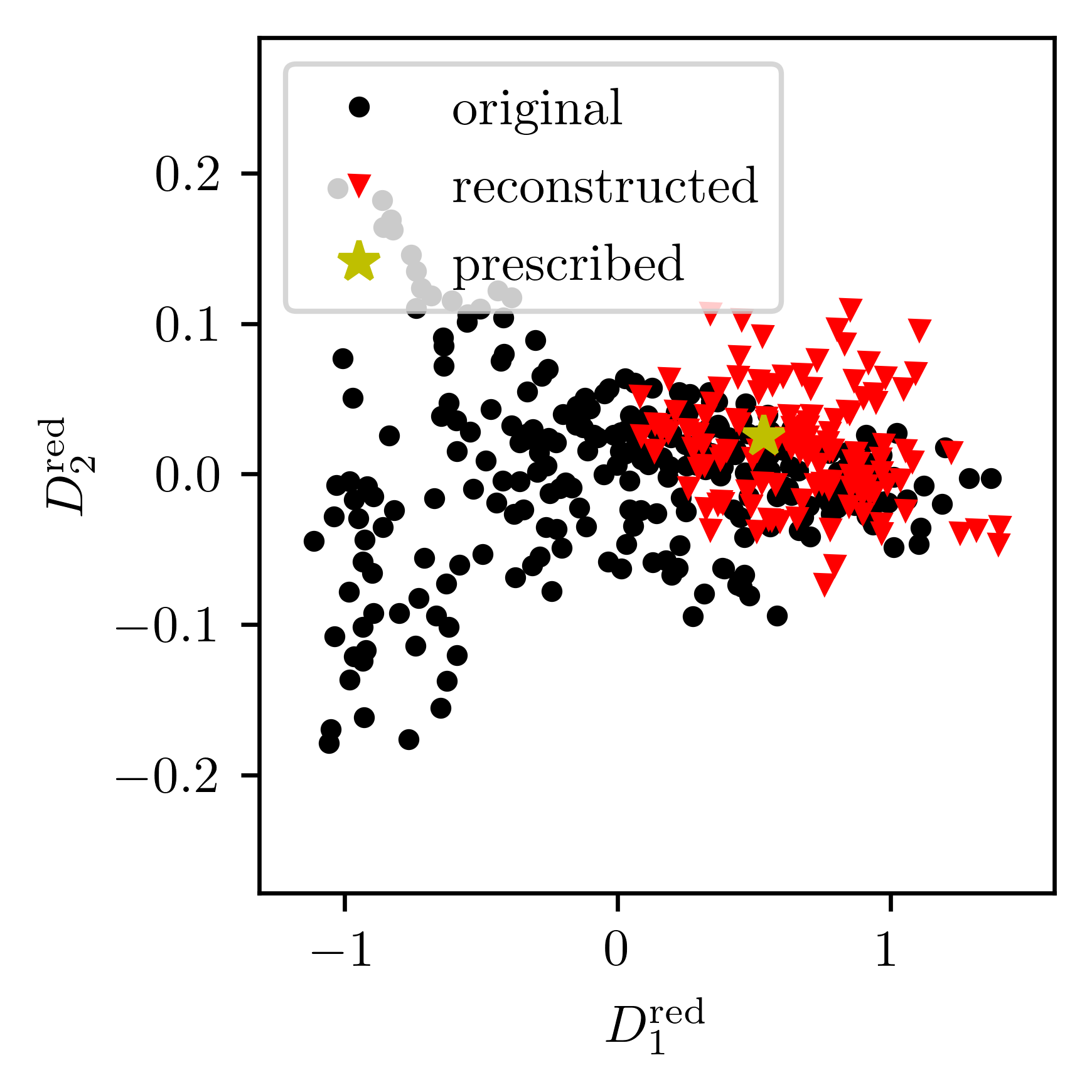}}
	\hfill
	\subfloat[Slicing in $x_2$-direction]{\includegraphics[height=0.3\textwidth]{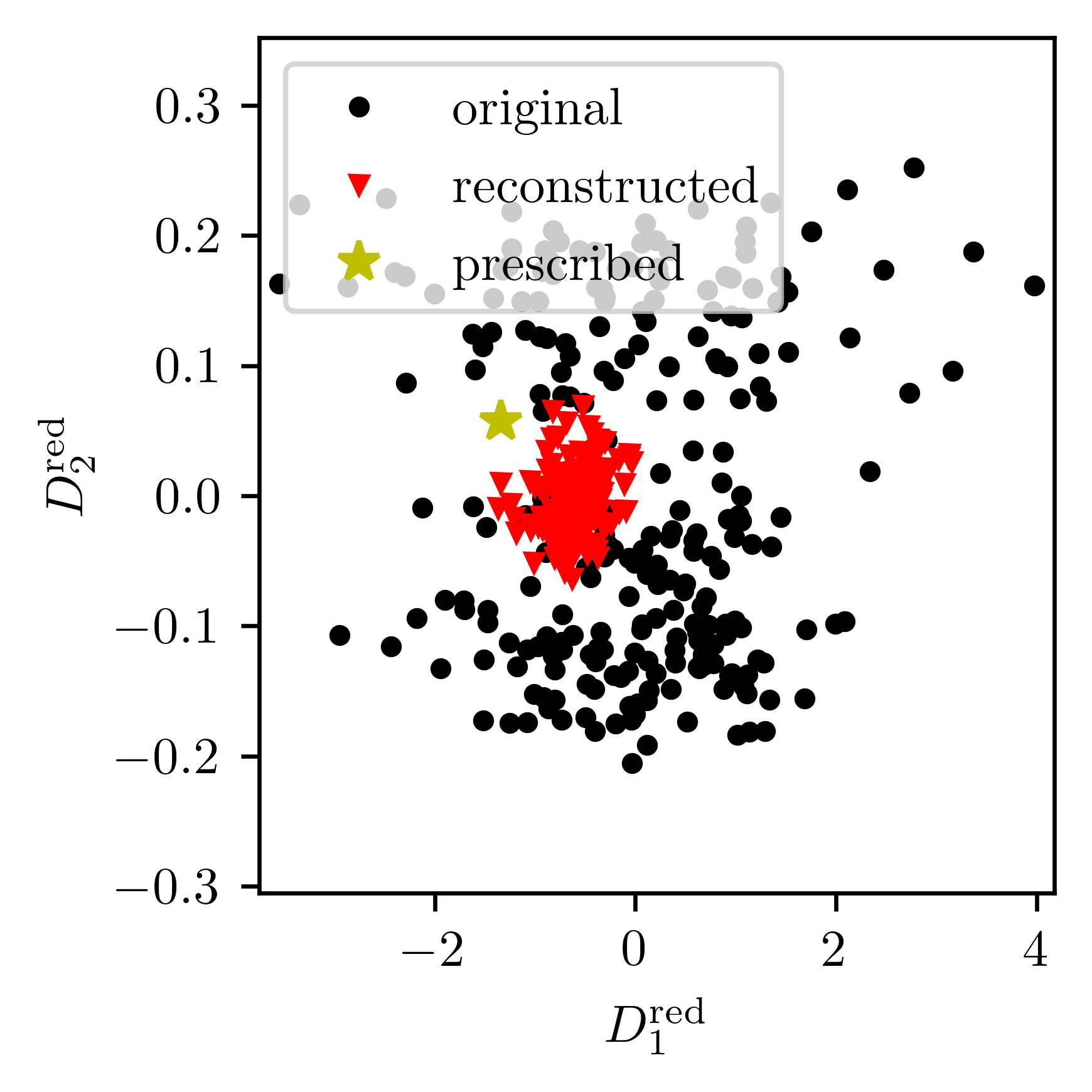}}
	\hfill
	\subfloat[Slicing in $x_3$-direction]{\includegraphics[height=0.3\textwidth]{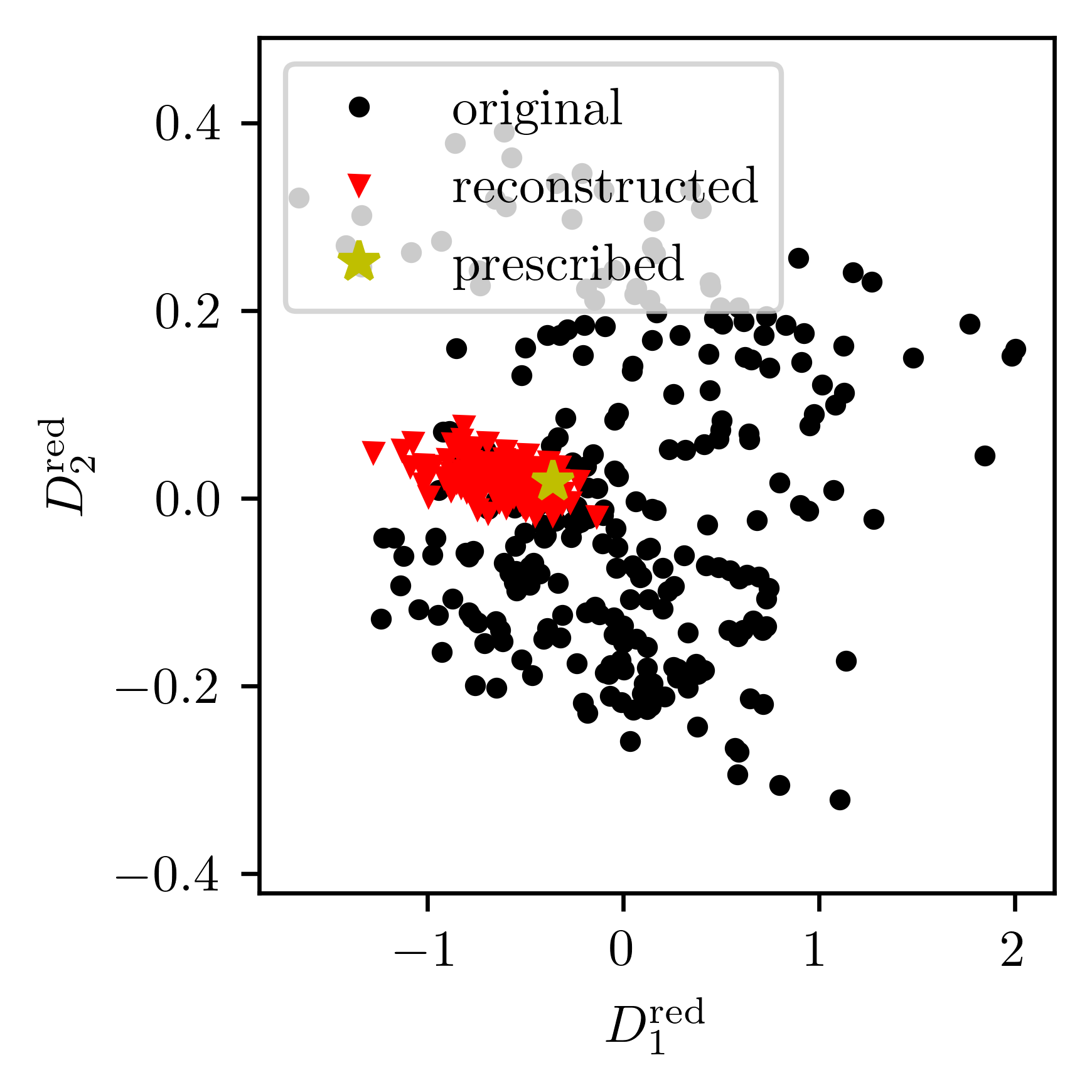}}
	\caption{Low-dimensional analysis of the descriptor distribution over the slices of the original and reconstructed Ti-Fe structure. Herein, every marker denotes a descriptor value, which is either prescribed during optimization or computed on a single slice of the original or reconstructed structure.\label{fig:descriptorerrors}}
\end{figure}

In order to visually analyze the high-dimensional two-point correlations, a principal component analysis (PCA) is used to reduce them to two dimensions.
In Figure~\ref{fig:descriptorerrors}, the original microstructure is exemplarily compared to the first reconstructed structure shown in Figure~\ref{fig:recexamples} in terms of the first two modes of the slice descriptors.
It is worth noting that the 2D-to-3D reconstruction is not informed about the entire distribution of original descriptors, which would describe the entire 3D structure, but only one randomly chosen value which corresponds to the single original 2D slice.
Based on this discrepancy, three phenomena are discussed based on simple volume fractions and are identified in Figure~\ref{fig:descriptorerrors} for spatial correlations:
\begin{enumerate}
    \item[\textit{(i)}] \emph{Descriptor concentration}: Due to random variations in the microstructure, the volume fraction of a single slice might be higher or lower than that of the entire microstructure. The expected magnitude of these fluctuations decreases as the microstructure size increases. Since the reconstruction is not based on the statistical distribution of volume fractions, but on a single value, there is no control of the fluctuations. Analogously, for generic descriptors, the descriptors in the reconstructed slices is observed to scatter less than in the original structure in Figure~\ref{fig:descriptorerrors}.
    \item[\textit{(ii)}] \emph{Descriptor difference}: Due the same variations in the microstructure, the prescribed value of the volume fraction is not optimal. A similar phenomenon can be observed for generic descriptors. In Figure~\ref{fig:descriptorerrors} (a), the prescribed descriptor is close to the boundary of the original descriptor point cloud. Therefore, although the descriptors of the reconstructed structure are centered around the prescribed value, a discrepancy is observed.
    \item[\textit{(iii)}] \emph{Descriptor incompatibility}: Moreover, the random fluctuations of the slice descriptors can lead to contradictions between orthogonal slices. For example, it is impossible to achieve a volume fraction of~$v_\mathrm{f}= 5 \%$ on all slices in $x_1$-direction while at the same time requiring~$v_\mathrm{f} = 6\%$ on all slices in $x_2$-direction. The average of the loss function over all slices and dimensions in Eq.~(\ref{eq:slicing}) leads to a compromise between incompatible descriptors, which manifests itself as offset between the mean descriptor of the reconstructed slices and the prescribed value. For generic descriptors, this phenomenon can be observed in Figures~\ref{fig:descriptorerrors}~(b) and~(c).
\end{enumerate}

Because the purpose of Figure~\ref{fig:descriptorerrors} is a discussion of qualitative trends, only a single reconstructed structure is shown.
Similar trends can be observed for all reconstructed structures.
While the principal components are well-suited for discussing qualitative trends in point clouds, the magnitude of the errors can not be assessed very well.
The absolute and relative descriptor error between original (og) and reconstructed (rec) structure are measured given by
\begin{equation}
    \mathcal{E}_D^\text{abs} = || \textbf{S}_2^\text{3D}(\textbf{M}^\text{rec}) - \textbf{S}_2^\text{3D}(\textbf{M}^\text{og}) ||_\text{MSE}
\end{equation}
and
\begin{equation}
    \mathcal{E}_D^\text{rel} = \mathcal{E}^\text{abs} / || \textbf{S}_2^\text{3D}(\textbf{M}^\text{og}) ||_\text{MSE} .
\end{equation}
Note that this comprises the full 3D correlations computed with the software \emph{pyMKS}~\cite{brough_materials_2017} and not a slice-wise comparison.
The maximum and average error of all 20 reconstructed microstructures is given in Table~\ref{tab:descriptorerrors}.
With less than~$0.1 \%$ deviation, the 3D two-point correlation is captured very well.
However, it presently not clear how small a descriptor error should be in order to guarantee similar effective properties.
Therefore, the homogenized material response is analyzed in the following.
\begin{table*}[h]
	\centering
	\caption{Errors of the full 3D two-point correlations.}
	\label{tab:descriptorerrors}
	\begin{tabular}{l | c  c  c  c }
\toprule
& $\underset{\alpha}{\text{max}} \; \mathcal{E}_D^\text{abs}(\textbf{M}_\alpha^\text{rec})$ & 
$\frac{1}{20} \sum_\alpha
\mathcal{E}_D^\text{abs}(\textbf{M}_\alpha^\text{rec}) $ & $\underset{\alpha}{\text{max}} \; \mathcal{E}_D^\text{rel}(\textbf{M}_\alpha^\text{rec}) $ & $\frac{1}{20} \sum_\alpha \mathcal{E}_D^\text{rel}(\textbf{M}_\alpha^\text{rec})$ \\ 
\midrule
Alloy & $3.01 \cdot 10^{-5}$ & $2.84 \cdot 10^{-5}$ & $0.071 \%$ & $0.067 \%$\\ 
Columnar & $2.66 \cdot 10^{-5}$ & $2.52 \cdot 10^{-5}$ & $0.043 \%$ & $0.040 \%$\\ 
Lamellar & $2.94 \cdot 10^{-5}$ & $2.88 \cdot 10^{-5}$ & $0.046 \%$ & $0.045 \%$\\ 
\bottomrule
	\end{tabular}
\end{table*}

\subsection{Effective alloy properties}
\label{sec:resultsalloy}
The effective properties of the titanium alloy are quantified in terms of the elastic and plastic behavior.

\begin{figure}[t]
	\centering
	\subfloat[Original structure]{\includegraphics[height=0.25\textwidth]{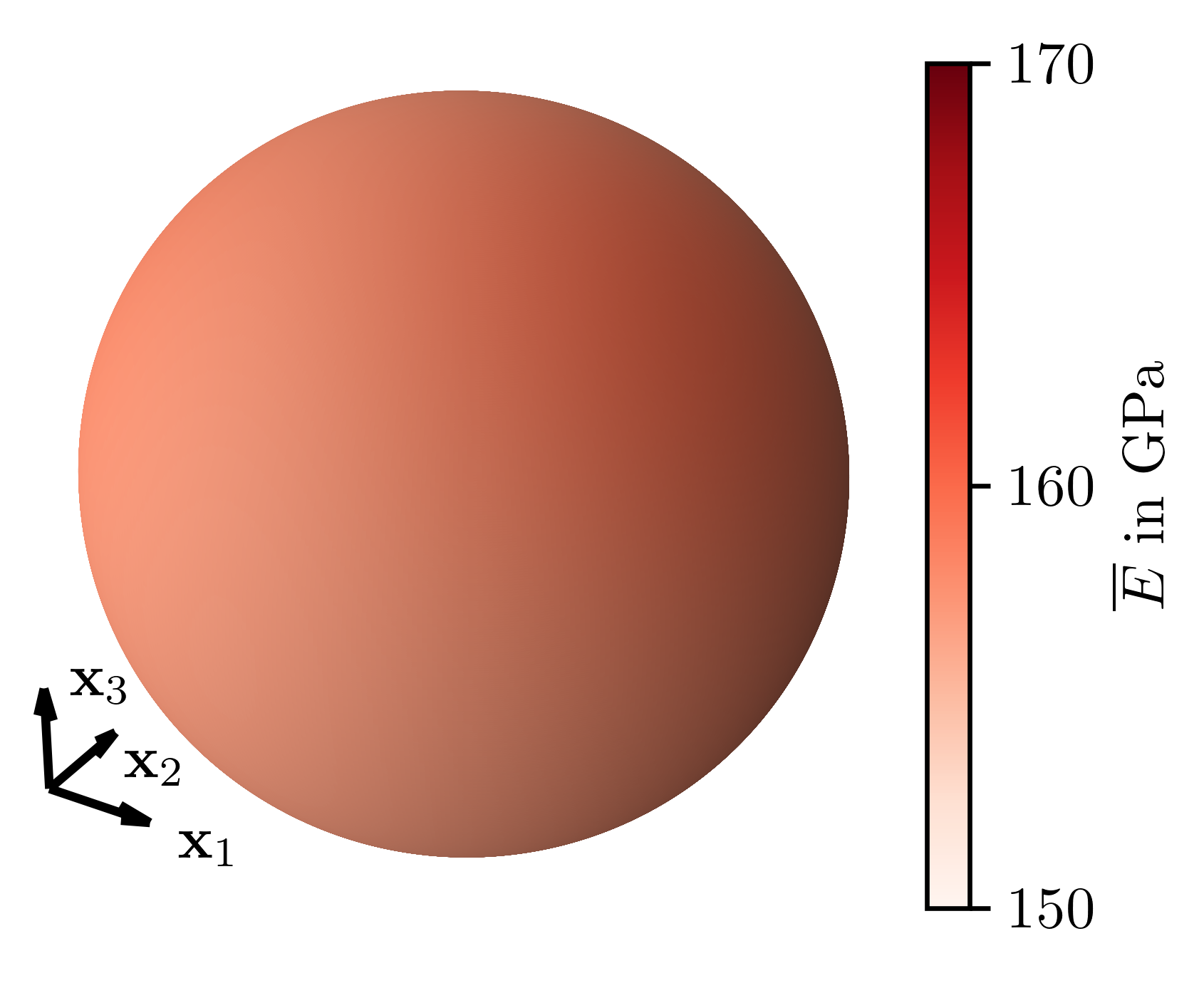}}
	\hfill
	\subfloat[Reconstructed structure]{\includegraphics[height=0.25\textwidth]{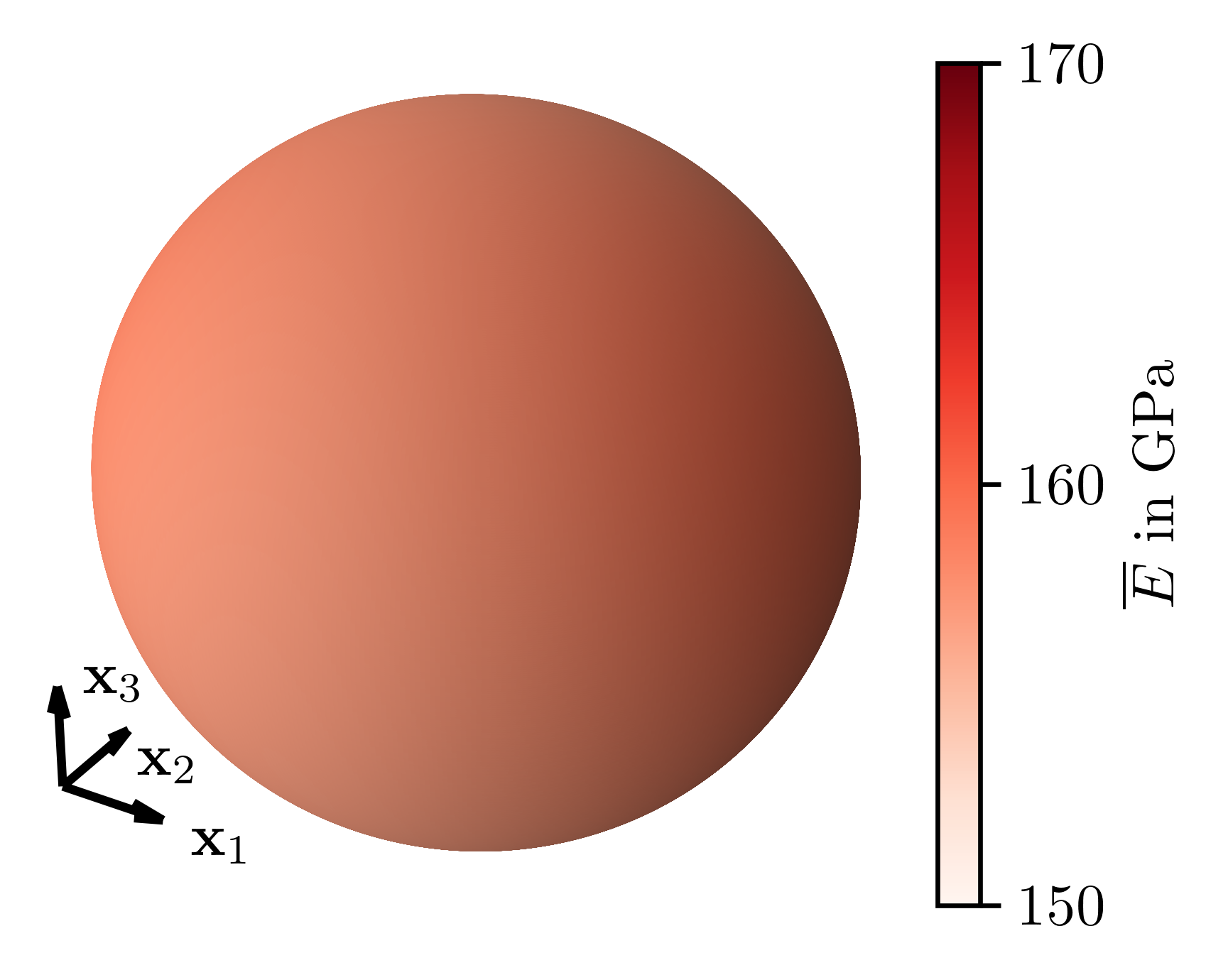}}
	\hfill
	\subfloat[Difference]{\includegraphics[height=0.25\textwidth]{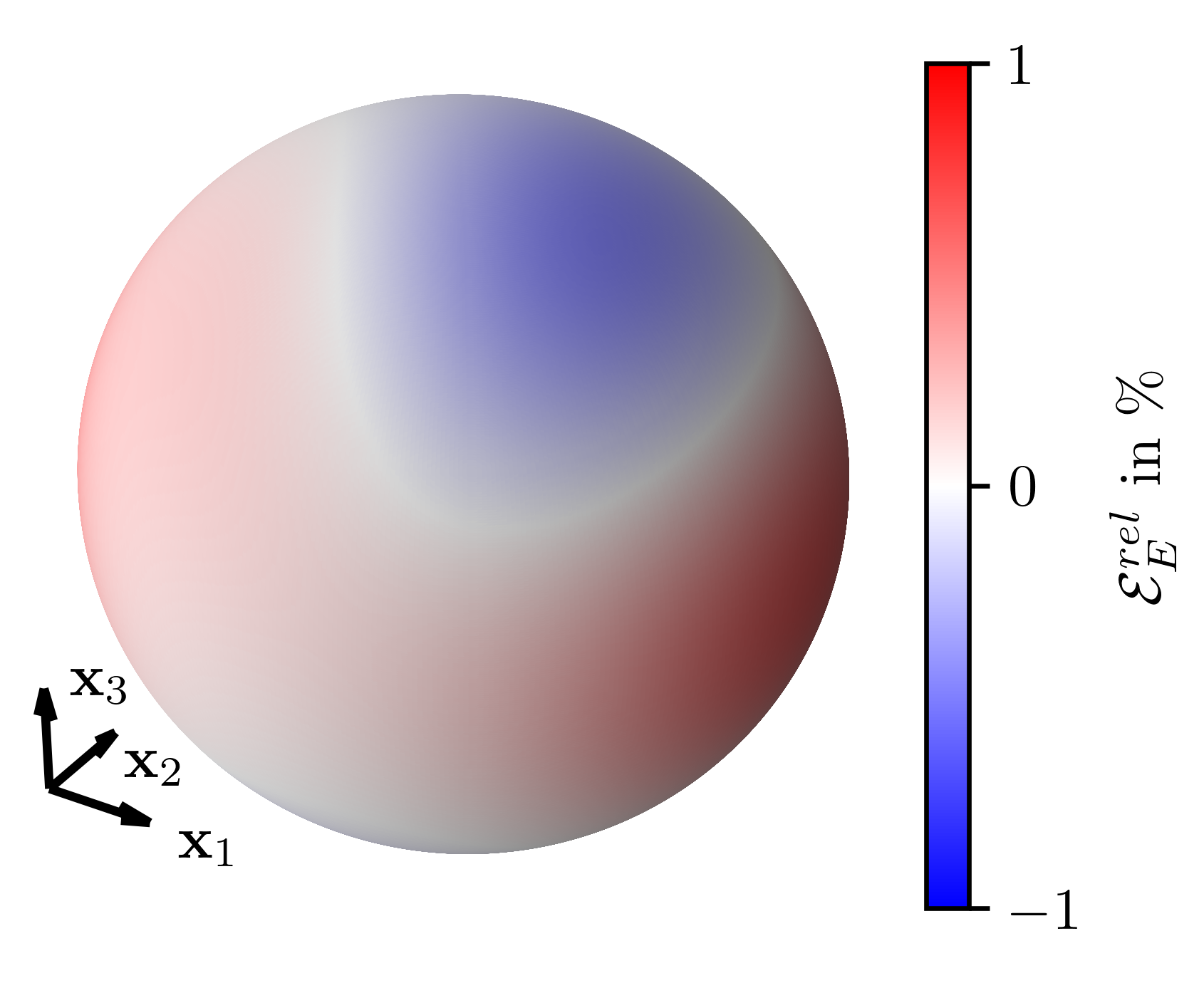}}
	\caption{Elastic surface plots of the original and generated Ti-Fe structure. \label{fig:stiffnesserrors}}
\end{figure}

The elastic behavior is captured by the full stiffness tensor, which is visualized by means of \emph{elastic surface} or \emph{YMS} plots~\cite{bohlke_graphical_2001,nordmann_visualising_2018}, see Section~\ref{sec:simulation}.
A comparison between the simulations based on the original CT scan and the first reconstructed structure is given in Figure~\ref{fig:stiffnesserrors}.
Table~\ref{tab:propertiestife} enables a quantitative comparison of~$\bar{E}$ in~$x_1$-, $x_2$- and $x_3$-direction.
It can be seen that the range of the Young's modulus as well as the degree of anisotropy is captured very well.
The exact direction of the anisotropy slightly differs from the correct value.
However in Table~\ref{tab:propertiestife}, for each reconstructed structure, the effective Young's modulus in a certain direction is closer to the reference modulus in the same direction than to the reference modulus in another direction.
Overall, the relative error
\begin{equation}
    \mathcal{E}_{E}^\text{rel} = \dfrac{\bar{E}^\text{rec} - \bar{E}^\text{og}}{\bar{E}^\text{og}}
\end{equation}
of the directional Young's modulus is limited to approximately~$\pm0.5 \%$ with a maximum error of~$0.61\%$, which is extremely small considering a phase contrast of $E^\text{TiFe} / E^{\beta-\text{Ti}} \approx 2$.
As a comparison, for the same material properties, simple homogenization by Voigt's and Reuss' formulae yield an upper and lower bound of $163.3$ GPa and $151.3$ GPa, respectively.
\begin{table*}[h]
	\centering
	\caption{Errors of the effective directional Young's modulus and yield strength for the Ti-Fe structure.}
	\label{tab:propertiestife}
	\begin{tabular}{c  c  c  c  c  c  c }
\toprule
Direction & $\bar E^\text{ref}$ in GPa & $\bar E^\text{rec}$ in GPa & $\mathcal{E}_\text{E}$ in \% & $\bar \sigma_\text{y}^\text{ref}$ in MPa & $\bar \sigma_\text{y}^\text{rec}$ in MPa & $\mathcal{E}_{\sigma_\text{y}}$ in \%\\ 
\midrule
x & 158.383 & 159.346 $\pm$ 0.006 & 0.61 & 1417.69 & 1400.53 $\pm$ 0.18 & -1.21\\ 
y & 156.713 & 156.864 $\pm$ 0.005 & 0.10 & 1455.76 & 1455.73 $\pm$ 0.23 & -0.00\\ 
z & 157.825 & 157.515 $\pm$ 0.006 & -0.20 & 1428.05 & 1439.74 $\pm$ 0.21 & 0.82\\ 
\bottomrule
	\end{tabular}
\end{table*}

The plastic material response is quantified by the effective yield strength~$\bar{\sigma}_\text{y}$, whereby this work is limited to the~$x_1$-, $x_2$- and $x_3$-direction for computational efficiency.
The results are summarized in Table~\ref{tab:propertiestife}.
Again, all relative errors
\begin{equation}
    \mathcal{E}_{\sigma_y}^\text{rel} = \dfrac{\bar{\sigma}_y^\text{rec} - \bar{\sigma}_y^\text{og}}{\bar{\sigma}_y^\text{og}}
\end{equation} 
are very small, and the anisotropy of the effective properties is captured very well.

\begin{figure}[t]
    \centering
    \includegraphics[width=0.65\textwidth]{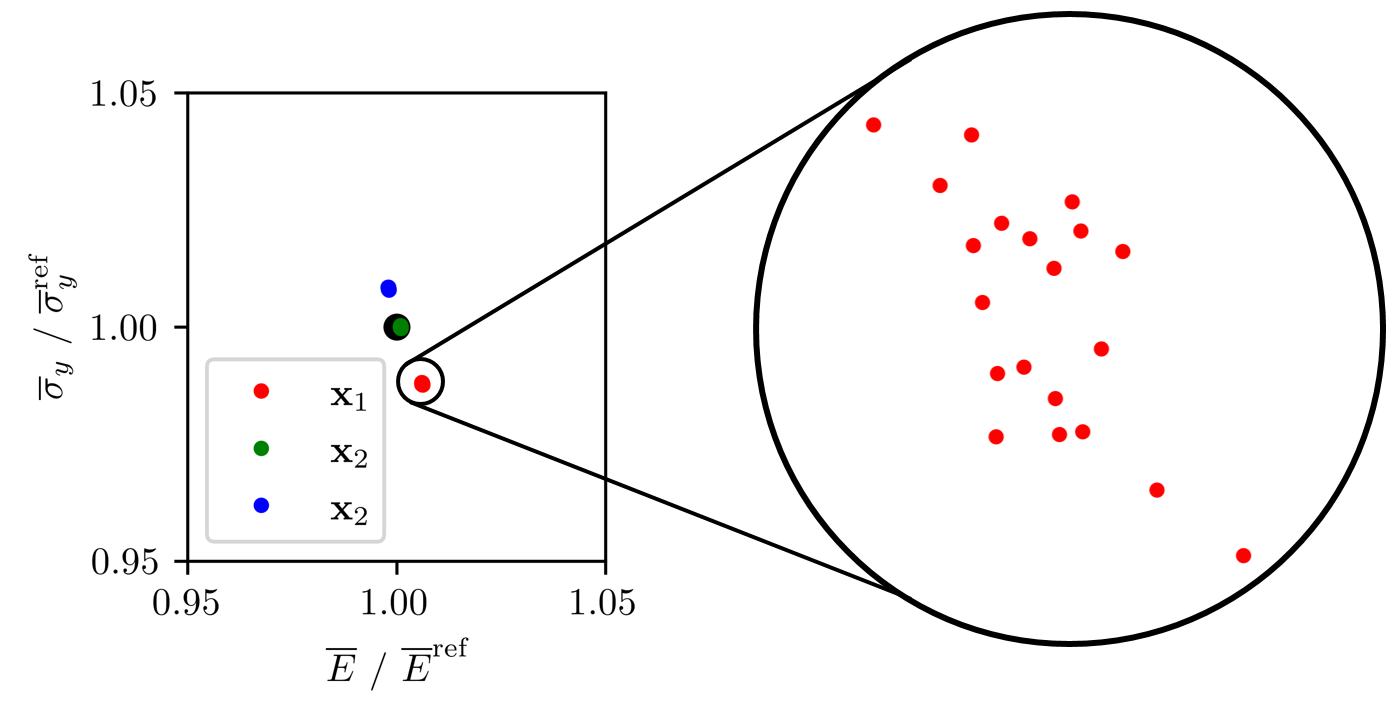}
    \caption{Relative error of the effective directional Young's modulus and yield strength for the Ti-Fe structures. It can be seen that the scatter over different realizations is much smaller than the systematic error. Note that all values are divided by the reference, but the reference is different in each direction. The numerical values are given in Table~\ref{tab:propertiestife}. \label{fig:results_pl}}
\end{figure}

Figure~\ref{fig:results_pl} shows the scatter of the effective isotropic elasto-plastic properties compared to the reference.
Note that the reference is different for each spatial direction, as can be seen in Table~\ref{tab:propertiestife}.
However, as shown in Table~\ref{tab:descriptorerrors}, the differences within the reconstructed structures are much smaller than the deviation between the reconstructed and original structure.
This is likely attributable to the systematic descriptor deviations mentioned in Section~\ref{sec:resultsdescriptors} and is discussed further in the following section with a higher degree of anisotropy.

\begin{figure}[t]
	\centering
	\subfloat[Applied load case]{\includegraphics[height=0.24\textwidth]{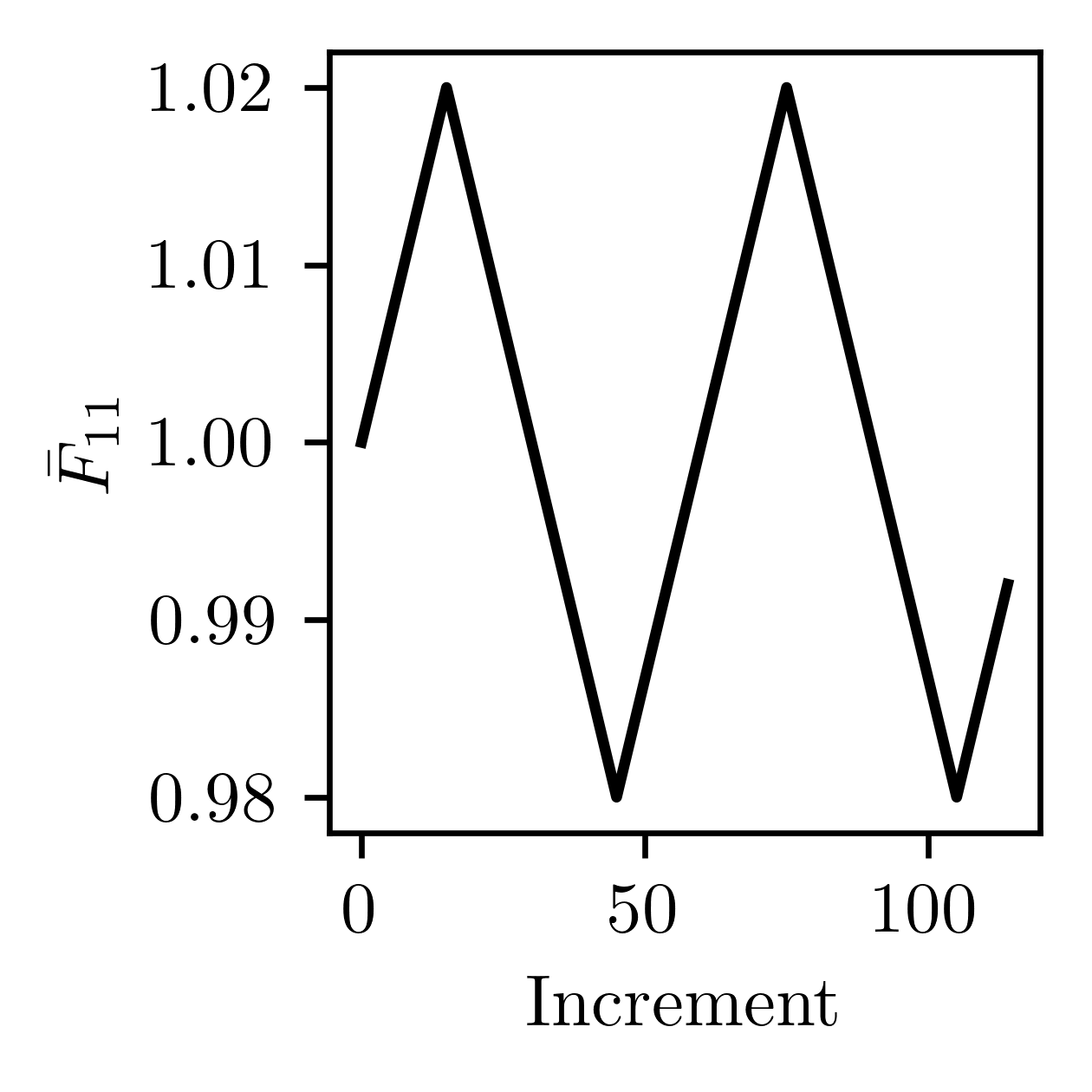}}
	\hfill
	\subfloat[Displacement with magnification of 30]{\includegraphics[height=0.23\textwidth]{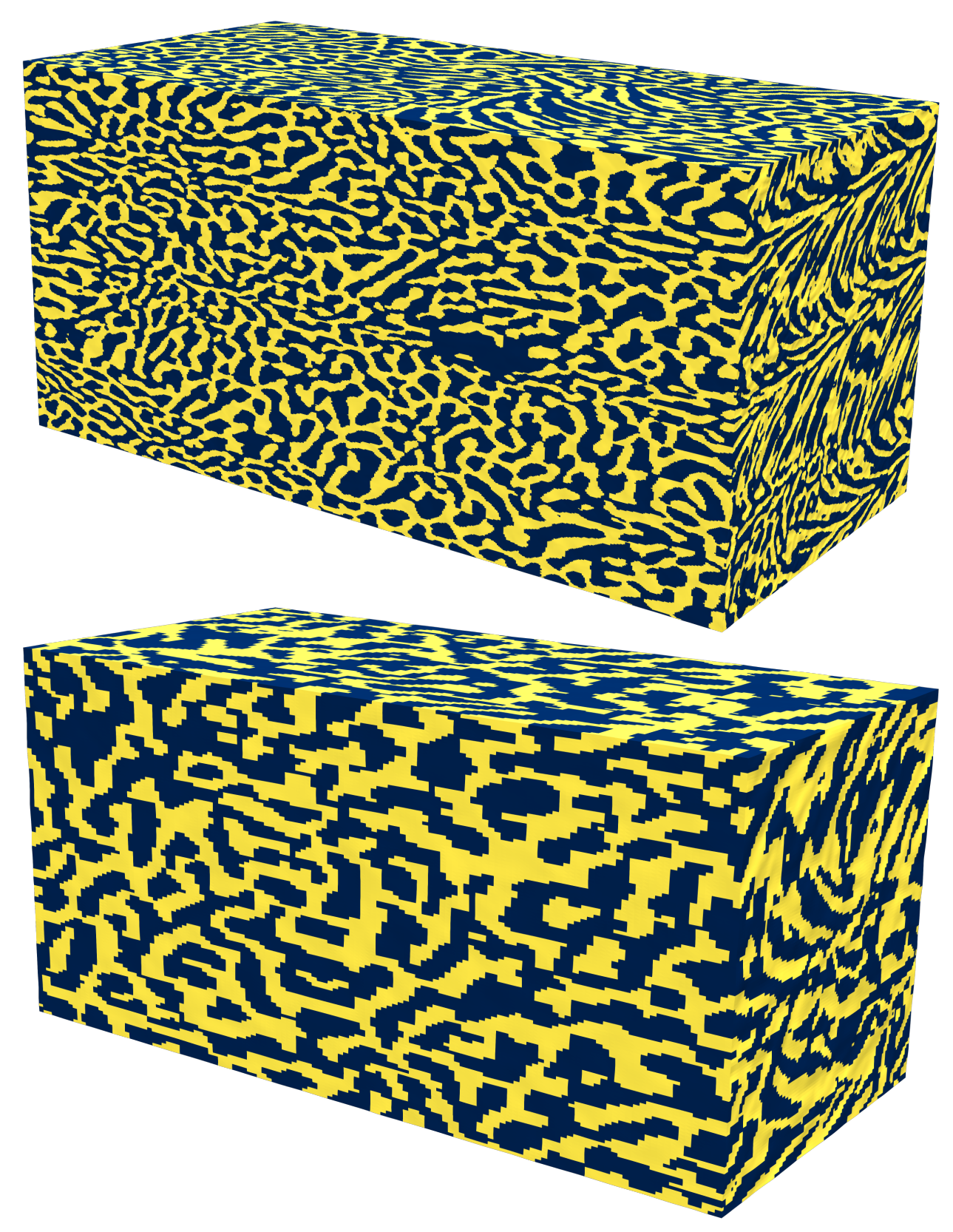}}
	\hfill
	\subfloat[Material response]{\includegraphics[height=0.24\textwidth]{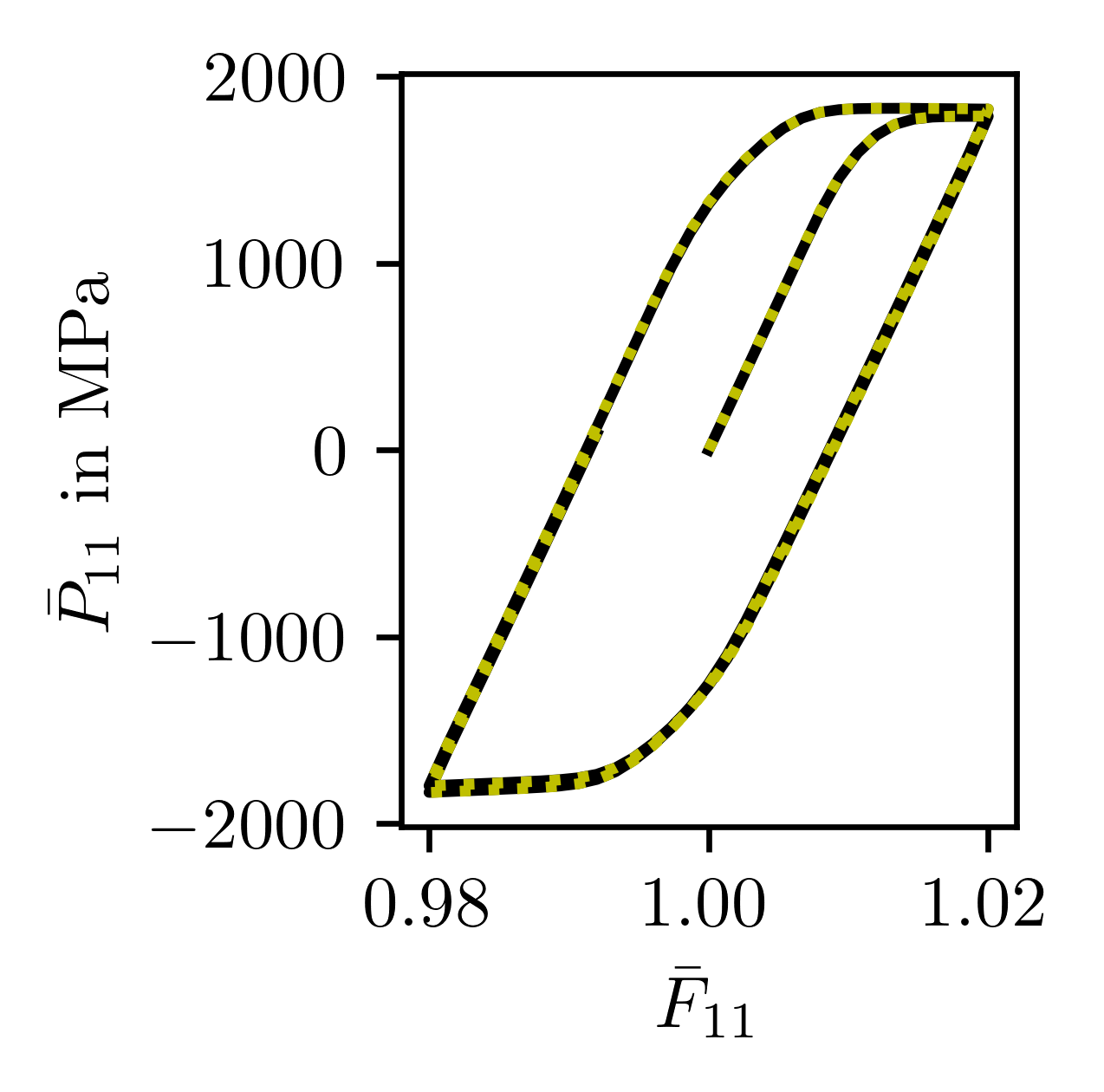}}
	\hfill
	\subfloat[Zoom into (c)]{\includegraphics[height=0.24\textwidth]{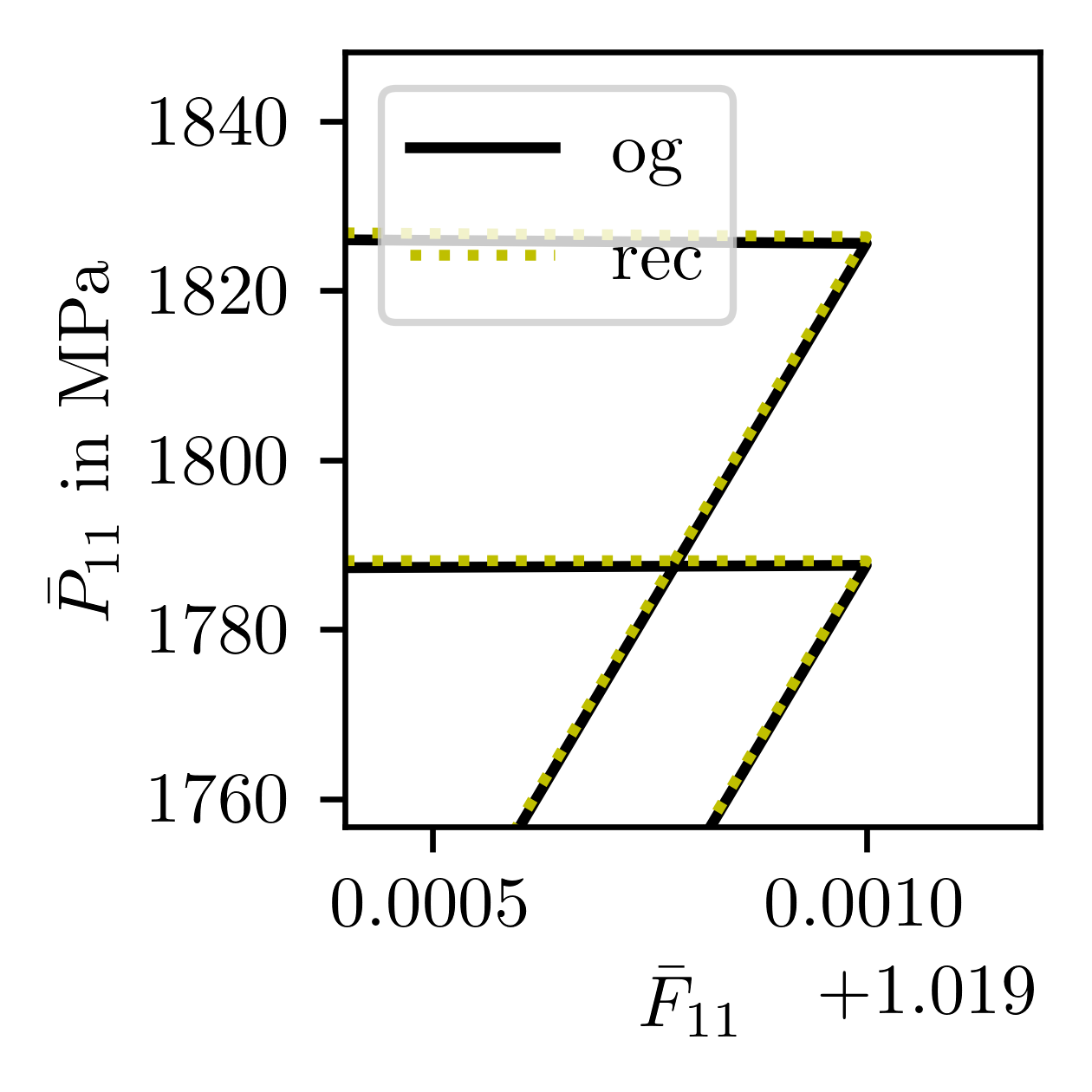}}
	\caption{Material response after two load cycles. The original structure (top, black) is compared to the first reconstruction result (bottom, yellow), illustrating the same qualitative and quantitative behavior. \label{fig:cycles}} 
\end{figure}

Finally, Figure~\ref{fig:cycles} shows the material response of the original structure and the first reconstruction result under two load cycles.
Also in this case, it can be seen that the effective behavior is captured very well and even at a large magnification, the difference between the curves is very small.
The response under load in the other directions is predicted similarly well.

\subsection{Properties of spinodoid metamaterial}
\label{sec:resultsspinodal}
Based on the effective properties to the Ti-Fe system identified in the previous section, the elasto-plastic material model introduced in Subsec.~\ref{subsec:constitutive} is parametrized as given in Table~\ref{tab:materialparameters}.
It is used to determine the macroscopic properties of the spinodoid, "bone-like" materials.
This poses several challenges because with increasing phase contrast and anisotropy, 
(i) both, original as well as reconstructed structures, need to be increasingly large to adequately capture the morphology, 
(ii) small errors in reconstructing highly stressed parts of the structure can lead to spurious local stress concentrations and hence strongly affect the effective behavior, and 
(iii) for the same reason, noise that is inherent to 2D-to-3D reconstruction~\cite{bostanabad_reconstruction_2020,seibert_descriptor-based_2022} becomes increasingly problematic.
The remainder of this section shows the results, which are surprisingly accurate considering the above challenges.
Further investigations on the effect of insufficiently large volume elements and the effect of smoothing on the properties are given in Appendices~\ref{sec:appendixslicesize} and~\ref{sec:appendixsmoothingeffect}, respectively.

\begin{figure}[t]
	\centering
	\subfloat[Original columnar]{\includegraphics[height=0.25\textwidth]{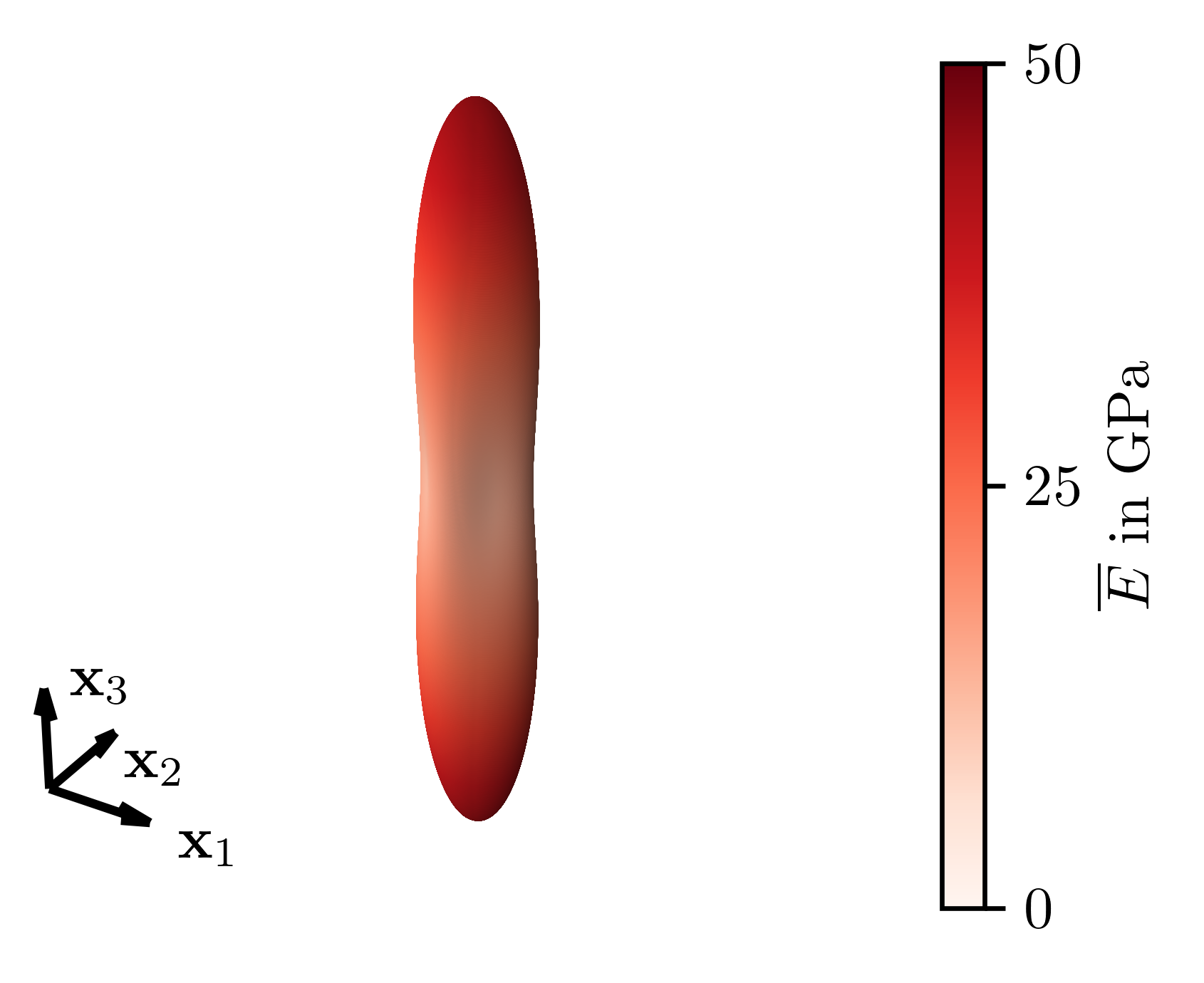}}
	\hfill
	\subfloat[Reconstructed columnar]{\includegraphics[height=0.25\textwidth]{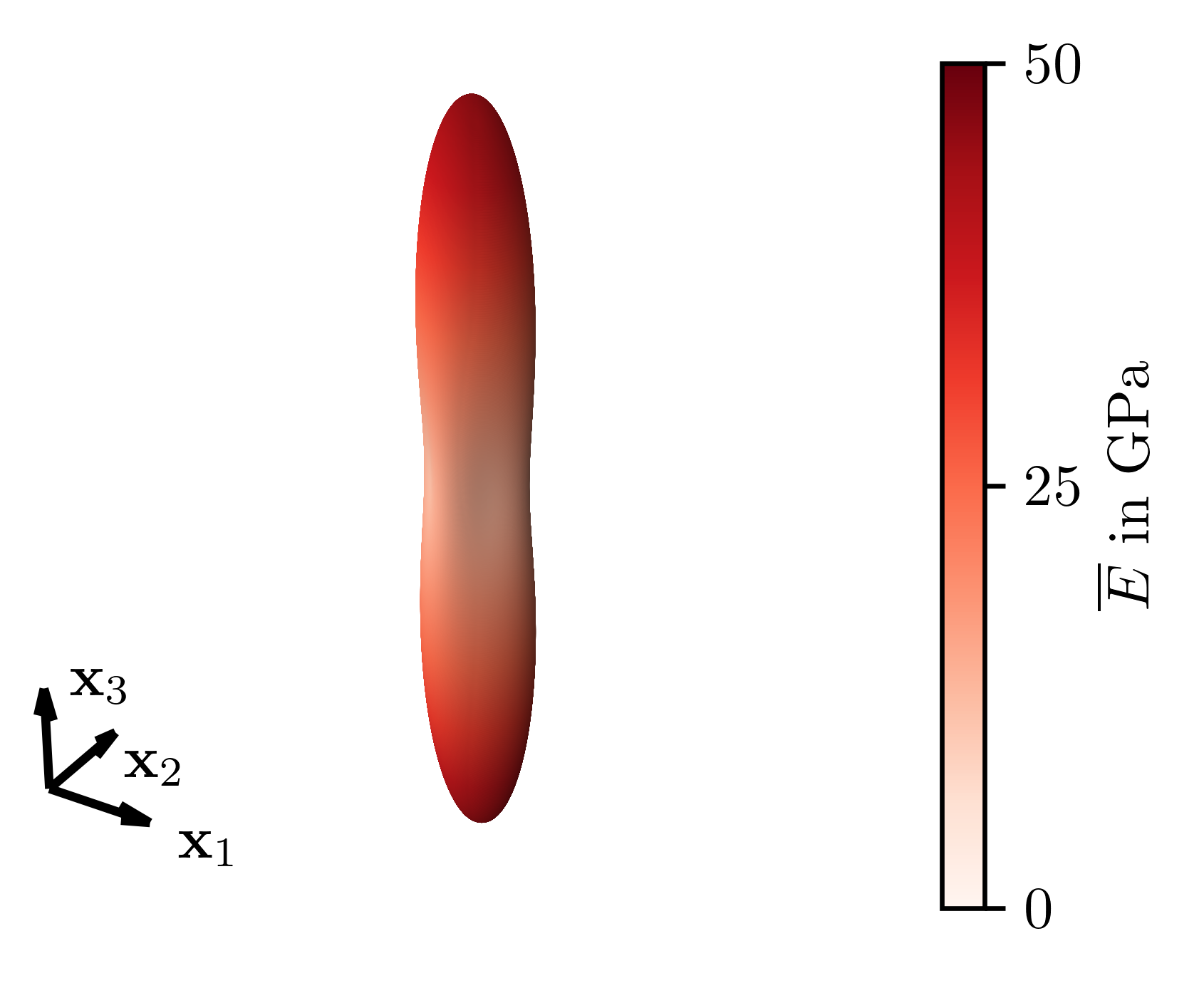}}
	\hfill
	\subfloat[Difference]{\includegraphics[height=0.25\textwidth]{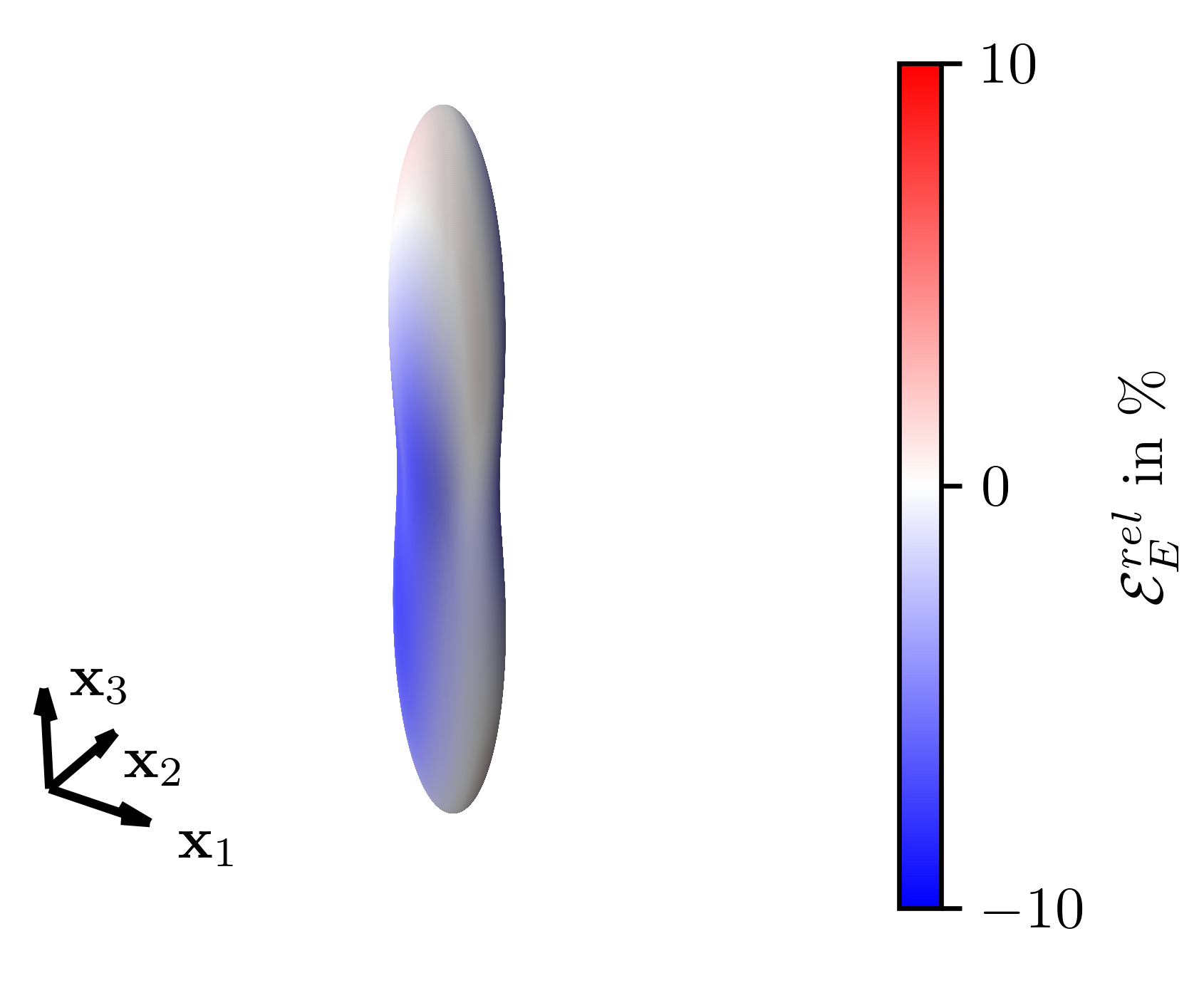}}
	\vfill
	\subfloat[Original lamellar]{\includegraphics[height=0.25\textwidth]{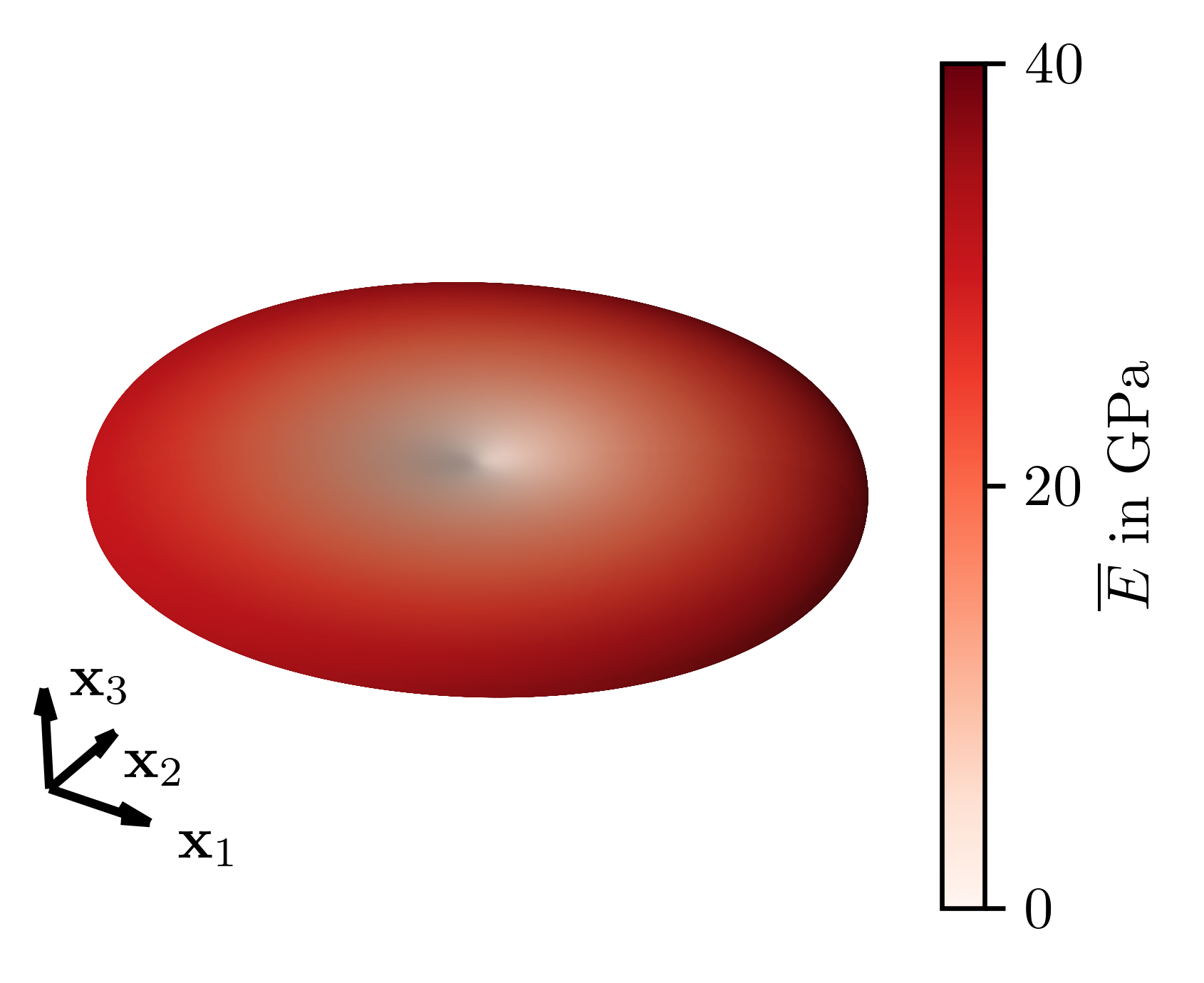}}
	\hfill
	\subfloat[Reconstructed lamellar]{\includegraphics[height=0.25\textwidth]{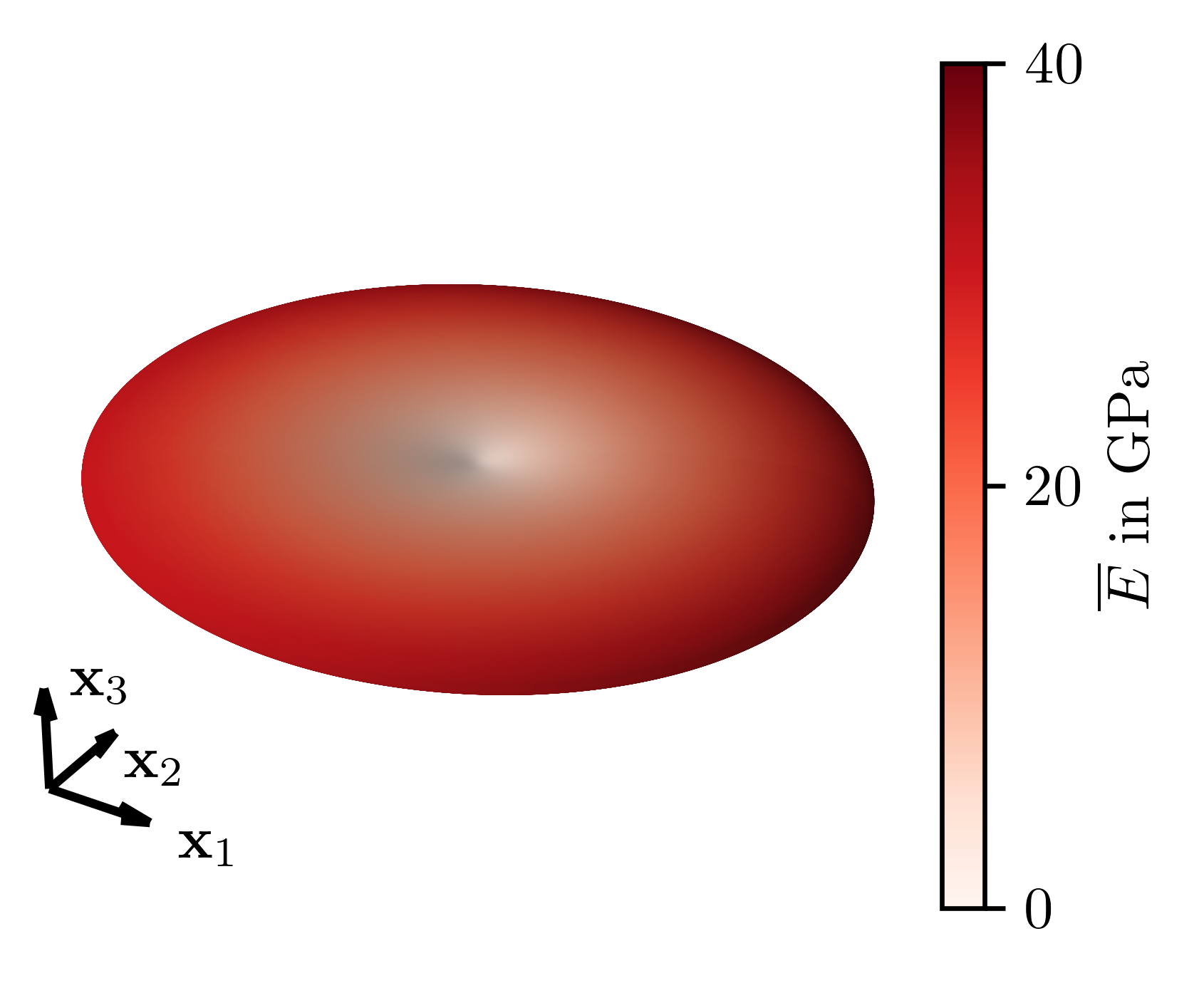}}
	\hfill
	\subfloat[Difference]{\includegraphics[height=0.25\textwidth]{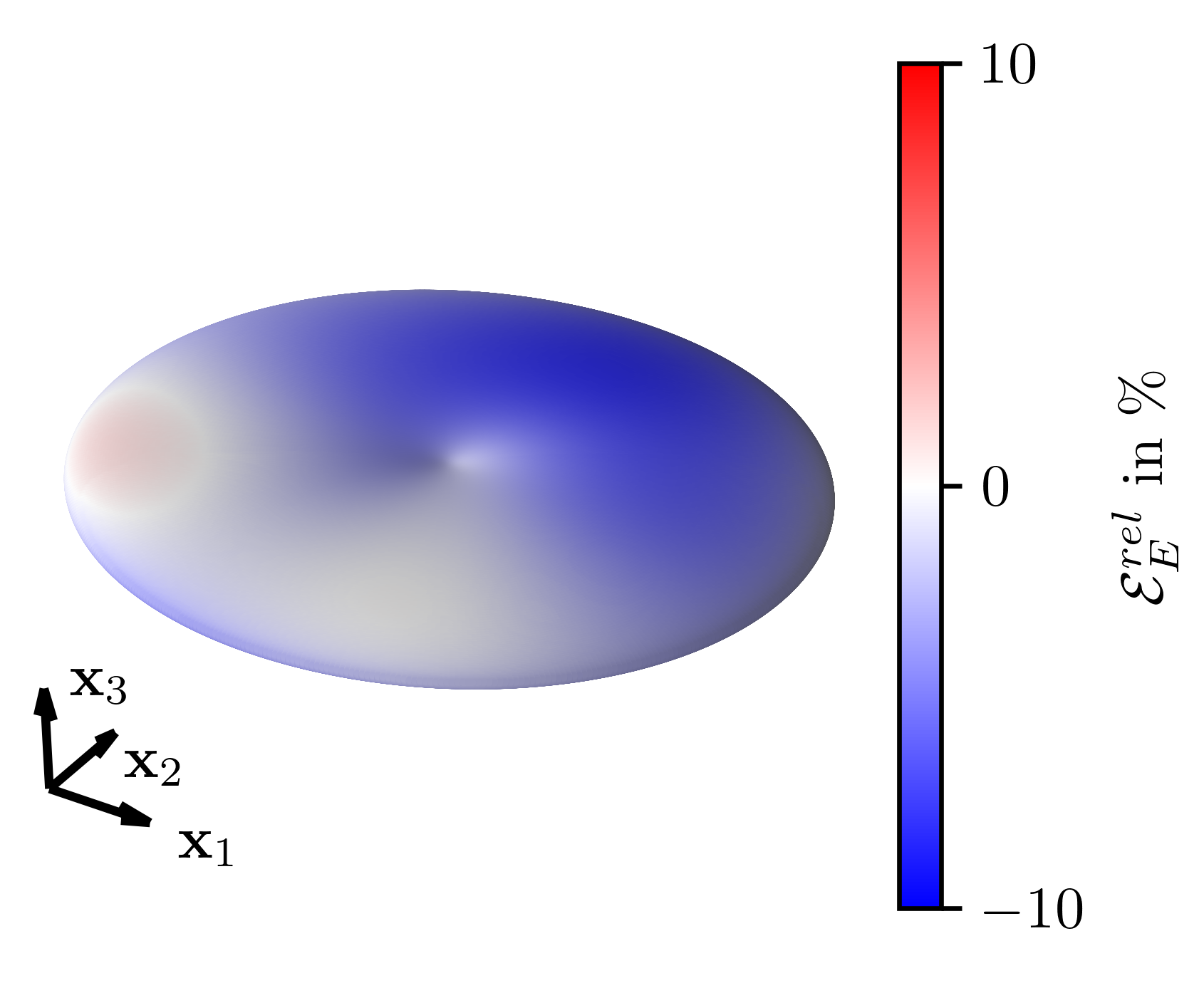}}
	\caption{Elastic surface plots of the original and reconstructed spinodoid structures. \label{fig:stiffnesserrorsspinodal}}
\end{figure}

\begin{figure}[t]
	\centering
	\subfloat[Columnar, $\varphi = 0$]{\includegraphics[height=0.25\textwidth]{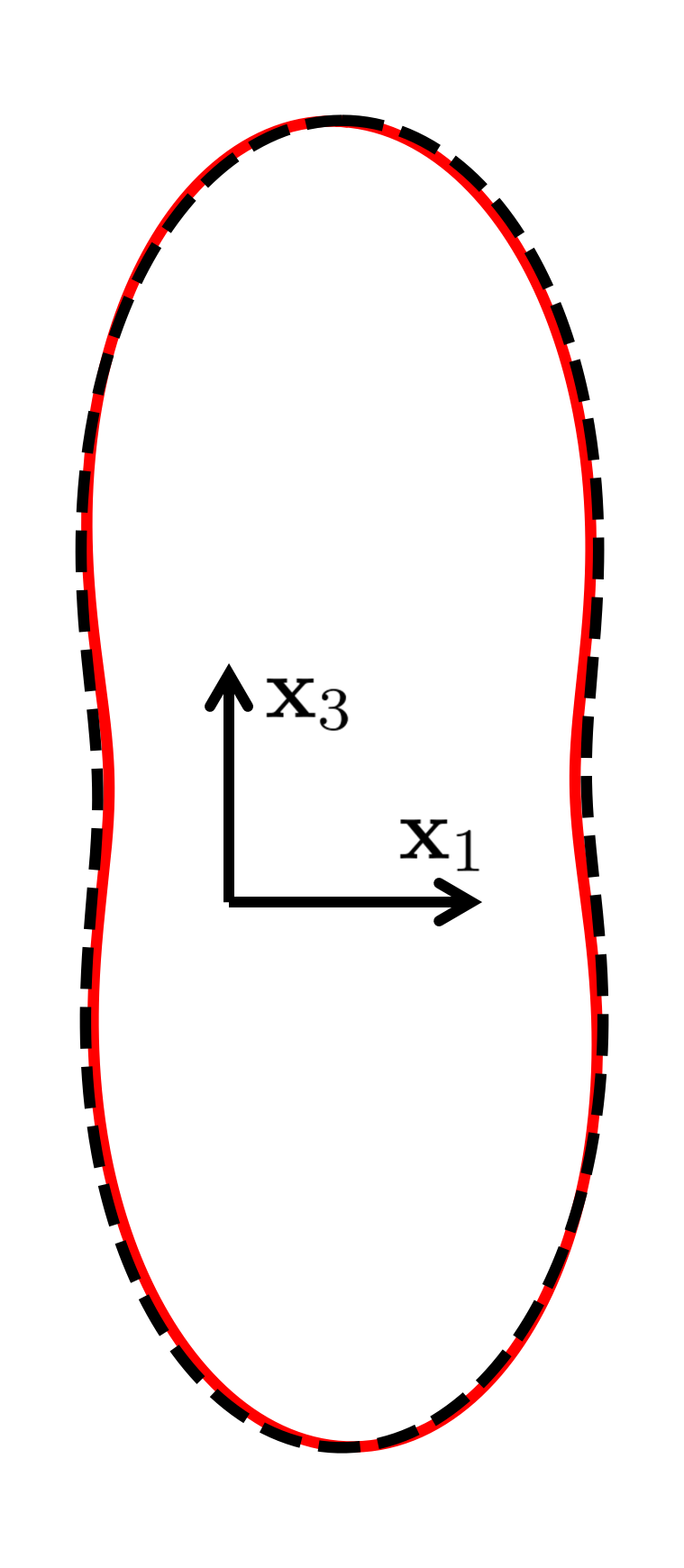}}
	\hspace{0.2cm}
	\subfloat[Columnar, $\varphi = \pi / 2$]{\includegraphics[height=0.25\textwidth]{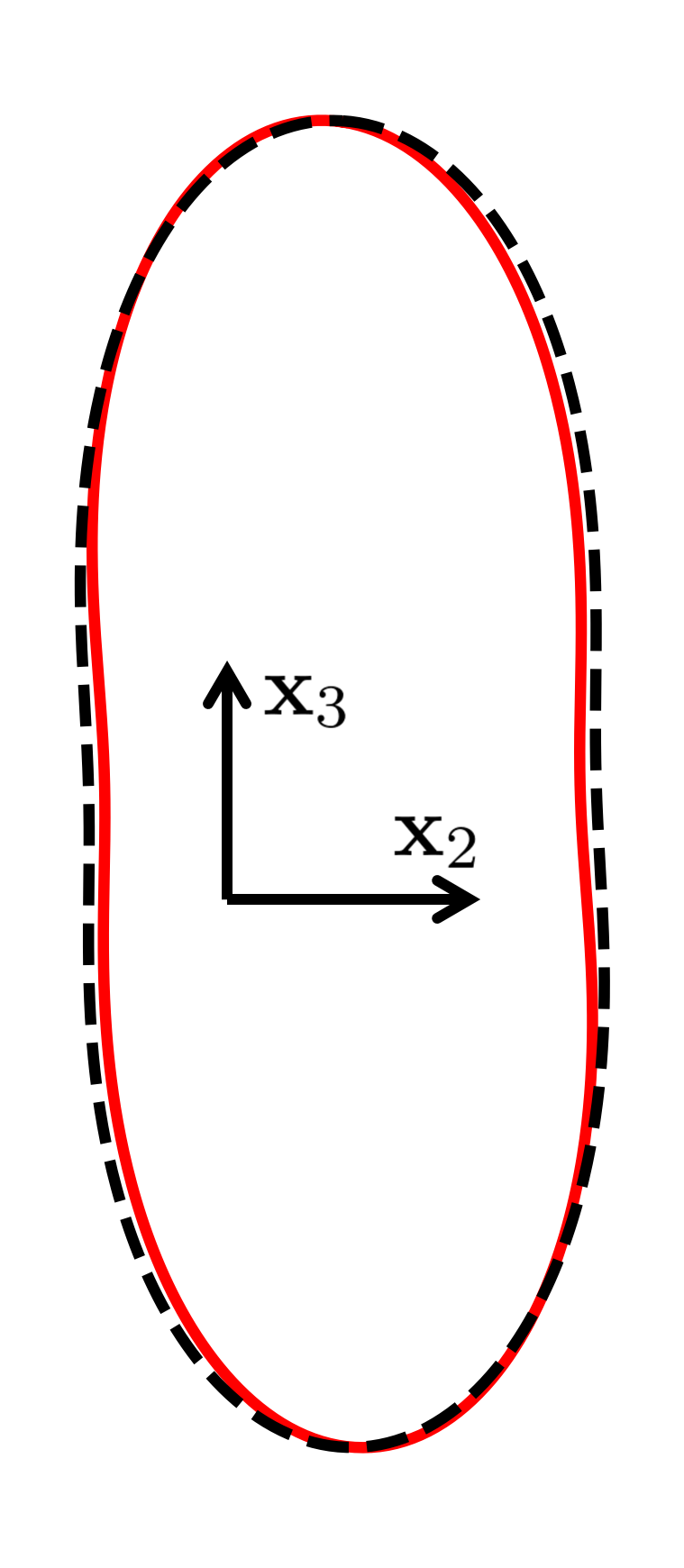}}
	\hspace{0.2cm}
	\subfloat[Lamellar, $\varphi = 0$]{\includegraphics[height=0.2\textwidth]{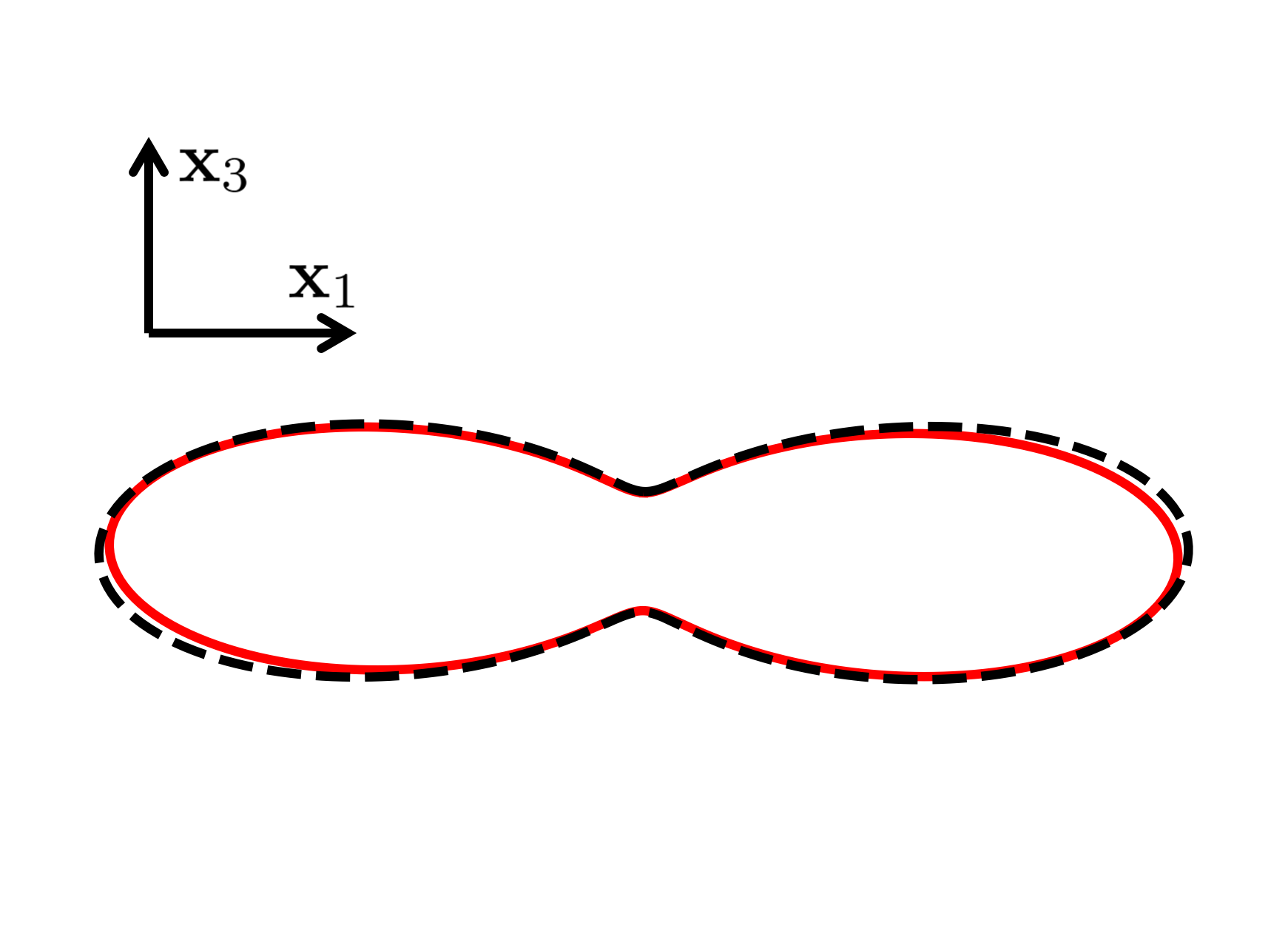}}
	\hspace{0.2cm}
	\subfloat[Lamellar, $\varphi = \pi / 2$]{\includegraphics[height=0.2\textwidth]{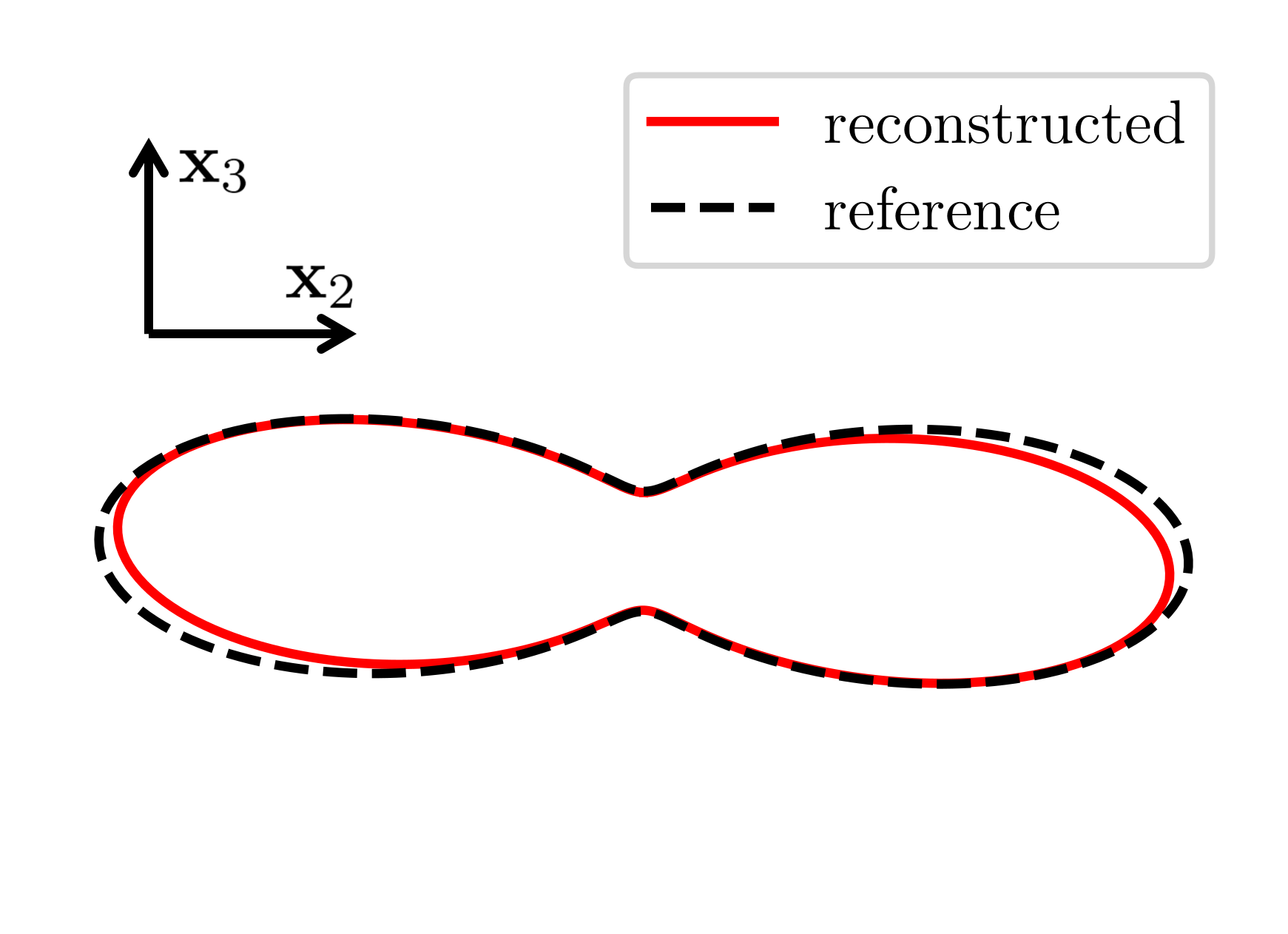}}
	\caption{Cuts through Figure~\ref{fig:stiffnesserrorsspinodal} (c) and (f) at $\varphi = 0$ and $\varphi = \pi / 2$ reveal that a slightly wrong orientation of the elastic surface leads to large deviations in the radial direction. \label{fig:stiffnesserrorsspinodalcut}}
\end{figure}

The elastic surfaces of the original structure in Figure~\ref{fig:ct} and one reconstruction result (first column of Figure~\ref{fig:recexamples}) are compared in Figure~\ref{fig:stiffnesserrorsspinodal}.
As opposed to the Ti-Fe system in Figure~\ref{fig:stiffnesserrors}, the elastic response of the spinodoid materials is highly anisotropic, making the result extremely sensitive to small errors in the reconstruction process.
Nevertheless, the "rod-like" and "disc-like" shape of the elastic surfaces as well as their maximum radius are captured very well.
The largest errors do no occur in the stiffest direction, but at an angle close to the stiffest direction.
The reason for these errors is visualized by means of cuts through the elastic surfaces in Figure~\ref{fig:stiffnesserrorsspinodalcut}: 
Due to the very steep gradient of the elastic surfaces with respect to the load orientation, even small deviations in the orientation lead to large errors in the effective stiffness.
The same applies to a slightly "thinner rod" in the columnar case or a slightly "flatter disk" in the lamellar case.

\begin{figure}[t]
	\centering
	\subfloat[Columnar structure]{\includegraphics[width=0.3\textwidth]{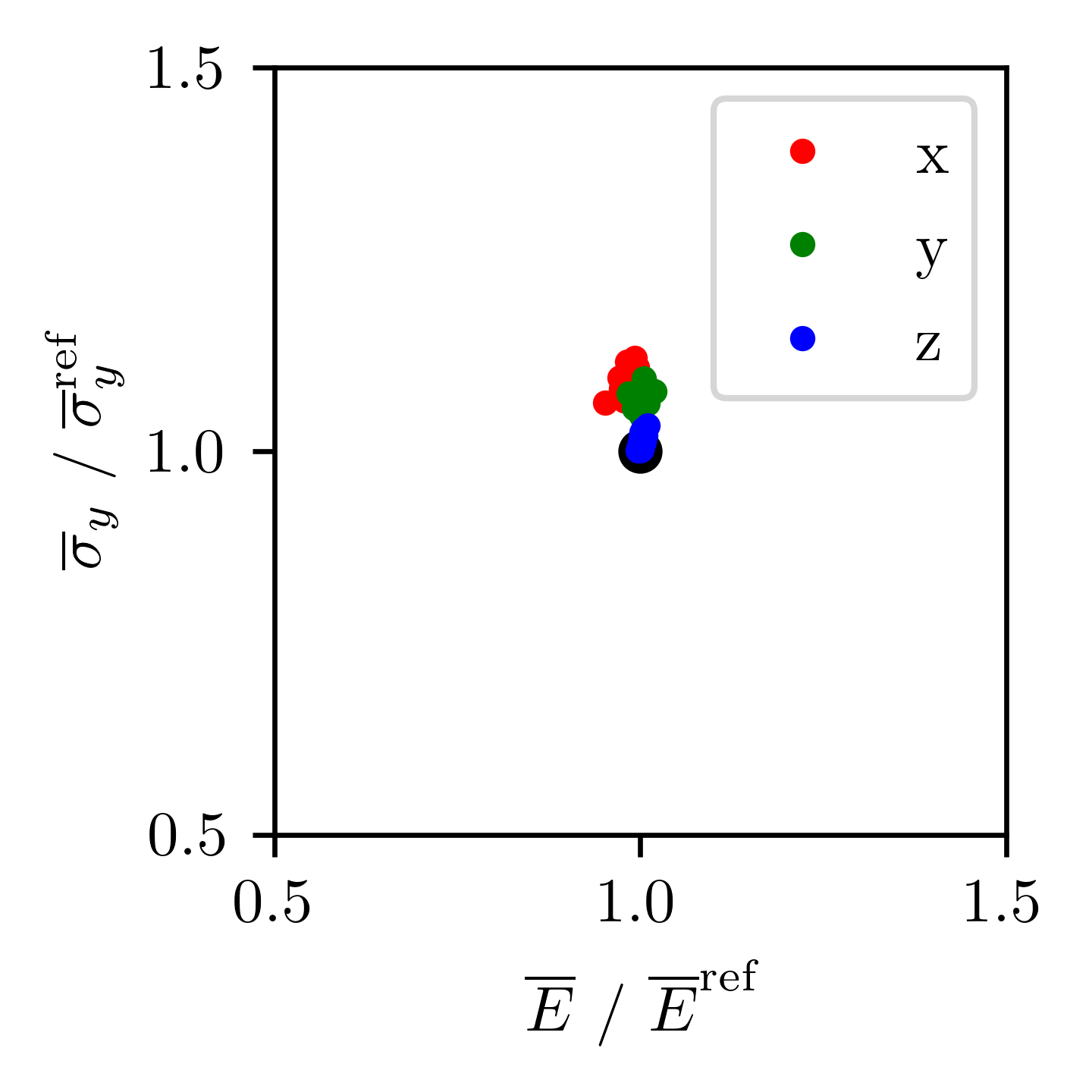}}
	\hspace{1cm}
	\subfloat[Lamellar structure]{\includegraphics[width=0.3\textwidth]{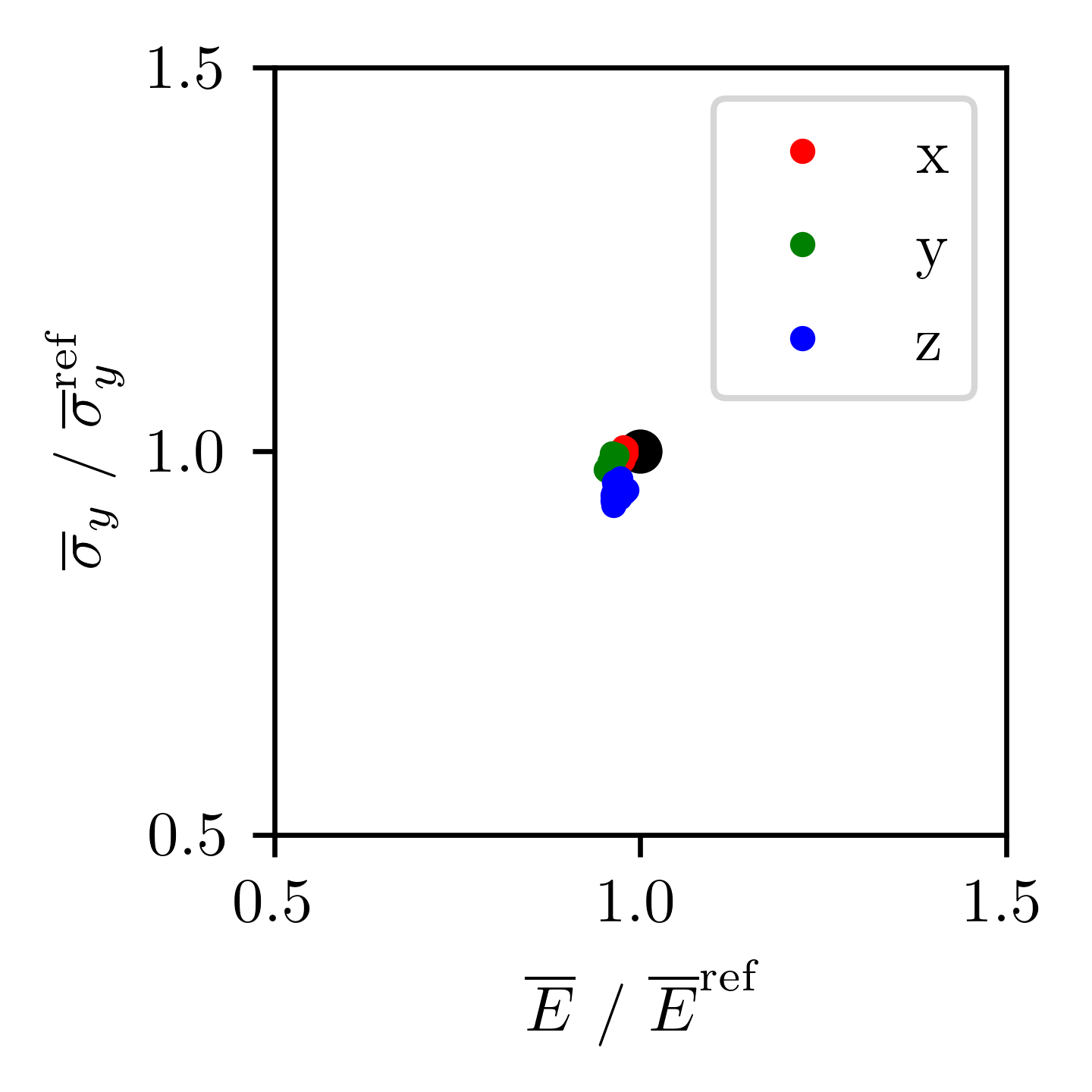}}
	\caption{Relative error of the effective directional Young's modulus and yield strength for the spinodoid structures. The zoom as in Figure~\ref{fig:results_pl} is omitted to avoid redundancy. Note that all values are divided by the reference, but the reference is different in each direction. The numerical values are given in Table~\ref{tab:propertiesspinodal}.\label{fig:alldirectionsspinodal}}
\end{figure}

A quantitative comparison of the effective Young's modulus and yield stress is given in Table~\ref{tab:propertiesspinodal} and Figure~\ref{fig:alldirectionsspinodal}.
It can be seen that in~$x_1$-, $x_2$- and $x_3$-direction, $\mathcal{E}_E$ stays within $\pm 5 \%$ and the anisotropy is captured very well.
As a comparison, for the same material properties, simple homogenization by Voigt's and Reuss' formulae yield an upper and lower bound of $80$ GPa and $0.4$ GPa\footnote{It should be mentioned that this lower bound is a purely numerical value, since the Fourier-based solver used in this work requires a non-zero stiffness to be assigned to the void phase as shown in Table~\ref{tab:materialparameters}. The correct lower bound by Reuss' formula is of course $0$ GPa.}, respectively, irrespective of loading direction.
\begin{table*}[h]
	\centering
	\caption{Errors of the effective directional Young's modulus and yield strength for the spinodoid structures.}
	\label{tab:propertiesspinodal}
	\begin{tabular}{l  c  c  c  c  c  c  c }
\toprule
& Direction & $\bar E^\text{ref}$ in GPa & $\bar E^\text{rec}$ in GPa & $\mathcal{E}_\text{E}$ in \% & $\bar \sigma_\text{y}^\text{ref}$ in MPa & $\bar \sigma_\text{y}^\text{rec}$ in MPa & $\mathcal{E}_{\sigma_\text{y}}$ in \%\\ 
\midrule
\multirow{ 3}{*}{\begin{turn}{90} col. \end{turn} } 
& x & 15.53 & 15.3 $\pm$ 0.15 & -1.7 & 111.8 & 122 $\pm$ 1.8 & 9.3\\ 
& y & 15.56 & 15.6 $\pm$ 0.12 & 0.23 & 113.4 & 121 $\pm$ 1.7 & 6.9\\
& z & 43.03 & 43.1 $\pm$ 0.14 & 0.27 & 331.8 & 337 $\pm$ 3.1 & 1.5\\ 
\midrule
\multirow{ 3}{*}{\begin{turn}{90} lam. \end{turn} } 
& x & 32.06 & 31.2 $\pm$ 0.12 & -2.7 & 257.8 & 256  $\pm$ 1.4 & -0.6\\
& y & 32.05 & 30.8 $\pm$ 0.11 & -4.0 & 259.0 & 255  $\pm$ 1.5 & -1.6\\ 
& z & 3.55  & 3.4  $\pm$ 0.02 & -3.1 & 45.99 & 43.5 $\pm$ 0.4 & -5.4\\
\bottomrule
	\end{tabular}
\end{table*}

In summary, the extreme phase contrast and strong anisotropy of the "bone-like" spinodoid structures naturally lead to larger errors compared to the real CT scan of the Ti-Fe system.
Despite being synthetically constructed, the columnar and lamellar structure thus demonstrate the robustness of the reconstruction procedure.

\section{Conclusions and Outlook}
\label{sec:summary}
In this work, a two-stage optimization procedure is proposed to reconstruct 3D microstructures from 2D slices based on statistical descriptors.

As a \emph{first} step, the recently developed differentiable microstructure characterization and reconstruction (DMCR) framework~\cite{seibert_reconstructing_2021-1,seibert_microstructure_2022} is utilized to quickly obtain a solution that minimizes the sum of descriptor errors over all microstructure slices.
While this approach converges quickly and scales favorably, two problems remain: 
The given volume fractions are not guaranteed to be fulfilled after the real-valued reconstruction result~$\textbf{M}^\text{rec} \in \mathcal{M}^\text{3D}$ is projected to the integer-valued~$\hat{\textbf{M}} \in \mathcal{\hat{M}}^\text{3D}$ by element-wise rounding.
Furthermore, despite employing the variation descriptor for noise reduction as highlighted in~\cite{bostanabad_reconstruction_2020,seibert_descriptor-based_2022}, the remaining amount of noise cannot be neglected, motivating a tailored post-processing procedure.

As a \emph{second} step, the microstructure is iteratively adjusted to the correct volume fractions and smoothed, whereby a custom algorithm distinguishes between sharp microstructural features and spurious noise by considering the previously computed microstructure descriptors.
Starting from the Yeong-Torquato algorithm, a custom different-phase-neighbor sampling distribution and a zero-tolerance acceptance criterion are defined.
These measures are implemented in order to account for the fact that the input data can be expected to already exhibit the desired morphology.
Thus, the two-stage procedure makes use of the gradient-based optimization in DMCR while also employing microstructure descriptors to eliminate noise without degrading the microstructure quality. 

A strong focus is placed on validating the 2D-to-3D reconstruction in terms of morphology and effective properties.
For this purpose, a computed tomography (CT) scan of a recently developed $\beta$-Ti/TiFe alloy as well as morphologically extremely anisotropic "bone-like" spinodoid metamaterials are used as 3D reference structures.
From each of these structures, three orthogonal slices are extracted to mimic the situation that only 2D information is available.
Based on \emph{only} these slices, statistical descriptors are computed and used for reconstructing 20 independent 3D realizations of each structure.
The synthetic structures are visually very similar to their original counterpart and an in-depth quantitative analysis is carried out.

A detailed discussion of errors in the descriptor space classifies the sources of systematic deviations as descriptor \emph{concentration}, \emph{difference} and \emph{incompatibility}.
While the former is inherent to the formulation of the optimization problem and might be addressed in the future by formulating advanced cost functions, the latter two are expected to vanish as the structure for characterization becomes infinitely large.
All phenomena are visualized by means of a scatter plot in the first two principal components of the descriptor space and the full 3D correlation errors are given for reference.

While there is no easy criterion to determine which magnitude of descriptor errors is acceptable, the effective properties are a practical indicator of whether the reconstructed structures match the reference.
For this reason, elasto-plastic simulations are performed and a numerical homogenization is carried out in order to determine the full stiffness tensor as well as the effective yield strength in three directions.
The errors of the real Ti-Fe system are extremely small.
For the synthetic spinodoid structures, the extreme anisotropy and high phase contrast naturally lead to higher deviations, however, a good prediction quality is reached and the anisotropy is captured very well.

In summary, the utility of the presented approach in reconstructing realistic 3D microscale domains from 2D slices is confirmed.
This is an essential step towards making modern multiscale simulations more applicable to computational materials engineering.
In this context, it is worth noting that all algorithms will be made freely available in the open-source software \emph{MCRpy}.
Thus, the excellent results in this study motivate further initiatives across disciplines in applying numerical multiscale simulations for computational materials engineering. \\

\section*{Declaration of competing interest}
The authors declare no competing interest.

\section*{Acknowledgements}
The authors thank Tobias Strohmann for his valuable input and for establishing the cooperation between the German Aerospace Center (DLR) and Technische Universität Dresden (TUD) on this subject.
The authors appreciate the European Synchroton (ESRF), especially Da Silva and Cloetens at Beamline \emph{ID16A} for their contributions.
Furthermore, the authors thank Michał Basista and his colleagues at the Polish Academy of Sciences for providing data, which was unfortunately not suitable for this publication due to a volume fraction gradient.
In this context, the authors also thank Ulrike Gebhardt at Technische Universität Dresden~(TUD) for segmenting the same data.
Finally, we thank Jörg Brummund~(TUD) for the discussions and support regarding material modeling.

The group of M. Kästner thanks the German Research Foundation DFG which supported this work under Grant number KA 3309/18-1.
The authors are grateful to the Centre for Information Services and High Performance Computing [Zentrum für Informationsdienste und Hochleistungsrechnen (ZIH)] TU Dresden for providing its facilities for high throughput calculations.

\printcredits

\appendix

\section{Descriptor-based smoothing}
\label{sec:descriptorbasedsmoothing}
\begin{figure}[t]
	\centering
	\subfloat[Original structure]{\includegraphics[width=0.23\textwidth]{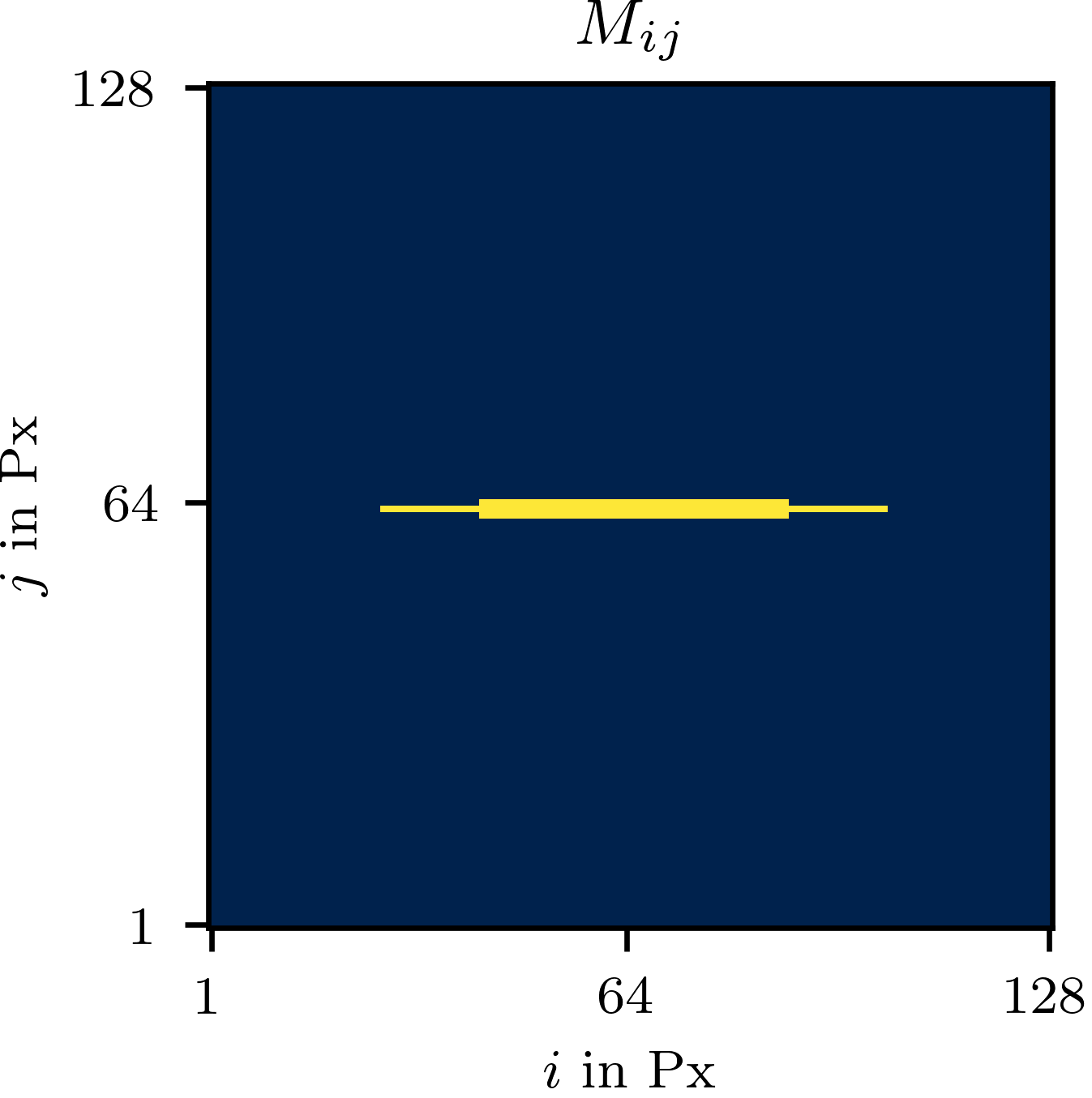}}
	\hfill
	\subfloat[Noisy version of (a)]{\includegraphics[width=0.23\textwidth]{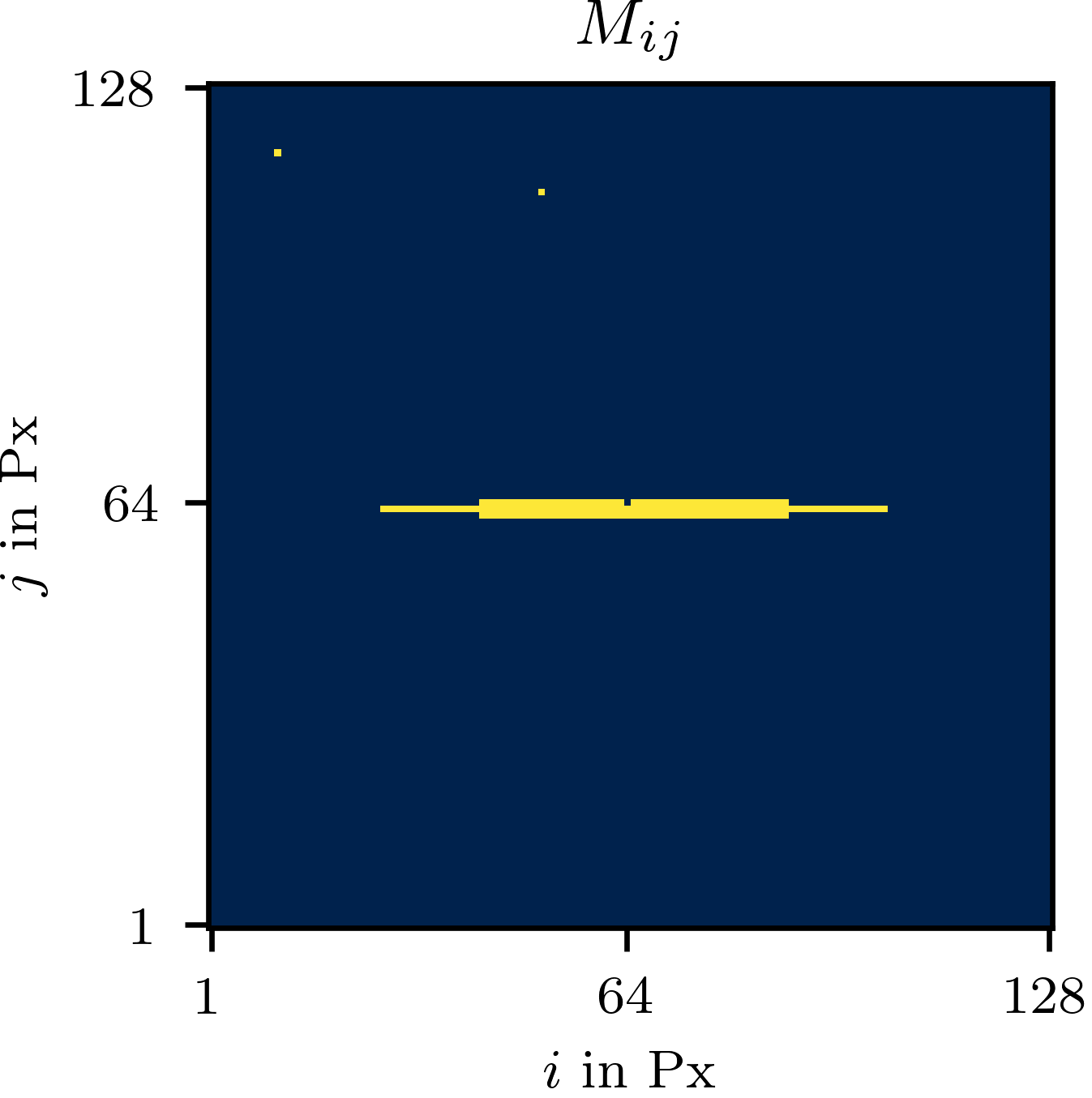}}
	\hfill
	\subfloat[Standard Gaussian smoothing and thresholding of (b)]{\includegraphics[width=0.23\textwidth]{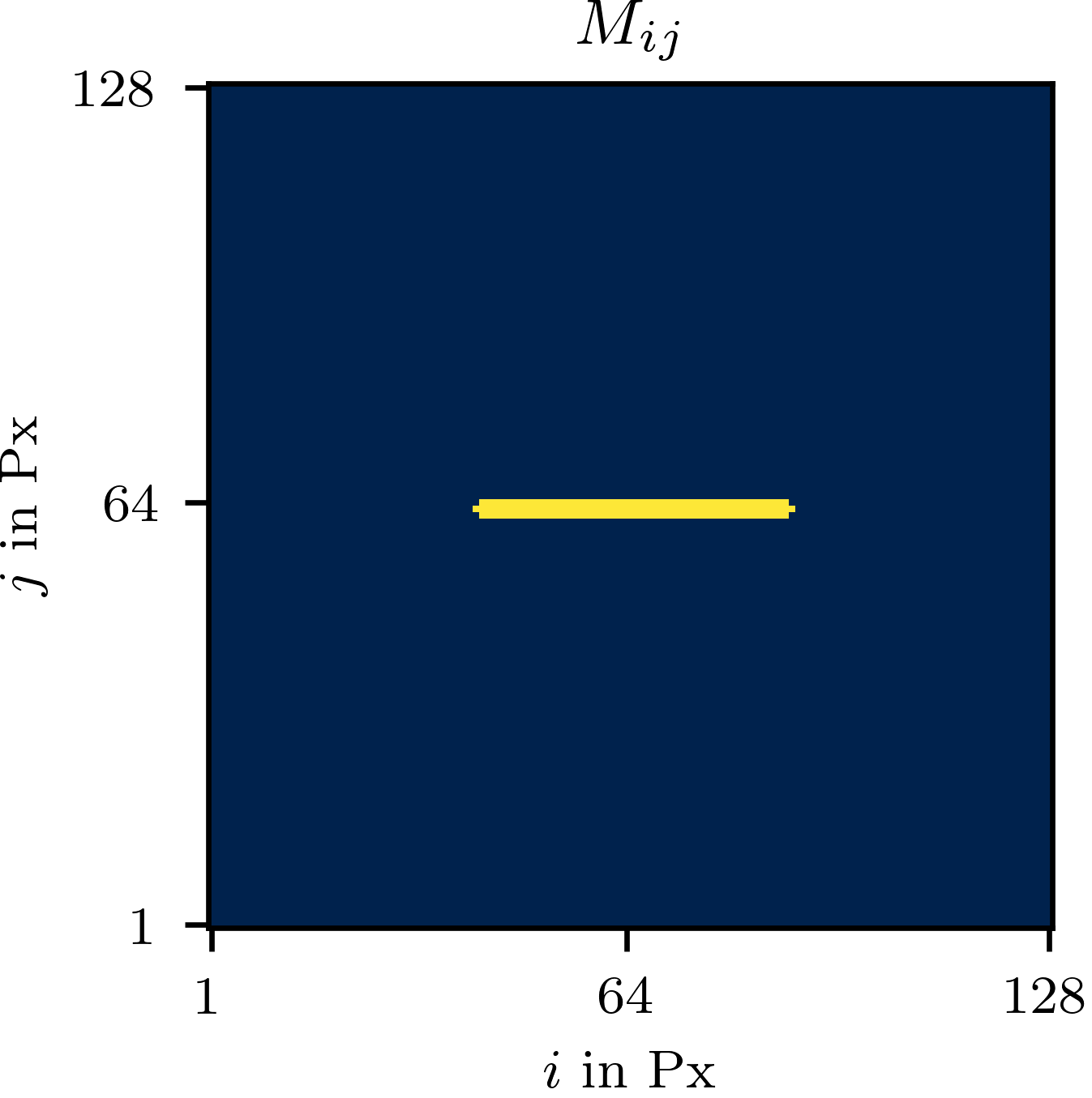}}
	\hfill
	\subfloat[Suggested descriptor-based smoothing of (b)]{\includegraphics[width=0.23\textwidth]{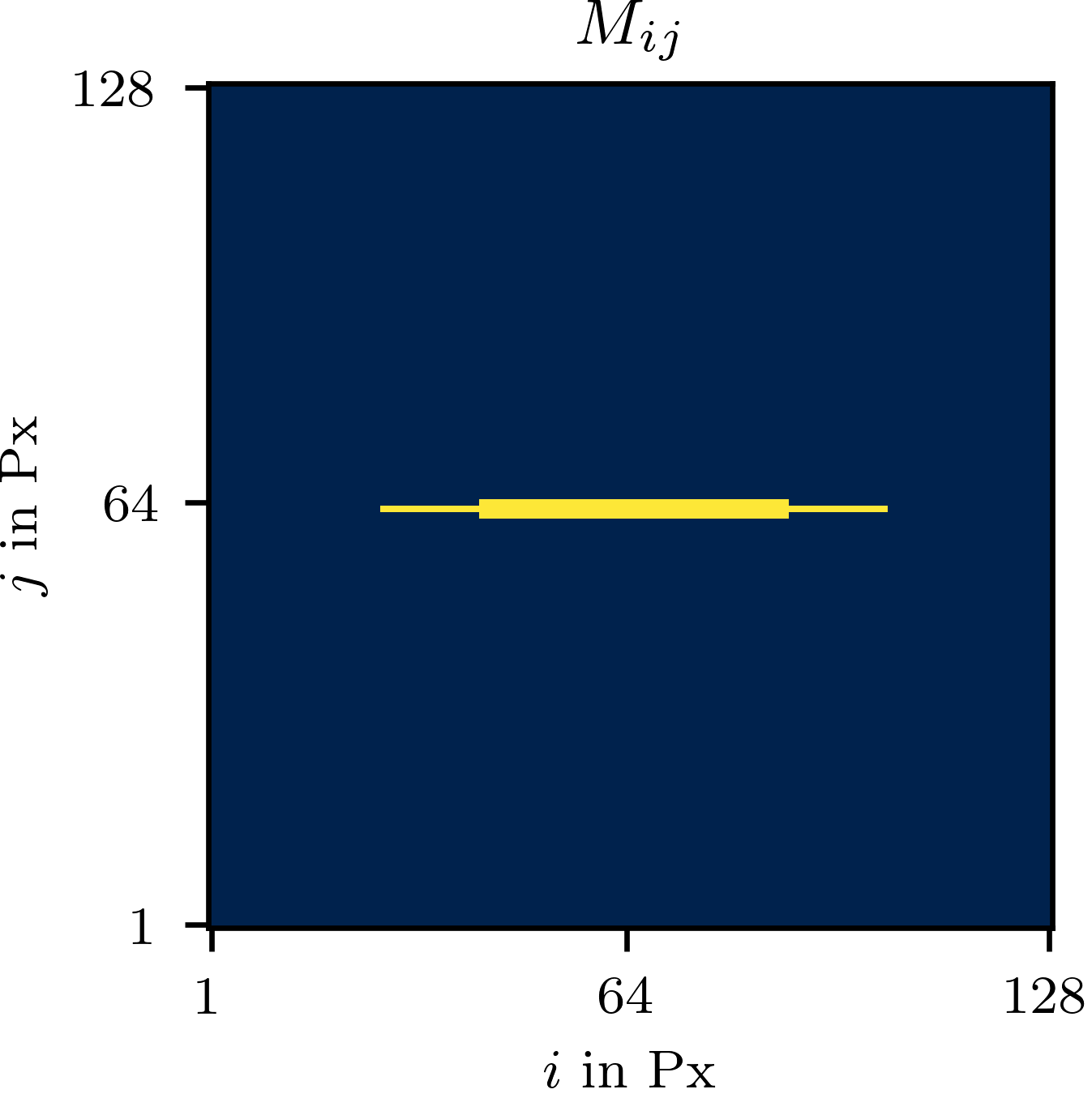}}
	\caption{Motivation of descriptor-based smoothing: Sharp corners in the original structure (a) cannot be distinguished from noise (b) by a simple Gaussian smoothing and significant errors occur (c). In contrast, the descriptor-based algorithm recovers the original structure (d).\label{fig:descriptorbasedsmoothing}}
\end{figure}

The difference between the descriptor-based smoothing suggested in Section~\ref{sec:methods} and simple Gaussian smoothing followed by thresholding is not very pronounced in the investigated structures.
However, this is not always the case.
In this section, a simple academic example is constructed to illustrate the differences between the methods.
Consider a structure with a horizontal slit as shown in Figure~\ref{fig:descriptorbasedsmoothing} (a), which is to be recovered from a noisy version in Figure~\ref{fig:descriptorbasedsmoothing} (b).
This noisy version could, for example, stem from a reconstruction algorithm.
As can be seen in Figure~\ref{fig:descriptorbasedsmoothing} (c), simple Gaussian smoothing followed by thresholding fails to differentiate between noisy pixels and true sharp corners, hence the smoothing significantly alters the microstructure and its effective properties.
In contrast, the suggested descriptor-based smoothing procedure exactly recovers the original structure in very few iterations\footnote{The number of iterations varies but is mostly less than 10.}.
In conclusion, although simple smoothing procedures \emph{can} work for real microstructures, there is no warranty that they do.
It is therefore recommended to use descriptor-based smoothing procedures whenever possible, despite the higher computational cost.

\section{Size of original slice}
\label{sec:appendixslicesize}
\begin{figure}[t]
	\centering
	\subfloat[Columnar structure]{\includegraphics[width=0.3\textwidth]{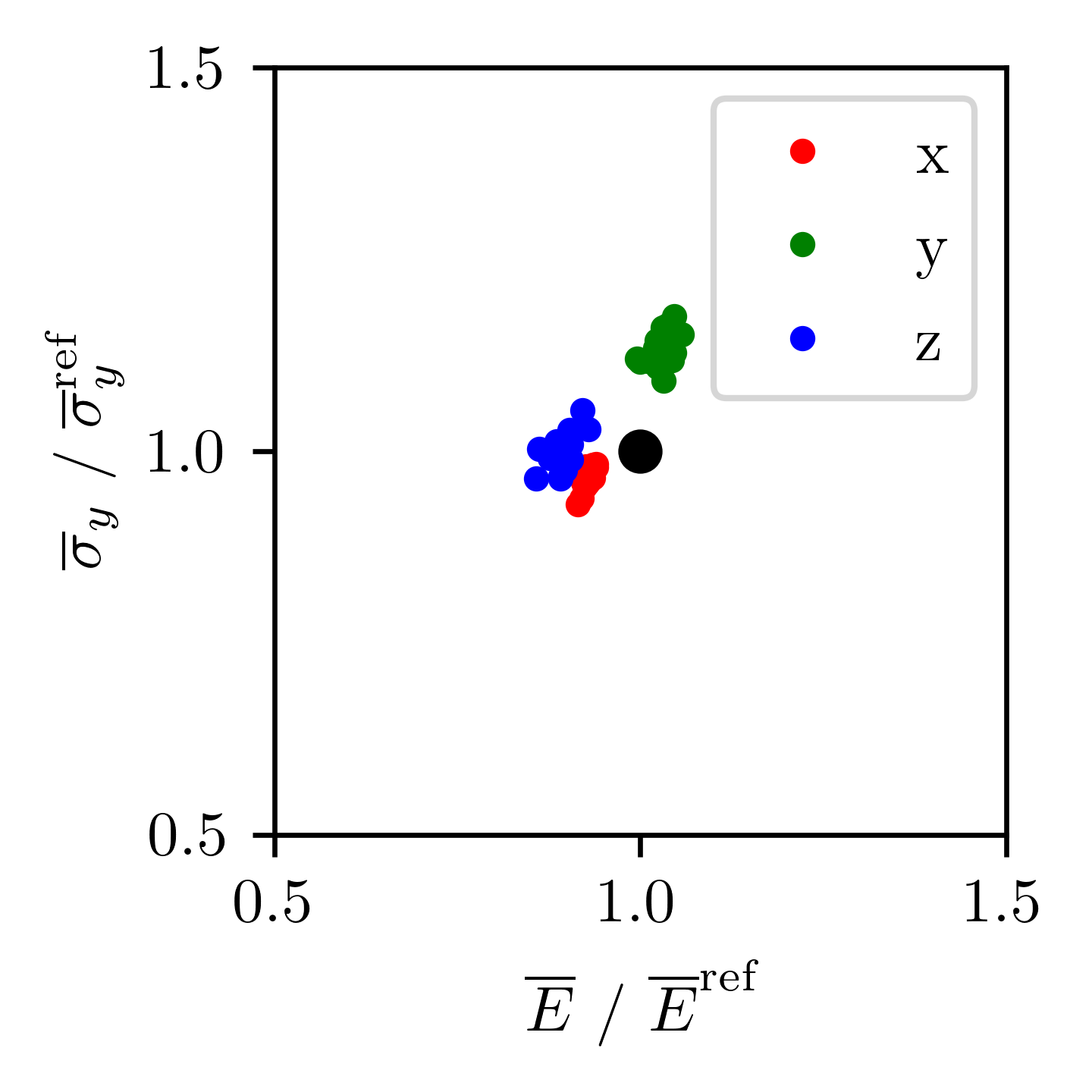}}
	\hspace{1cm}
	\subfloat[Lamellar structure]{\includegraphics[width=0.3\textwidth]{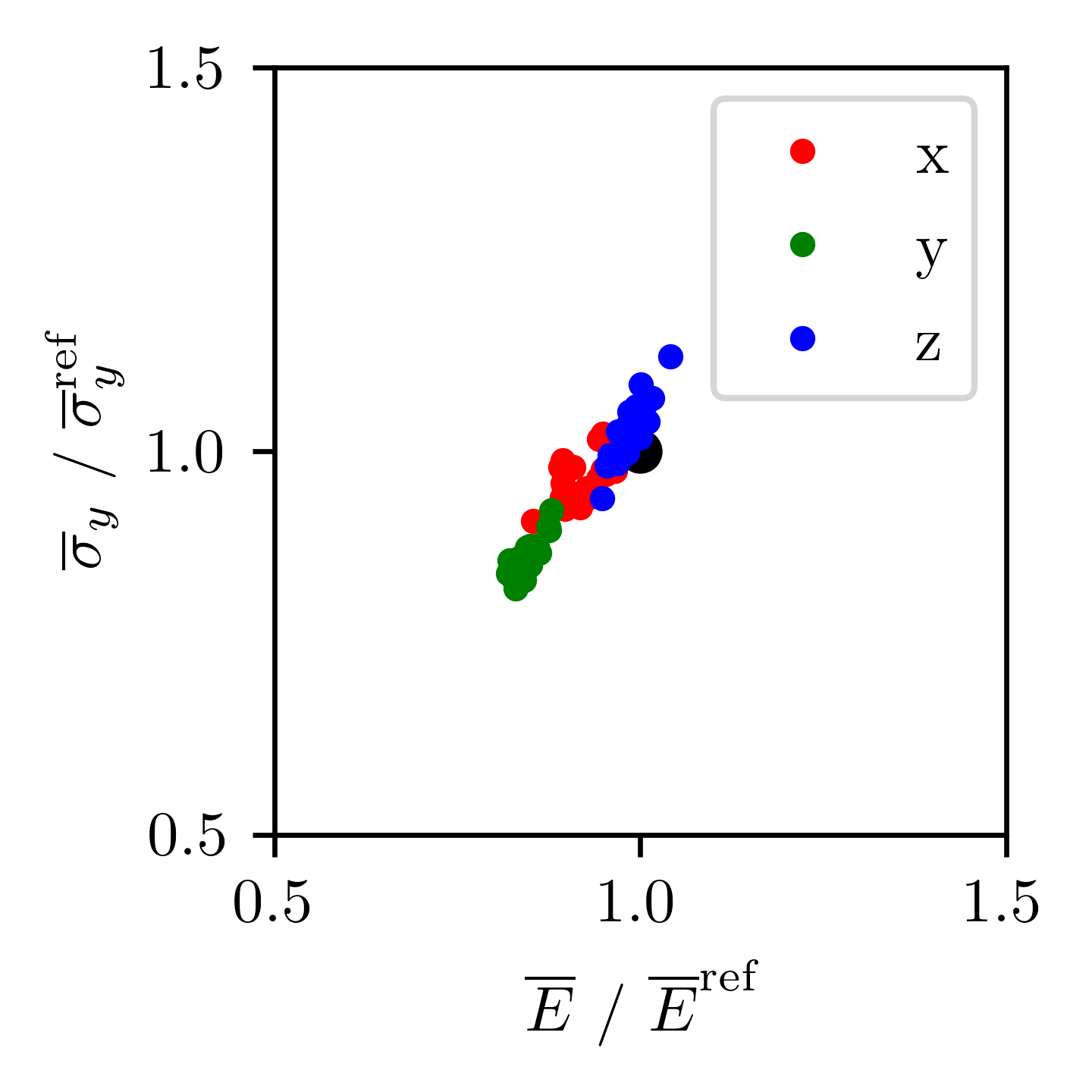}}
	\caption{Relative error of the effective directional Young's modulus and yield strength for the spinodoid structures when using an insufficiently-sized structure for computing the desired descriptor. Compared to Figure~\ref{fig:alldirectionsspinodal}, a higher scatter is observed and the error is larger in general.\label{fig:alldirectionsspinodalsmall}}
\end{figure}

If the original 2D microscopy image or slice for the computation of the desired descriptor in Eq.~(\ref{eqn:optgeneral}) is small, then, as discussed in Section~\ref{sec:resultsdescriptors}, the descriptor is most likely not representative for the entire structure. 
For example, in a matrix structure with inclusions, as the sample size decreases, if becomes increasingly likely that a random 2D slice does not cut a single inclusion.
In this extreme case, the result of the characterization would be an inclusion volume fraction of $0 \%$.
To demonstrate this effect, a smaller cuboid with $128^3$ voxels is cut out of the columnar and lamellar spinodoid structures and used for validation in analogy to Figure~\ref{fig:schema}.
Figure~\ref{fig:alldirectionsspinodalsmall} summarizes the error in the effective elastic and plastic properties.
In direct comparison to the results from the full original structure in Figure~\ref{fig:alldirectionsspinodal}, a significantly increased scatter can be observed as well as a larger error in general.
In theory, for increasing the representativeness of a descriptor, averaging descriptors over multiple slices of a small sample should have the same effect as using a single slice of a larger sample.
This is, however, less relevant in practice, since as discussed in the introduction, 3D data is often significantly more cost- and time-intensive to obtain than 2D data.
Nevertheless, the authors averaged the descriptor over all slices from the smaller version of the original structure and used this value for an additional reconstruction and homogenization.
To keep the computational effort for this a simple experiment low, only a single example was reconstructed in this manner.
Indeed, averaging descriptors over multiple small slices has a similar effect as using a single large slice as can be seen in Table~\ref{tab:results_largevolumevsaveraging}.
This shows the possibility to use multiple 2D microscopy images if this is of interest.
\begin{table*}[h]
	\centering
	\caption{The effect of averaging descriptors over multiple slices on the effective Young's modulus under tension in $\boldsymbol{x}$-direction compared to a larger single slice. The $\boldsymbol{y}$- and $\boldsymbol{z}$-direction are similar}
	\label{tab:results_largevolumevsaveraging}
	\begin{tabular}{c | c  c }
\toprule
$\mathcal{E}_{\overline{E}_x}$ & Columnar & Lamellar \\ 
\midrule
Small sample, single slice & $-7.8 \%$ & $-11 \%$\\ 
Small sample, averaged slices & $-0.5 \%$ & $-8.4 \%$\\ 
Large sample, single slice & $-2.1 \%$ & $-2.0 \%$\\ 
\bottomrule
	\end{tabular}
\end{table*}

\section{Effect of smoothing}
\label{sec:appendixsmoothingeffect}
The impact of smoothing on the effective properties is outlined in Table~\ref{tab:results_effectofsmoothing}, where the unsmoothed and smoothed version of the same reconstructed structure are compared regarding their effective properties.
In general, a lower stiffness is observed for unsmoothed structures.
This can be explained as follows.
Consider a connected domain of the stiffer phase.
Random noise has the effect of erroneously assigning the weaker phase to some voxels within that domain, thus weakening its effective stiffness.
On the other side, in a connected domain of the more compliant phase, single voxels with higher stiffness do not contribute much to the overall behavior as long as they are not connected.
Similarly, spurious noise acts as a stress concentrator and thus promotes strain localization at lower stress levels. 
Hence, it is plausible that smoothing increases the effective yield strength.
An even larger impact is expected to occur if fatigue indicator parameters are to be computed as in~\cite{rasloff_accessing_2021}, which further underlines the need for microstructure post-processing.
\begin{table*}[h]
	\centering
	\caption{Effect of smoothing on the effective Young's modulus and shear stress for the spinodoid structures.}
	\label{tab:results_effectofsmoothing}
	\begin{tabular}{c | c  c  c  c}
\toprule
& \multicolumn{2}{c}{ $\mathcal{E}_{\overline{E}}$} & \multicolumn{2}{c}{$\mathcal{E}_{\overline{\sigma}_y}$} \\ 
& unsmoothed & smoothed & unsmoothed & smoothed \\
\midrule
Columnar & $-1.3 \%$ & $0.1 \%$ & $-0.9 \%$ & $0.5 \%$\\ 
Lamellar & $-4.1 \%$ & $-2.0 \%$ & $-1.9 \%$ & $-0.3 \%$\\ 
\bottomrule
	\end{tabular}
\end{table*}

\bibliographystyle{cas-model2-names}

\bibliography{main}

\end{document}